\documentclass[longauth]{aa}
\usepackage{graphicx}
\usepackage{xurl}
\usepackage{hyperref}
\usepackage{amsmath}
\usepackage{xcolor}
\usepackage{lscape}

\UseRawInputEncoding

\usepackage{txfonts}
\begin{document} 
\newcommand{\bhb}{[BHB2007] 11~}
\newcommand{\chtoh}{CH$_3$OH~}

   \title{Hot methanol in the \bhb protobinary system: hot corino versus shock origin? : FAUST V}

   \author{C. Vastel \inst{1}
	\and
          F. Alves \inst{2}
	\and 
	  C. Ceccarelli \inst{3}
	\and
	  M. Bouvier \inst{3}
	\and
	  I. Jim\'enez-Serra \inst{4}
	\and
	  T. Sakai \inst{5}
	\and
	  P. Caselli \inst{2}
	\and
	  L. Evans \inst{1,6}
	\and
	  F. Fontani\inst{6}
	\and
	  R. Le Gal \inst{3,12}
	\and
	  C. J. Chandler \inst{7}
	\and
	  B. Svoboda \inst{7}
	\and
	  L. Maud \inst{8}
	\and
	  C. Codella \inst{3,6}	
	\and
	  N. Sakai \inst{9}
	\and
	  A. L\'{o}pez-Sepulcre\inst{3,12}
	\and
	  G. Moellenbrock \inst{7}
	\and
	  Y. Aikawa\inst{13}
	\and
	  N. Balucani\inst{14}
	\and
	  E. Bianchi\inst{3}  
	\and
	  G. Busquet\inst{34}
	\and
	  E. Caux\inst{1}
	\and
	  S. Charnley\inst{15}
	\and
	  N. Cuello\inst{3}
	\and
	  M. De Simone\inst{3}  
	\and
	  F. Dulieu\inst{16}
	\and
	  A. Dur\'{a}n\inst{17}
	\and
	  D. Fedele\inst{6}
	\and
	  S. Feng\inst{18}
	\and
	  L. Francis\inst{20,21}
	\and
	  T. Hama\inst{22}
	\and
	  T. Hanawa\inst{23}
	\and
	  E. Herbst\inst{24}
	\and
	  T. Hirota\inst{19}
	\and
	  M. Imai\inst{10}
	\and
	  A. Isella\inst{25}
	\and
	  D. Johnstone\inst{20,21}
	\and
	  B. Lefloch\inst{3}
	\and
	  L. Loinard\inst{17,26}
	\and
	  M. Maureira\inst{2}
	\and
	  N. M. Murillo\inst{9}
	\and
	  S. Mercimek\inst{6}
	\and
	  S. Mori\inst{34}
	\and
	  F. Menard\inst{3}
	\and
	  A. Miotello\inst{8}
	\and
	  R. Nakatani\inst{9}
	\and
	  H. Nomura\inst{19}
	\and
	  Y. Oba\inst{27}
	\and
	  S. Ohashi\inst{9}
	\and
	  Y. Okoda\inst{10}
	\and
	  J. Ospina-Zamudio\inst{3}
	\and
	  Y. Oya\inst{10,11}
	\and
	  J. E. Pineda\inst{2}
	\and
	  L. Podio\inst{6}
	\and
	  A. Rimola\inst{28}
	\and
	  D. Segura Cox\inst{2}
	\and
	  Y. Shirley\inst{29}
	\and
	  L. Testi\inst{6,8}
	\and
	  S. Viti\inst{30,31}
	\and
	  N. Watanabe\inst{27}
	\and
	  Y. Watanabe\inst{32}
	\and
	  A. Witzel\inst{3}
	\and
	  C. Xue\inst{24}
	\and
	  Y. Zhang\inst{9}
	\and
	  B. Zhao\inst{2}
	\and
	  S. Yamamoto\inst{10,11}
          }
   \institute{IRAP, Universit\'e de Toulouse, CNRS, UPS, CNES, 31400 Toulouse, France, \email{cvastel@irap.omp.eu}\label{inst1}
	\and
	  Max-Planck-Institut f\"ur extraterrestrische Physik, Gie{\ss}enbachstra{\ss}e 1, D-85748 Garching, Germany\label{inst2}
	\and
	  Universit\'{e} Grenoble Alpes, CNRS, Institut de Plan\'{e}tologie et d'Astrophysique de Grenoble (IPAG), 38000 Grenoble, France\label{inst3}
	\and
	  Centro de Astrobiologia (CSIC-INTA), Ctra. de Torrejon a Ajalvir, km 4, 28850, Torrejon de Ardoz, Spain\label{inst4}
	\and
	  Graduate School of Informatics and Engineering, The University of Electro-Communications, Chofu, Tokyo 182-8585, Japan\label{inst5}
	\and
	  INAF-Osservatorio Astrofisico di Arcetri, Largo E. Fermi 5, I-50125, Florence, Italy\label{inst6}
	\and
	  National Radio Astronomy Observatory, PO Box O, Socorro, NM 87801, USA\label{inst7}
  	\and
	  European Southern Observatory, Karl-Schwarzschild Str. 2, 85748 Garching bei M\"{u}nchen, Germany\label{inst8}
	\and
	  RIKEN Cluster for Pioneering Research, 2-1, Hirosawa, Wako-shi, Saitama 351-0198, Japan\label{inst9}
	\and
	  Department of Physics, The University of Tokyo, 7-3-1, Hongo, Bunkyo-ku, Tokyo 113-0033, Japan\label{inst10}
	\and
	  Research Center for the Early Universe, The University of Tokyo, 7-3-1, Hongo, Bunkyo-ku, Tokyo 113-0033, Japan\label{inst11}
	\and
	  Institut de Radioastronomie Millim\'{e}trique, 38406 Saint-Martin d'H\`{e}res, France\label{inst12}
	\and
	  Department of Astronomy, The University of Tokyo, 7-3-1 Hongo, Bunkyo-ku, Tokyo 113-0033, Japan\label{inst13}
	\and
	  Department of Chemistry, Biology, and Biotechnology, The University of Perugia, Via Elce di Sotto 8, 06123 Perugia, Italy\label{inst14}
	\and
	  Astrochemistry Laboratory, Code 691, NASA Goddard Space Flight Center, 8800 Greenbelt Road, Greenbelt, MD 20771, USA\label{inst15}
	\and
	  Cergy Paris Universit\'e, Sorbonne Universit\'e, Observatoire de Paris, PSL University, CNRS, LERMA, F-95000, Cergy, France\label{inst16}
	\and
	  Instituto de Radioastronom\'{i}a y Astrof\'{i}sica, Universidad Nacional Aut\'{o}noma de M\'{e}xico, A.P. 3-72 (Xangari), 8701, Morelia, Mexico\label{inst17}
	\and
	  Department of Astronomy, Xiamen University, Zengcuo'an West Road, Xiamen, 361005, Peoples Republic of China\label{inst18}
	\and
	  National Astronomical Observatory of Japan, Osawa 2-21-1, Mitaka-shi, Tokyo 181-8588, Japan\label{inst19}
	\and
	  NRC Herzberg Astronomy and Astrophysics, 5071 West Saanich Road, Victoria, BC, V9E 2E7, Canada\label{inst20}
	\and
	  Department of Physics and Astronomy, University of Victoria, Victoria, BC, V8P 5C2, Canada\label{inst21}
	\and
	  Komaba Institute for Science, The University of Tokyo, 3-8-1 Komaba, Meguro, Tokyo 153-8902, Japan\label{inst22}
	\and
	  Center for Frontier Science, Chiba University, 1-33 Yayoi-cho, Inage-ku, Chiba 263-8522, Japan\label{inst23}
	\and
	  Department of Chemistry, University of Virginia, McCormick Road, PO Box 400319, Charlottesville, VA 22904, USA\label{inst24}
	\and
	  Department of Physics and Astronomy, Rice University, 6100 Main Street, MS-108, Houston, TX 77005, USA\label{inst25}
	\and
	  Instituto de Astronom\'{i}a, Universidad Nacional Aut\'{o}noma de M\'{e}xico, Ciudad Universitaria, A.P. 70-264, Cuidad de M\'{e}xico 04510, Mexico\label{inst26}
	\and
	  Institute of Low Temperature Science, Hokkaido University, N19W8, Kita-ku, Sapporo, Hokkaido 060-0819, Japan\label{inst27}
	\and
	  Departament de Qu\'{i}mica, Universitat Aut\`{o}noma de Barcelona, 08193 Bellaterra, Spain\label{inst28}
	\and
	  Steward Observatory, 933 N Cherry Ave., Tucson, AZ 85721 USA\label{inst29}
	\and
	  Leiden Observatory, Leiden University, PO Box 9513, 2300 RA Leiden, The Netherlands\label{inst30}
	\and
	  Department of Physics and Astronomy, University College London, Gower Street, London, WC1E 6BT, UK\label{inst31}
	\and
	  Materials Science and Engineering, College of Engineering, Shibaura Institute of Technology, 3-7-5 Toyosu, Koto-ku, Tokyo 135-8548, Japan\label{inst32}
	 \and
	   Departament de Física Qu\`antica i Astrofísica, Institut de Ci\`encies del Cosmos, Universitat de Barcelona (IEEC-UB), Mart\' i Franqu\`es, 1, E-08028 Barcelona, Catalunya, Spain\label{inst33}
	 \and
	    Astronomical Institute, Graduate School of Science, Tohoku University, 6-3 Aoba, Aramaki, Aoba-ku, Sendai, Miyagi, 980-8578, Japan, \label{inst34}
             }

   \date{Received March 4th, 2022; accepted June 17th, 2022}

 
  \abstract
{}
{Methanol is a ubiquitous species commonly found in the molecular interstellar medium. It is also a crucial seed species for the building-up of the chemical complexity in star forming regions.
Thus, understanding how its abundance evolves during the star formation process and whether it enriches the emerging planetary system is of paramount importance. }
{We used new data from the ALMA Large Program FAUST (Fifty AU STudy of the chemistry in the disk/envelope system of Solar-like protostars) to study the methanol line emission towards the \bhb protobinary system (sources A and B), where a complex structure of filaments connecting the two sources with a larger circumbinary disk has been previously detected.}
{Twelve methanol lines have been detected with upper energies in the range [45--537] K along with one $^{13}$CH$_3$OH transition and one methyl formate (CH$_3$OCHO) line blended with one of the methanol transitions.
The methanol emission is compact (FWHM $\sim$ 0.5$^{\prime\prime}$) and encompasses both protostars, which are separated by only 0.2$^{\prime\prime}$ (28 au).
In addition, the overall methanol line emission presents three velocity components, not spatially resolved by our observations. 
Nonetheless, a detailed analysis of the spatial origin of these three components suggests that they are associated with three different spatial regions, with two of them close to 11B and the third one associated with 11A.
A non-LTE radiative transfer analysis of the methanol lines gives a kinetic temperature of [100-140] K, an H$_2$ volume density of 10$^6$--10$^7$ cm$^{-3}$ and column density of a few 10$^{18}$ cm$^{-2}$ in all three components with a source size of $\sim$ 0.15$^{\prime\prime}$.
Thus, this hot and dense gas is highly enriched in methanol with an abundance as high as 10$^{-5}$.
Using previous continuum data, we show that dust opacity can potentially completely absorb the methanol line emission from the two binary objects.}
{Although we cannot firmly exclude other possibilities, we suggest that the detected hot methanol is resulting from the shocked gas from the incoming filaments streaming towards \bhb A and B, respectively. Higher spatial resolution observations are necessary to confirm this hypothesis.}
{}


 \maketitle


\section{Introduction}
\label{sec: Intro}
The rich chemistry observed in Solar-type protostars is triggered at earlier stages of the protostellar evolution \citep[e.g.][]{Ceccarelli2007,CaselliCeccarelli2012}. 
Methanol (CH$_3$OH) is a key ingredient in the enhancement of this chemical complexity from prestellar cores to protoplanetary disks \citep{Vastel2014,Bizzocchi2014,Maret2005,Walsh2016,Spezzano2020,Spezzano2022,Punanova2021}.
It is the seed from which interstellar Complex Organic Molecules \citep[iCOMs:][]{Ceccarelli2017} sprout \citep[e.g.][]{Charnley1992,Garrod2008,balucani2015,Taquet2016,Aikawa2020,Jin2020}.

iCOMs have been detected towards prestellar cores \citep{Bacmann2012,Vastel2014,Jimenez2016,Jimenez-Serra2021,Scibelli2020}, where their presence in the gas has been explained by non-thermal desorption processes such as photodesorption from UV photons \citep{Bertin2016}, chemical desorption \citep{Minissale2016}, sputtering caused by a gentle shock \citep{Flower1995}, or cosmic-ray induced chemistry \citep{Shingledecker2018,Wakelam2021} and reactive desorption of methanol followed by gas-phase reactions \citep{Vasyunin2017}.
iCOMs have also been found at a later stage, the Class 0/I protostars \citep{Andre1993}. Here, the central protostar is embedded in a dense envelope as the parental dense core undergoes gravitational collapse. The iCOMs are observed in the lukewarm region where the dust temperature is between the CO and H$_2$O sublimation temperatures ($\sim$ 30--100 K \citep[e.g.][]{Jaber2014} and in the compact hot corino region where the dust temperatures are larger than $\sim$100 K and densities larger than $\sim 10^7$ cm$^{-3}$ \citep[e.g.][]{Ceccarelli2000b,Cazaux2003,Bottinelli2004,Jorgensen2016,Yang2021,Chahine2022}. 
Finally, iCOMs are also detected in protoplanetary disks \citep{Oberg2015,Oberg2021,Favre2018,Lee2019}.

The gas phase is therefore chemically enriched in iCOMs, due to either direct release from dust mantles or gas phase formation from simpler molecules released from the mantles. 
Two possibly distinct classes of Solar-like Class 0/I protostars have been found: the hot corinos \citep{Ceccarelli2007} and the warm carbon chain chemistry sources \citep[WCCC:][]{Sakai2013}, although intermediate situations are also possible \citep[e.g.][]{Yang2021}. An intermediate (hybrid) source was also found by \citet{Oya2017}.



Since the chemical composition of the planetary systems depends on the chemical evolution starting from the earliest phases of the protostar formation, it is crucial to understand how the molecular complexity is transferred from the large-scale envelope (a few thousand au) to the small-scale structures of the protostellar system, and how it is maintained in the inner disk system ($\le$ 50 au).

In this context, the ALMA (Atacama Large Millimeter/submillimeter Array) Large Program FAUST (Fifty AU STudy of the chemistry in the disk/envelope system of Solar-like protostars)\footnote{\url{http://faust-alma.riken.jp}} is designed to survey the chemical composition of a sample of 13 Class 0/I protostars at the scale of about 50 to 1000 au (all sources have a distance lower than 250 pc). 
The description of the project is reported in \citet{Codella2021}.
The selected sources represent the protostellar chemical diversity observed at large scales  and have been observed in three frequency setups chosen to study both continuum and line emission from specific molecules: 85.0--89.0 GHz, 97.0--101.0 GHz, 214.0--219.0 GHz, 229.0--234.0 GHz, 242.5--247.5 GHz, and 257.5--262.5 GHz. 
The FAUST survey provides a uniform sample in terms of frequency setting, angular resolution and sensitivity. 
We report the  first results obtained towards the \bhb Class 0/I protostar system focused on the most simple iCOM, methanol.

\section{The \bhb protobinary system}
\label{sec: Sources}

The  \bhb source was originally identified as a single object in the Barnard 59 (B59) molecular cloud, which is located at a distance of 163 $\pm$ 5 pc \citep[Gaia second data release:][]{Dzib2018}. B59 hosts a proto-cluster of low-mass young stellar objects (YSOs) at different evolutionary stages as first shown by \citet{Onishi1999} using the 1--0 transition of $^{12}$CO, $^{13}$CO, and C$^{18}$O with the NANTEN telescope with a 2.7$^{\prime}$ beam. Following this, the proto-cluster was extensively observed in infrared bands \citep{Brooke2007,Forbrich2009,Covey2010,Roman2010,Sandell2021}, ammonia (NH$_3$) emission \citep{Rathborne2008,Redaelli2017} and X-rays \citep{Forbrich2010}. The protostellar nature of the youngest and most embedded member of the cluster, \bhb, was confirmed by several studies reporting bipolar outflows \citep{Riaz2009,Duarte2012}, hints of dynamical infall \citep{Hara13,Alves2017} and a Spectral Energy Distribution (SED) consistent with a Class 0/I protostar with bright far-infrared emission and $L_\mathrm{bol} \sim 4.4$ $L_\mathrm{\odot}$ \citep{Sandell2021}.


\citet{Alves2017} used ALMA in its extended configuration to observe the dust continuum emission at 1.3 mm and molecular transitions of CO, C$^{18}$O and H$_{2}$CO lines (angular resolution of 0.22$^{\prime\prime}$). The dust distribution reveals a $\sim 180$ au disk surrounded by a fainter, elongated structure tracing the inner component of the core envelope.  The spectral line data reveal a bipolar outflow launched at the edge of the inner disk at a radial distance of 90--130 au from the central source. The disk is rotationally supported, with Keplerian rotation revealed by H$_{2}$CO lines \citep{Alves2017,Alves2019}. 
At much higher angular resolution ($\sim 0.04\arcsec$), \citet{Alves2019} uncovered a young binary protostar system embedded in circumstellar disks that have radii of 2 to 3 au.  The protostars are named \bhb A (hereafter 11A) for the Northern source ($\alpha$ (2000) = 17h11m23.097s, $\delta$ (2000) = -27$^{\circ}$24$^{\prime}$32.85$^{\prime\prime}$) and \bhb B (hereafter 11B) for the Southern source ($\alpha$ (2000) = 17h11m23.09s, $\delta$ (2000) =  -27$^{\circ}$24$^{\prime}$33$^{\prime\prime}$). These systems are surrounded by a complex filamentary structure (the so-called streamers) connecting to the larger circumbinary disk. Note that streamers have recently been seen in other more evolved objects \citet{Alves2020} and modeled as mostly free-falling (\citealp{Pineda2020}; \citealp[see][for a review]{Pineda2022}). The binary mass ratio is $\gtrsim 1$, since the 11A disk is slightly more massive (of the order of a Jupiter mass) than the 11B disk and the system has a projected separation of 28 au ($\sim$ 0.2$^{\prime\prime}$). Source 11B is located between high velocity components of the CO emission (V$_{LSR}$ < -1.5 km~s$^{-1}$ and V$_{LSR}$ > 9 km~s$^{-1}$), which is interpreted as enhanced gas accretion towards 11B from the circumbinary disk. Both protostars are accreting from their individual circumstellar disks as inferred from centimetric continuum emission observed from both sources. The radio data are consistent with free-free emission from ionised jets launched by the disk-star system \citep{Alves2019}. The dynamical mass of the protobinary system has been estimated as an upper limit from the Keplerian model of the H$_{2}$CO data with a value of 2.25 $\pm$ 0.13 M$_{\odot}$. 

In this paper, we report high-$J$ methanol emission (45 K  < E$_{\mathrm{up}}$ < 537 K ) from \bhb obtained in the 
framework of the FAUST collaboration observed in a $\sim$ 0.4$^{\prime\prime}$ beam. These data unveil the hot corino nature of the source seen over multiple velocity components. From these data, we derive the hot gas properties (column density and temperature) for each
velocity component. 

\section{Observations}
\label{sec: obs}

\begin{figure}
    \centering
    \includegraphics[width=0.45\textwidth]{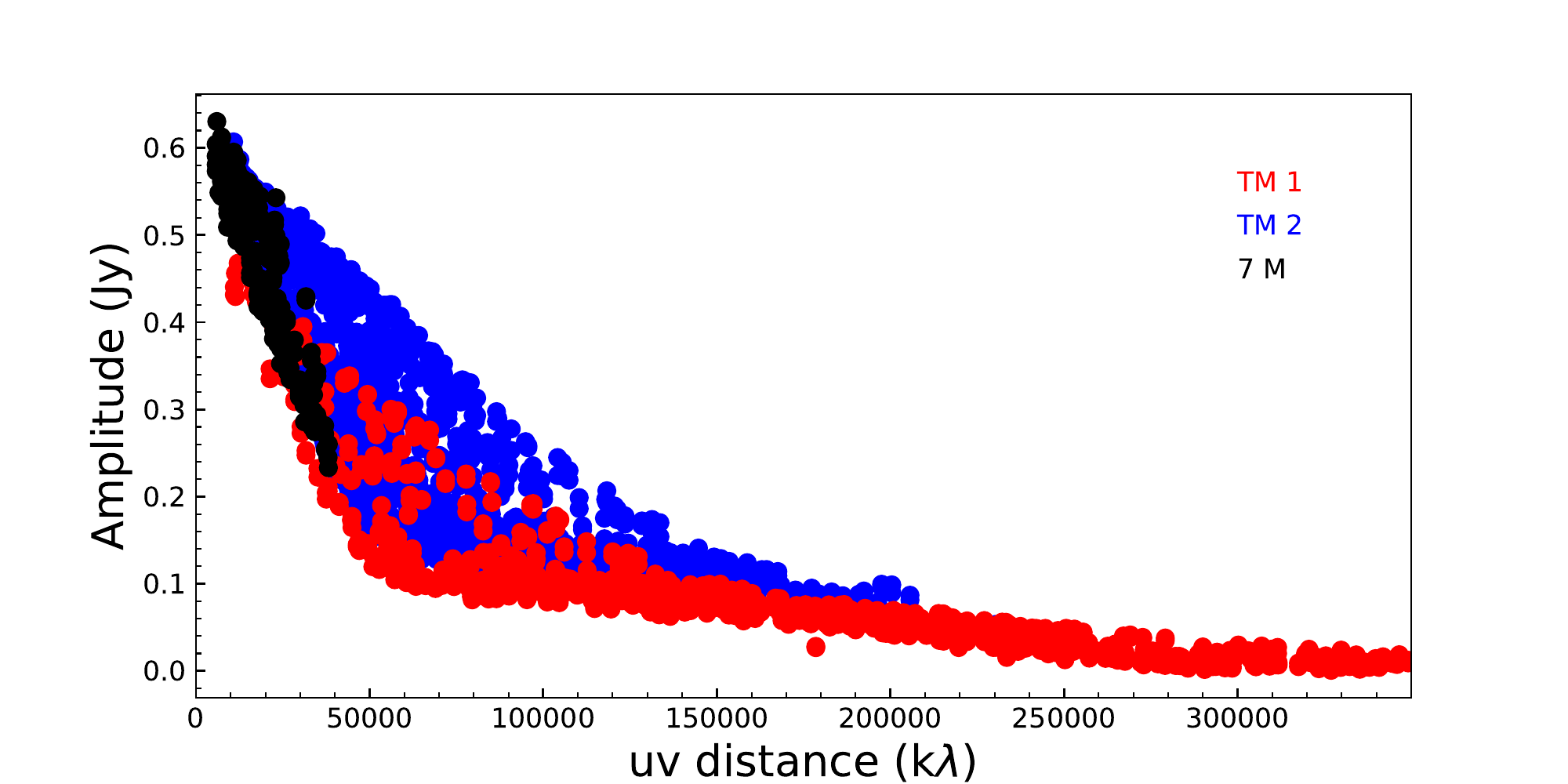} \ 
    \includegraphics[width=0.45\textwidth]{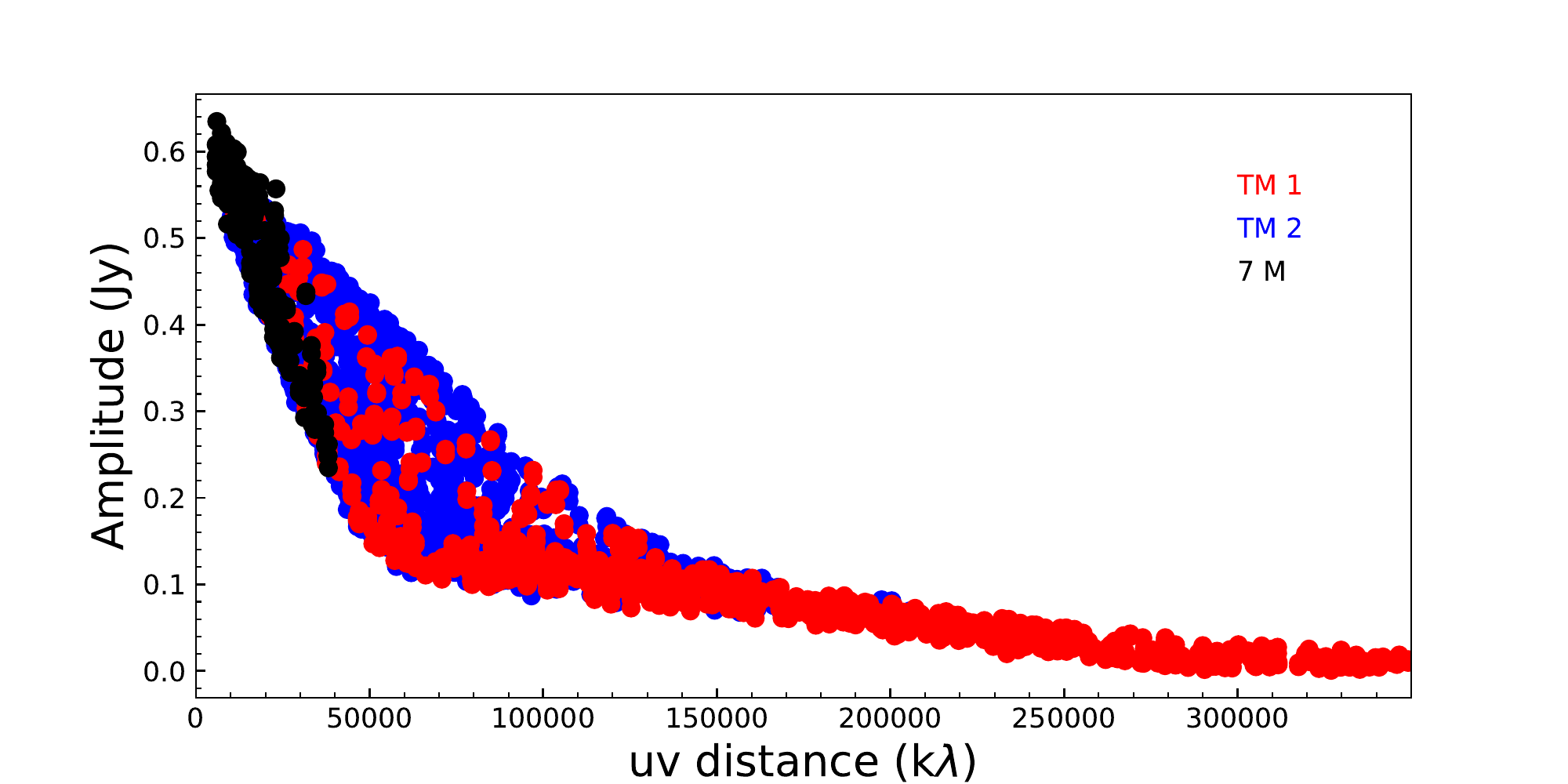} \
    \caption{Amplitude versus uv-distance plots, before (top) and after (below) self-calibration.}
    \label{fig: selfcal}
\end{figure}

\bhb has been observed with ALMA (FAUST Large Program 2018.1.01205.L) with two frequency setups in Band 6 (setup 1 and 2) and one frequency setup in Band 3 (setup 3). The data exploited here were acquired between 2018 and 2020 using the 12-m and 7-m array for setup 1 and 2, and the 12-m array for setup 3. The baseline length in the configurations of the 12-m array for setup 1 and 2 (C43-4 and C43-1) ranges from 15 m to 1.3 km, and the baseline length in the configuration of the 7-m array ranges from 8.9 m to 48.9 m. For setup 3, the baseline length in the configurations of the 12 m array (C43-6 and C43-3) ranges from 15 m to 3 km. The observations were centred at $\alpha$ (2000) = 17h11m23.125s, $\delta$ (2000) = -27$^{\circ}$24$^{\prime}$32.87$^{\prime\prime}$. J2056-4714 and J1427-4206 were used as bandpass calibrators for setup 2 (12-m array), and J1700-2610 as phase calibrator and flux calibrator. J1427-4206 and J2056-4714 were used as bandpass calibrator and flux calibrator for setup 1 (12-m array), and J1700-2610 as phase calibrator. \\
The data were reduced in the Common Astronomy Software Applications package (CASA) version 5.6.1 (McMullin et al. 2007) using a modified version of the ALMA calibration pipeline and an additional in-house calibration routine to correct for the system temperature and spectral line data normalisation\footnote{\url{https://help.almascience.org/kb/articles/what-errors-could-originate-from-the-correlator-spectral-normalization-and-tsys-calibration}}.
Self-calibration was carried out using carefully-selected, line-free, continuum channels for each configuration. The complex gain corrections derived from the self-calibration were then applied to all channels in the data, and the continuum model derived from the self-calibration was then subtracted from the data to produce continuum-subtracted line data. A self-calibration technique was also used to align both amplitudes and phases across the multiple configurations, to correct for any remaining offsets in the flux density scales, and to correct for systematic position offsets that may be caused by either source proper motion or other system issues.\\ 
For each setup, iterative phase self-calibration using a very long solution interval was then used to do the position alignment across multiple configurations. In addition, amplitude offsets between datasets needed to be corrected. To do this, a model derived from a very deeply-cleaned and masked image was used to obtain amplitude corrections per execution block. The dynamic range of the resulting self-calibrated image was improved by a factor of three for setup 1, without reduction in the peak flux density of the source. Fig. \ref{fig: selfcal} shows an amplitude versus uv-distance plot of the data from the different configurations before and after this amplitude alignment for setup 1 in the continuum spectral window (least noisy). The data have been averaged per baseline across the entire dataset. The improvement in the dynamic range was less significant for setup 2 (around 10\%), because for setup 2, the amplitude scale of the 12-m (TM 1 and TM 2) was much better aligned coming out of the calibration pipeline and per-configuration self-calibration process than they were for setup1.

The resulting continuum-subtracted line-cube has been made using a Briggs robust parameter of 0.5 and we obtained a synthesised beam of 0.44$^{\prime\prime}$ $\times$ 0.37$^{\prime\prime}$ Position Angle (PA) = -82$^{\circ}$ at 261.687 GHz, and 0.41$^{\prime\prime}$ $\times$ 0.39$^{\prime\prime}$ PA = 47$^{\circ}$ at 218 GHz.  
We estimate an overall systematic uncertainty of 10\% for each dataset. 
The observed spectral windows (spw) are centred on the frequencies listed in Table \ref{tab: spw}. The bandwidth is 59 MHz for 12-m (62 MHz for 7-m) except for the continuum window with a 1.875 GHz (2 GHz for 7-m) bandwidth.   
Spectral line imaging was performed with the CASA\footnote{\url{https://casa.nrao.edu/}} package, while the data analysis was performed using the IRAM/GILDAS\footnote{http://www.iram.fr/IRAMFR/GILDAS} package as well as the CASSIS\footnote{\url{http://cassis.irap.omp.eu}} package. 

\begin{table}
\small
\centering
\caption{Observed spectral windows in the observed FAUST data.}
\label{tab: spw}
\begin{tabular}{cccc}
\hline
Spectral window   & setup 1                         &  setup 2             & setup 3\\
                             &  (GHz)                              &  (GHz)                   &  (GHz)    \\   
\hline
spw 25   & 216.113                  &  243.916     & 93.181 \\
spw 27   & 216.563                  &  244.049     & 94.405\\
spw 29   & 217.105                  &  244.936     & 95.000 \\
spw 31   & 217.822                  &  245.606     &  107.014\\
spw 33   & 218.222                  &  246.700     &  108.040\\
spw 35   & 218.440                  &  260.189     & 104.239 \\
spw 37   & 219.560                  &  260.255     &  105.799\\
spw 39   & 219.949                  & 261.687     & \\
spw 41   & 231.061                  & 262.004     & \\
spw 43   & 213.221                  & 257.896     &  \\
spw 45   & 231.322                  & 258.256      & \\
spw 47   & 231.410                  & 258.549     &  \\
spw 49$^a$  &  233.796  & 259.035 \\
\hline
\end{tabular}
\tablefoot{$^a$: continuum window}
\end{table}

\begin{figure*}[ht!]
    \centering
    \includegraphics[width=\textwidth]{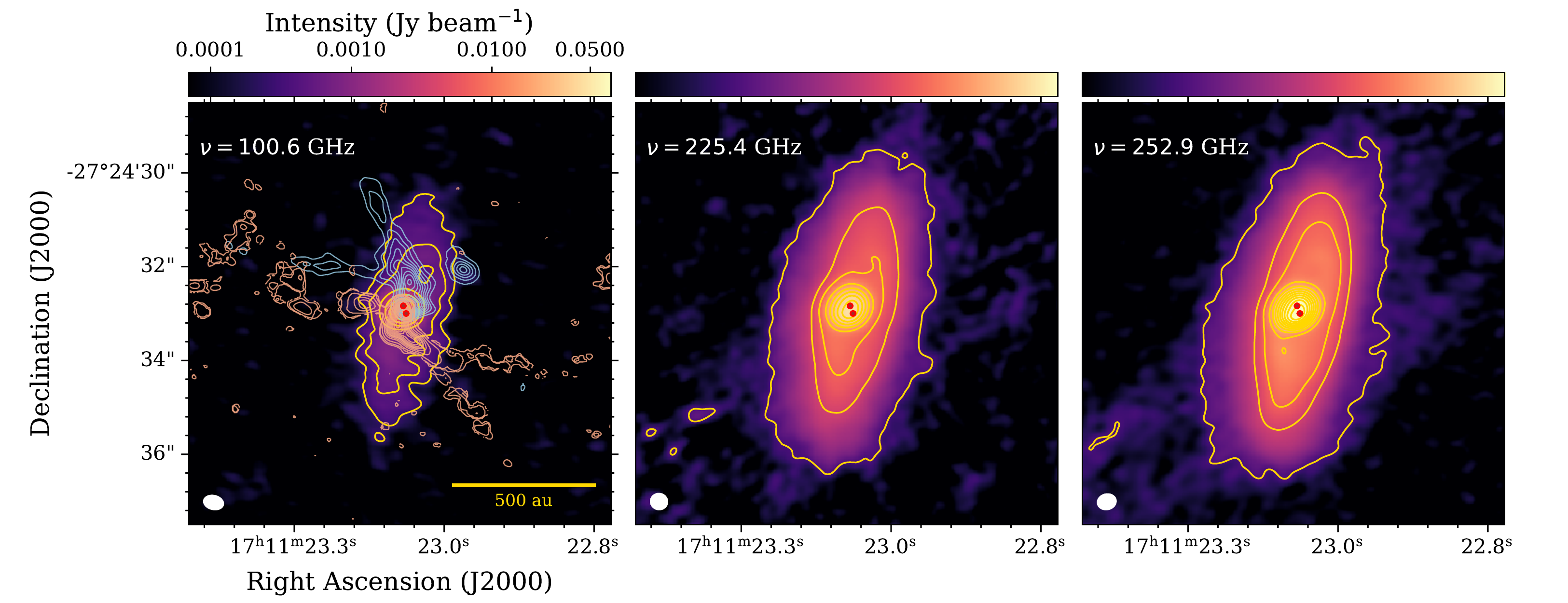}
    \caption{Dust continuum maps obtained in each setup. The {\it rms} noise level is 50 $\mu$Jy beam$^{-1}$ for Band 3 observations (setup 3, left panel), 90 $\mu$Jy beam$^{-1}$ for setup 1 in Band 6 (middle panel) and 70 $\mu$Jy beam$^{-1}$ for setup 2 in Band 6 (right panel). The intensity contour levels in the left panel are -5, 5, 10, 30, 50, 70, 90 $\times$ the noise level, while the intensity contours in the middle and right panels are -5, 5, 50 and from 100 to 1000 (in steps of 100) $\times$ 70 $\mu$Jy beam$^{-1}$, the noise level in setup 2. The colour scale at the top is the same for all panels. The red dots indicate the positions of 11A (North) and 11B (South) as reported by \citet{Alves2019}. The sources are not resolved with the present resolution. The left panel also shows the integrated intensity contours from CO ($2\rightarrow1$) emission from the bipolar outflow powered by \bhb \citep{Alves2017}. The synthesised beam is displayed in the bottom left corner of each panel ($0.36\arcsec \times 0.34\arcsec, 0.39\arcsec \times 0.33\arcsec$ and $0.42\arcsec \times 0.29\arcsec$ for Setups 3, 1 and 2, respectively).}
    \label{fig:dust}
\end{figure*}

\begin{figure}[ht!]
    \centering
    \includegraphics[width=\columnwidth]{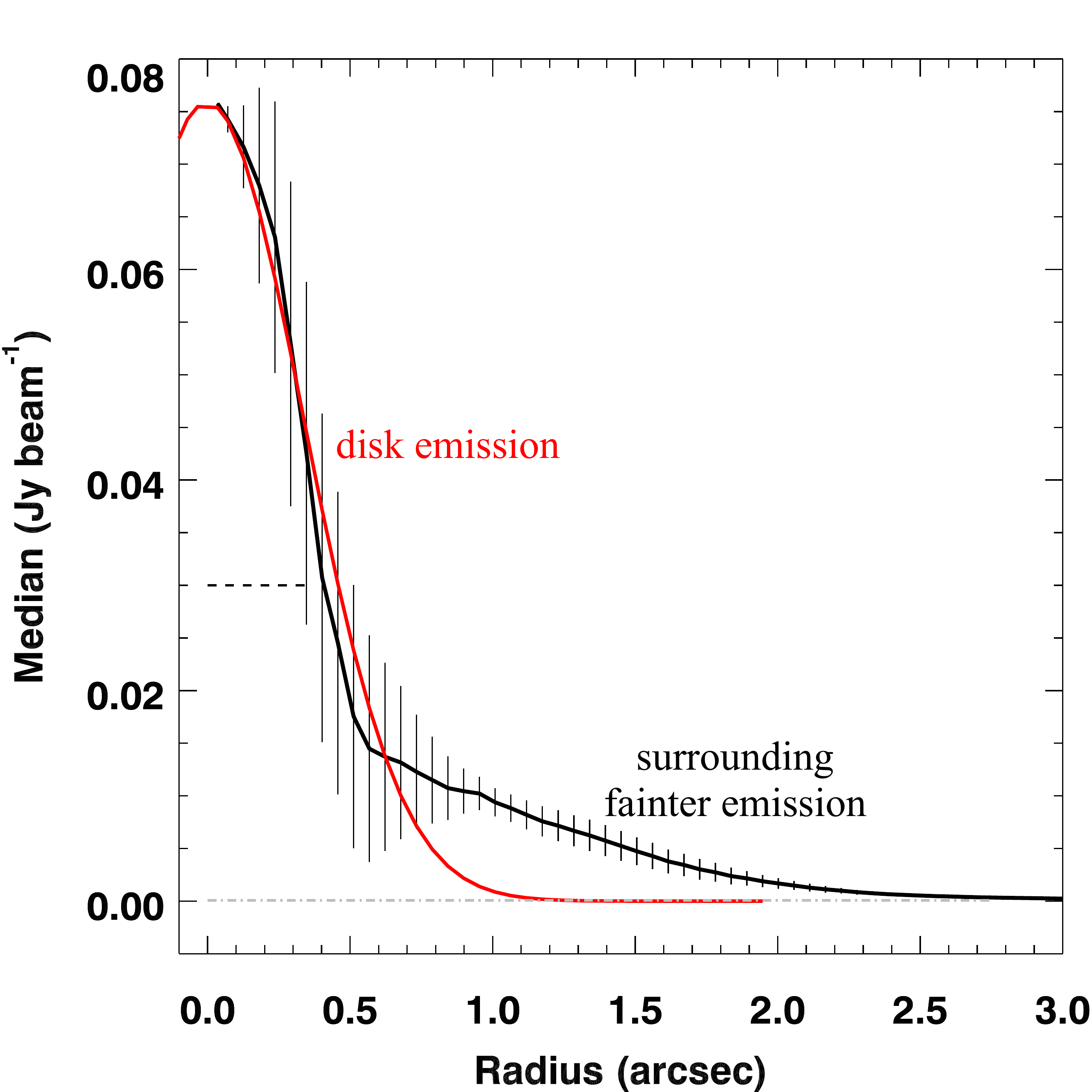}
        \caption{Radial profile of the dust continuum emission at 252 GHz (setup 2) towards \bhb. The profile is centred on the continuum peak. The red line shows a Gaussian fit to the disk brightness distribution. The dashed line indicates the synthesised beam size. The grey line shows the {\it rms} noise level of the continuum emission.}
    \label{fig:radprof}
\end{figure}

\begin{figure}[ht!]
    \centering
    \includegraphics[width=\columnwidth]{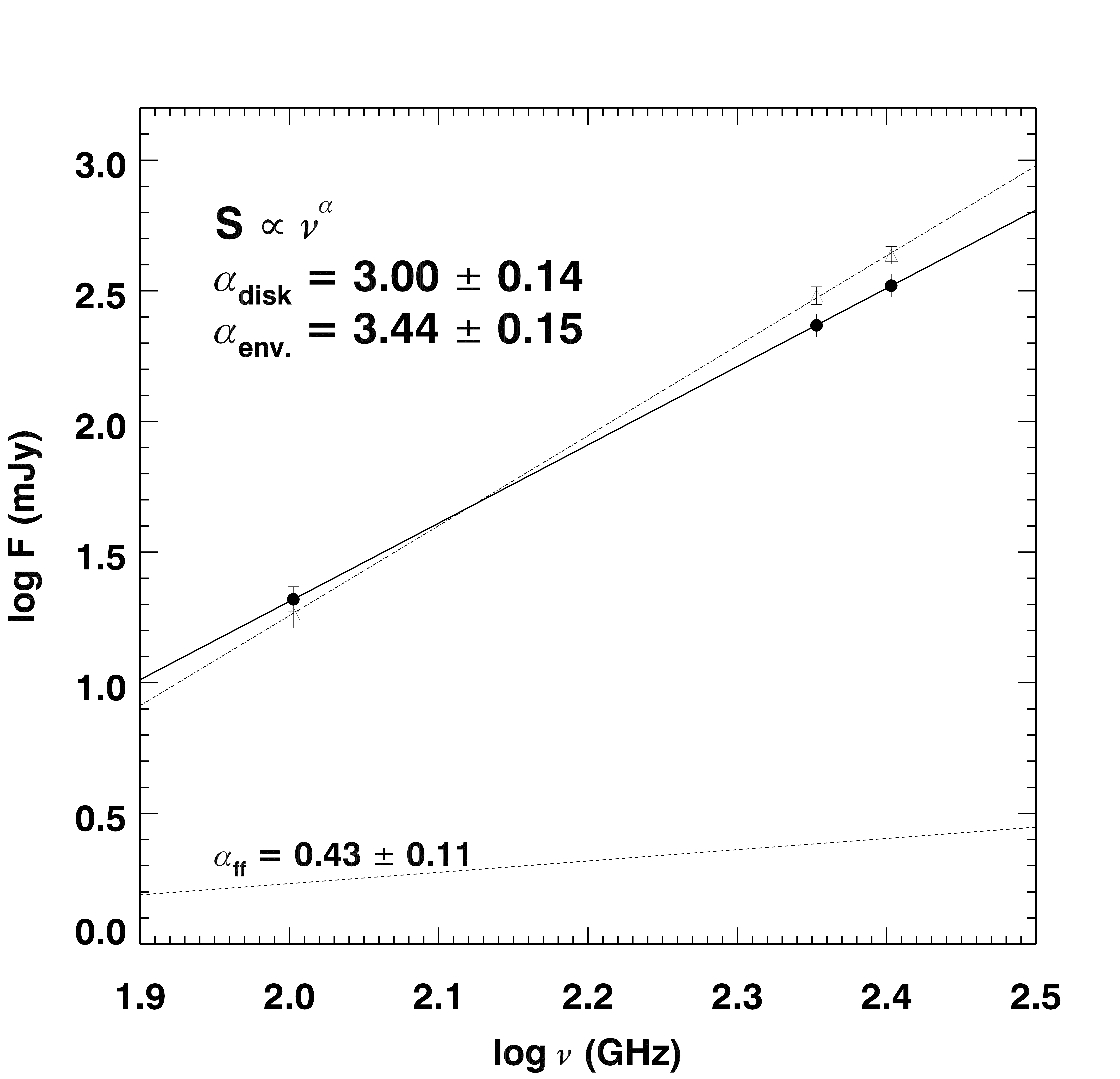}
        \caption{Spectral Energy Distribution determined for the dust emission within the circumbinary disk ($\alpha = 3.0$, black line) and in the envelope ($\alpha = 3.4$, dotted dashed line). The flux densities for disk and envelope are indicated as filled circles and open triangles, respectively. The errors are dominated by the uncertainty in the flux calibration, which is assumed to be $\sim 10\%$. The dashed line shows the SED determined from the free-free (centimetric) emission detected with the Very Large Array \citep{Alves2019}.}

    \label{fig:sed}
\end{figure}

\section{Line identification}
\label{sec: line-id}

The line identification has been performed using the  CASSIS software that connects to the JPL and CDMS databases. This is a standalone software, written in Java, freely delivered to the community to help with visualising, analysing and modelling observations from ground or space-based observatories. CASSIS has been developed at IRAP since 2005 and is part of the OVGSO\footnote{\href{https://ov-gso.irap.omp.eu/}{https://ov-gso.irap.omp.eu/}} data centre that aims at promoting the Virtual Observatory (VO) technology. CASSIS displays any spectrum (ASCII, FITS or GILDAS/CLASS format or the result from the query to any Simple Spectral Access Protocol (SSAP) VO or EPN-TAP\footnote{Europlanet-Table Access Protocol} VO service from IVOA\footnote{International Virtual Observatory Alliance} registries) and identifies atomic and molecular species through its link to the databases such as CDMS \citep{Muller2005} or JPL \citep{Pickett1998} via a SQLite database, or via a direct access to VAMDC\footnote{\href{http://www.vamdc.org/}{http://www.vamdc.org/}} for any available spectroscopic database. \\
The systemic velocity of \bhb with respect to the local standard of rest is V$\rm_{LSR}$ = 3.6 km~s$^{-1}$ \citep{Onishi1999}. We identified and detected twelve methanol transitions in the observed spectral windows of the three spectral setups (dust peak emission), and only one $^{13}$CH$_{3}$OH transition (234 GHz but not at 94.4 GHz where the rms reached 0.9 K). No deuterated counterparts such as CH$_{2}$DOH at 261.7 GHz have been detected. We list, in Table \ref{tab: spectro}, the spectroscopic parameters of the detected transitions using the CDMS database. The spectral resolutions are also quoted as $\delta$V (km~s$^{-1}$) for each detected transition. The methanol transitions cover a broad range in energy from E$\rm_{up}$ $\sim$ 45 to 537 K. 

\begin{table*}
\small
\centering
\caption{Spectroscopic Parameters of the methanol transitions (using the CDMS database) detected towards \bhb and results from the Gaussian fits.  }
\label{tab: spectro}
\begin{tabular}{cccccccccc}
\hline

                                  &                          &                                            &                               &       CH$_{3}$OH         &                           &                                      &                            &                               \\
\hline
Frequency                   &            QN                    &E$_\mathrm{up}$    & A$_\mathrm{ij}$     &  beam, PA &   rms     &     $\delta$V       &   T$_\mathrm{mb}$      &     FWHM            & V$_\mathrm{LSR}$\\
  (MHz)                        &                                      &          (K)                 &       (s$^{-1}$)    &  $^{\prime\prime} \times ^{\prime\prime}$,$^{\circ}$   &     (mK)    &  (km s$^{-1}$)  &        (K)                      & (km s$^{-1}$)      & (km s$^{-1}$)\\
\hline\hline
218440.063(0.013)      &{\small 4$_{- 2, 3}$ -- 3$_{-1,2}$ E}    & 45.46     &  4.69 $\times$ 10$^{-5}$  &   0.41 $\times$ 0.39, 47  & 158 & 0.17        & 1.57(0.06)  & 3.2(0.1)    & -2.0\\
                       &                                     &           &                           &                  &     &             &1.46(0.04)   & 6.1(0.3)    & 2.8  \\ 
                       &                                     &           &                           &                  &     &             &1.16(0.04)   & 5.4(0.2)    & 9.9  \\ 
243915.788(0.004)      &{\small 5$_{1,4}$ -- 4$_{1.3}$  A }         & 49.66     & 5.97 $\times$ 10$^{-5}$   &   0.47 $\times$ 0.40, -84 & 59  &  0.15       & 2.01(0.04)  & 3.2(0.1)    & -2.0\\
                       &                                     &           &                           &                  &     &             & 1.67(0.02)  & 5.9(0.1)    & 2.8    \\ 
                       &                                     &           &                           &                  &     &             & 1.41(0.02)  & 4.9(0.1)    & 9.9    \\ 
247228.587(0.013)      &{\small 4$_{2,2}$ -- 5$_{1,5}$  A}           & 60.92     & 2.12 $\times$ 10$^{-5}$   &   0.46 $\times$ 0.39, -80 & 34  &   1.18      &1.49(0.06)   & 2.9(0.2)    & -2.0\\
                       &                                     &           &                           &                  &     &             & 1.01(0.04)  & 5.6(0.0)    & 2.8    \\ 
                       &                                     &           &                           &                  &     &             & 0.99(0.04)  & 5.0(0.0)    & 9.9    \\ 
234683.37(0.012)       &{\small 4$_{2.3}$ -- 5$_{1,4}$  A}         &  60.92    & 1.87 $\times$ 10$^{-5}$   &   0.38 $\times$ 0.34, 70 &  74 &  0.62       & 1.32(0.07)  & 2.9(0.1)    & -2.0\\
                       &                                     &           &                           &                       &     &             & 0.92(0.04)  & 6.2(0.5)    & 2.8    \\ 
                       &                                     &           &                           &                       &     &             & 0.92(0.04)  & 5.3(0.3)    & 9.9    \\ 
234698.519(0.015)      &{\small 5$_{4.2}$ -- 6$_{3,3}$  E}         & 122.72    & 6.34 $\times$ 10$^{-6}$   & 0.38 $\times$ 0.34, 70   & 73  &  0.63       & 0.85(0.05)  & 3.1(0.2)    & -2.0\\   
                       &                                     &           &                           &                       &     &             & 0.53(0.03)  & 5.1(0.5)    & 2.8    \\ 
                       &                                     &           &                           &                       &     &             & 0.56(0.03)  & 5.7(0.5)    & 9.9    \\ 
232945.797(0.012)      &{\small 10$_{3,7}$ -- 11$_{2,9}$  E}       & 190.37   & 2.13 $\times$ 10$^{-5}$     & 0.38 $\times$ 0.34, 70   & 79  & 0.62        & 1.12(0.07)  & 3.1(0.2)    &-2.0\\
                       &                                    &          &                             &                        &     &             & 0.84(0.04)  & 5.9(0.6)    & 2.8    \\ 
                       &                                    &          &                             &                        &     &             & 0.91(0.04)  & 5.6(0.4)    & 9.9    \\ 
247161.95(0.015)       &{\small 16$_{-2,15}$ -- 15$_{3,13}$}  E     & 338.14   & 2.57 $\times$ 10$^{-5}$     & 0.46 $\times$ 0.39, -80   &  29 &       1.18  & 1.00(0.05)  & 3.0(0.2)   & -2.0\\
                       &                                    &          &                             &                  &     &             & 0.62(0.03)  & 5.6(0.0)    & 2.8    \\ 
                       &                                    &          &                             &                  &     &             & 0.72(0.03)  & 5.0(0.0)    & 9.9    \\ 

261704.409(0.015)     &{\small 12$_{-6,7}$ -- 13$_{-5,8}$}  E       & 359.77   & 1.78 $\times$ 10$^{-5}$     & 0.44 $\times$ 0.37, -82   & 104 & 0.14        & 1.41(0.05)  &  3.9(0.2)  & -2.0\\
                      &                                     &          &                             &                  &     &             & 0.53(0.05)  &  3.6(0.5)   & 2.8    \\ 
                      &                                     &          &                             &                  &     &             & 0.81(0.04)  & 5.2(0.3)   & 9.9    \\ 
233795.666(0.012)     &{\small 18$_{3,15}$ -- 17$_{4,14}$  A}      & 446.58   & 2.20 $\times$ 10$^{-5}$     & 0.38 $\times$ 0.34, 70   & 89  & 0.63        &  0.67(0.09) & 2.6(0.4)   & -2.0\\
                      &                                     &          &                             &                        &     &             & 0.48(0.05)  &  6.3(1.2)   & 2.8    \\ 
                      &                                     &          &                             &                        &     &             & 0.51(0.06)  & 4.4(0.7)   & 9.9    \\ 
247610.918(0.014)     &{\small 18$_{3,15}$ -- 18$_{2,16}$  A}      & 446.58   & 8.29 $\times$ 10$^{-5}$     & 0.46 $\times$ 0.39, -80 & 49  & 1.18        & 1.16(0.06)  & 2.9(0.2)    & -2.0\\
                      &                                     &          &                             &                &     &             & 0.67(0.04)  &  5.6(0.0)   & 2.8    \\ 
                      &                                     &          &                             &                &     &             & 0.82(0.04)  & 5.0(0.0)   & 9.9    \\ 
246873.301(0.015)     &{\small 19$_{3,16}$ -- 19$_{2,17}$  A}      & 490.65   & 8.27 $\times$ 10$^{-5}$     & 0.46 $\times$ 0.39, -80   &  19 & 1.18        & 1.01(0.06)  &  2.7(0.2)   & -2.0\\
                      &                                     &          &                             &                  &     &             & 0.62(0.04)  &  5.6(0.0)   & 2.8    \\ 
                      &                                     &          &                             &                  &     &             & 0.73(0.04)  & 5.0(0.0)   & 9.9    \\ 
246074.605(0.016)     &{\small 20$_{3,17}$ -- 20$_{2,18}$  A}      & 537.04   & 8.25 $\times$ 10$^{-5}$     & 0.46 $\times$ 0.39, -80   & 29  & 1.19        & 0.93(0.08)  & 3.2(0.3)    & -2.0\\
                      &                                     &          &                             &                  &     &             & 0.65(0.05)  & 5.6(0.0)    & 2.8 \\ 
                      &                                     &          &                             &                  &     &             & 0.66(0.05)  & 5.0(0.0)    & 9.9 \\ 
\hline
                                    &                          &                                            &        &        $^{13}$CH$_{3}$OH               &                &                           &                                      &                            &                               \\
\hline
234011.58(0.05)          &{\small 5$_{1,5}$ --  4$_{1,4}$   A$^{+}$}   & 48.25     & 5.27 $\times$ 10$^{-5}$   & 0.38 $\times$ 0.34, 70  &  76    & 0.62     & 0.41(0.05)  & 2.9(0.4)    & -2.0\\
                         &                                      &           &                           &                &        &          & 0.47(0.03)  & 5.6(0.0)    & 2.8    \\ 
                         &                                      &           &                           &                &        &          & 0.30(0.03)  & 5.0(0.0)    & 9.9    \\ 
    
\hline
\end{tabular}
\tablefoot{
The spectral resolution is indicated as $\delta V$. The 1$\sigma$ error are quoted in parenthesis for the main beam temperature, the full width half maximum and velocity in the standard of rest. The spectra were extracted over the region above 4$\sigma$ in the integrated intensity map for each transition.
}

\end{table*}

\section{Results}
\label{sec: results}

\subsection{Dust emission}

Figure \ref{fig:dust} shows the dust continuum maps obtained in the three setups (ACA+12m combined). The dust distribution is consistent with the previous maps of \citet{Alves2017}, showing bright emission from the circumbinary disk (inclined $\sim 56\degr$ with respect to the line-of-sight) at the centre of the elongated envelope emission. The apparent misalignment between the envelope major axis and the disk major suggests a warped disk. This is consistent with the velocity structure previously revealed by H$_2$CO lines, where low upper energy level (E$_\mathrm{up}$) transitions trace envelope rotation/infall misaligned with the disk rotation seen in high-E$_\mathrm{up}$\,transitions \citep{Alves2017,Alves2018}. At the present resolution, the proto-binary system reported by \citet{Alves2019} is not resolved, but faint, extended emission oriented in a north-west south-east direction is
detected (at a 3$\sigma$ level) in the $\sim 1$ mm bands (setups 1 and 2) beyond the envelope. The radial profile of the continuum emission, taken from increasing elliptical annuli oriented at a position angle of 167\degr\,East of North (the orientation of the inner envelope), is displayed in Fig. \ref{fig:radprof}. The contribution from the disk is distinguishable from the fainter structure that surrounds it, as the disk brightness is well represented by a Gaussian distribution whose wings do not encompass the extended emission.

With the continuum data aligned in position across the three FAUST setups, we reconstructed the continuum maps with a common clean beam of $0.35\arcsec$. We then computed the Spectral Energy Distribution (SED) from a linear fit (in log-log scale) over the fluxes observed in the three setups (Fig. \ref{fig:sed}). Note that the flux density (and error) used to construct the SED are derived from 2D Gaussian fits over the disk emission, where the emission is centrally peaked, while the envelope flux is derived from a contour excluding the disk.\footnote{The circumbinary disk emission over which the SED was computed is determined as the first closed contour around it (i. e., not encompassing the extended N-S emission): 15, 120 and 230 times the rms noise of the beam-matched maps in setups 3, 1, and 2, respectively. The SED for the envelope uses the integrated flux higher than 5 times the {\it rms} noise but lower than the contours listed above (i .e, it excludes the circumbinary disk).}
Table \ref{tab: flux} reports the peak flux and flux density for the circumbinary disk, for which we find a spectral index of $3.0 \pm 0.1$. The same index is reported by \citet{Alves2019}, whose observations cover ALMA Bands 3, 6 and 7. In contrast, the dust emission from the inner envelope has a spectral index of $3.4 \pm 0.2$, which is slightly lower than the canonical value of the Interstellar Medium \citep[$\alpha_{\mathrm{ISM}} = 3.7$,][]{Testi2014}, implying that the grains in the envelope are potentially mildly larger than in the ISM \citep[$\lesssim 10\,\mu$m,][]{Agurto2019}. 


\begin{table}
\small
\centering
\caption{Peak flux and flux density for the circumbinary disk.}
\label{tab: flux}
\begin{tabular}{cccc}
\hline
Setup    &   Frequency   &   Peak Flux   &   Flux Density   \\
             &      (GHz)        &  (mJ~beam$^{-1}$)   &   (Jy)\\
\hline
3           &      100.6        &         5.4 $\pm$  0.2       &   20.9 $\pm$ 0.9 \\
1           &      225.4        &       49 $\pm$ 1       &   233 $\pm$ 7 \\
2           &      253.0        &       76  $\pm$ 2      &   331 $\pm$ 12\\
\hline
\end{tabular}
\tablefoot{Uncertainties refer to the 2D Gaussian fit error.}
\end{table}

\subsection{Molecular emission}

\begin{figure}
    \centering
    \includegraphics[width = 0.45\textwidth,trim=0 0 0 0 clip]{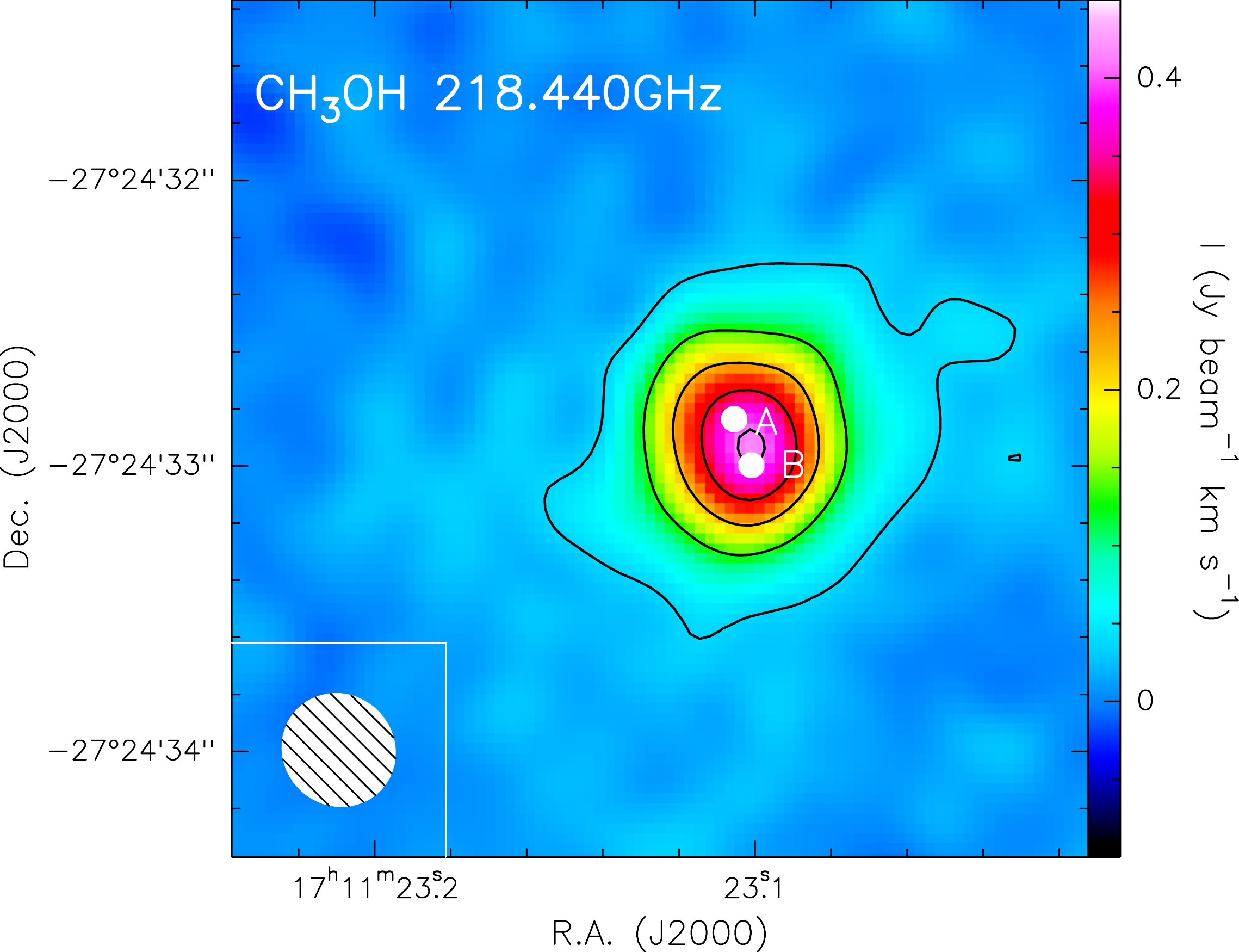}
    \includegraphics[width=0.45\textwidth,trim=0 0 0 0 clip]{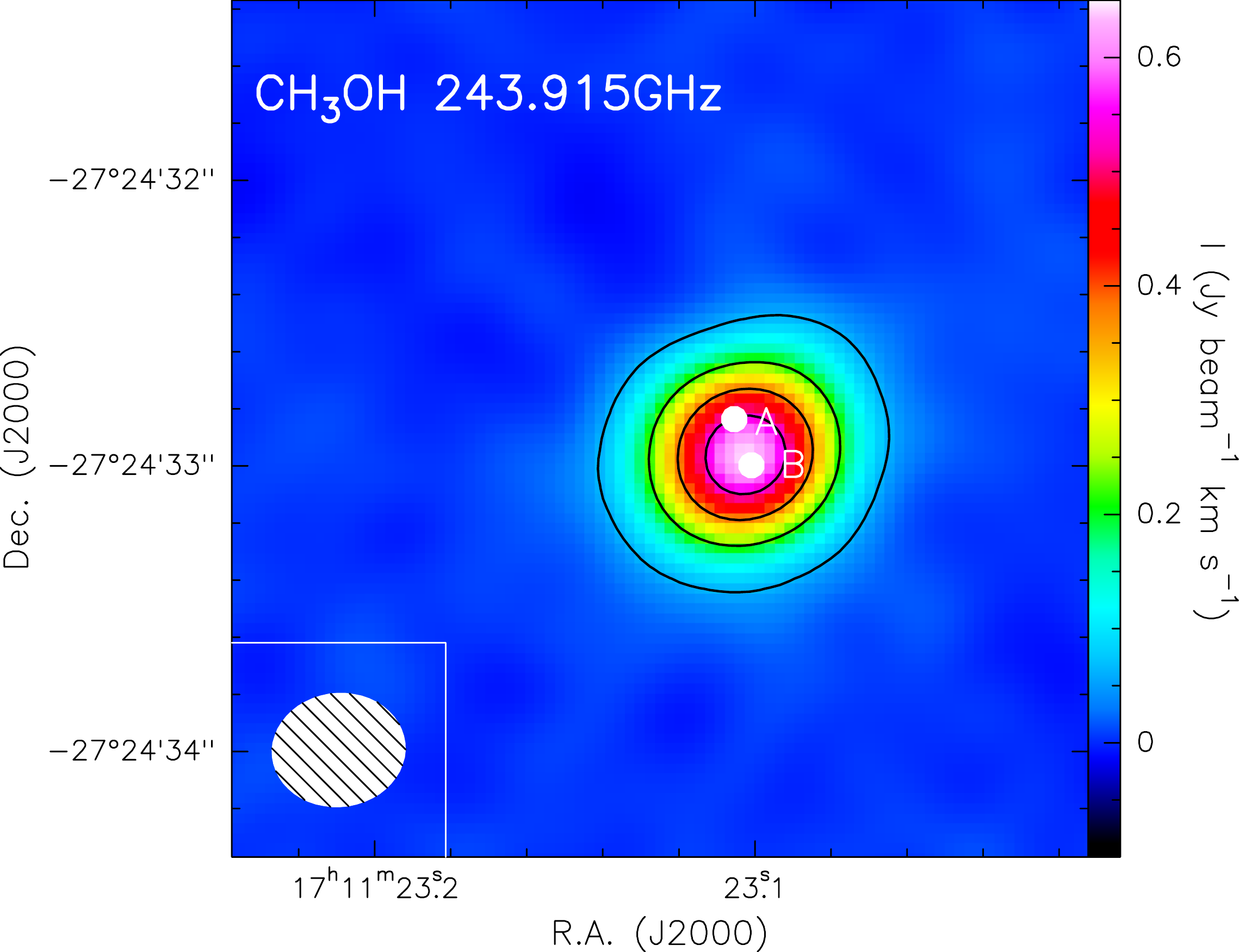} \ 
    \includegraphics[width=0.45\textwidth,trim=0 0 0 0 clip]{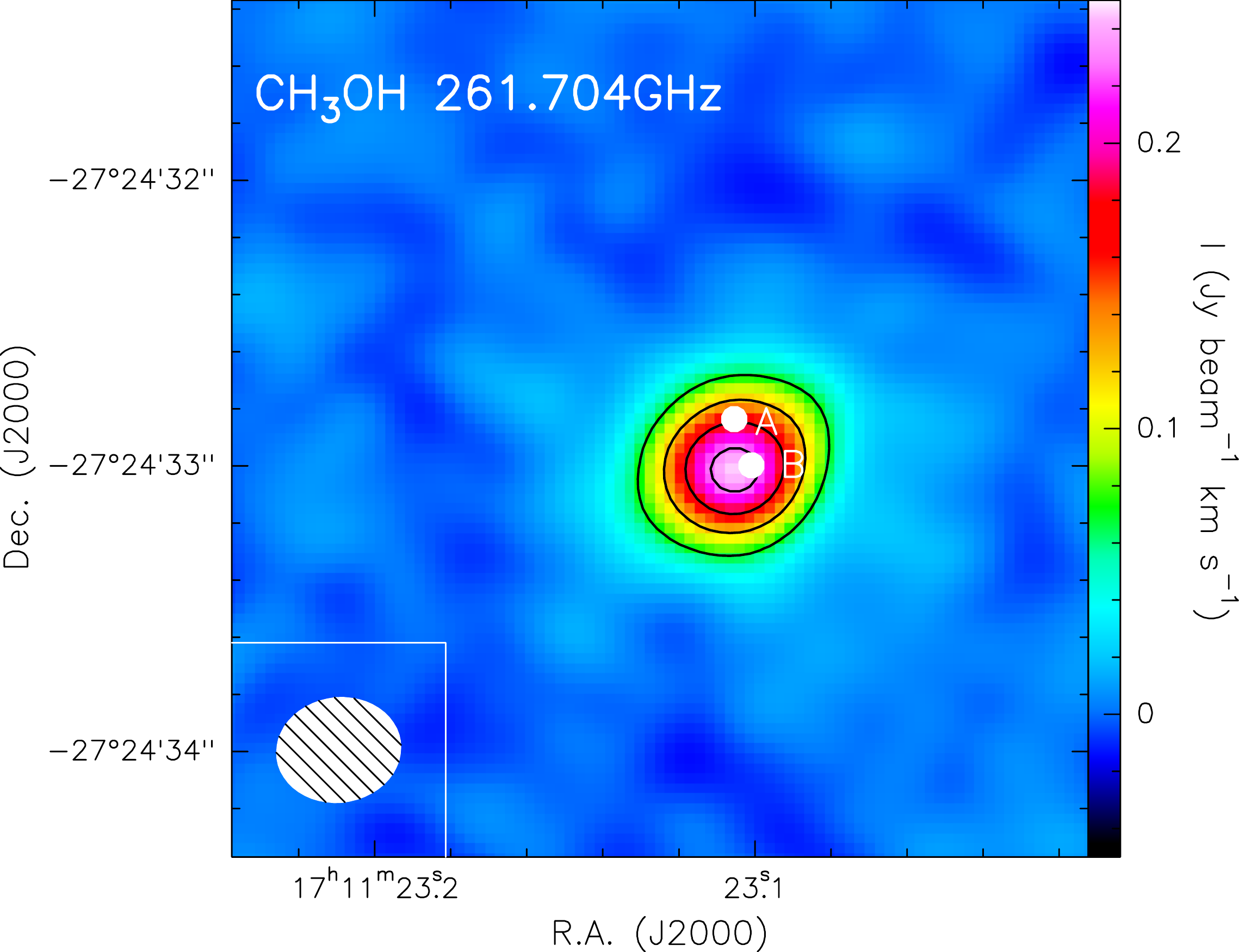} \ 
    \caption{Moment 0 map of the high spectral resolution methanol transitions at 218.440 GHz ($\rm E_{u}$=45.46 K), 243.916 GHz ($\rm E_{u}$=49.66 K) and 261.704 GHz ($\rm E_{u}$=359.77 K). Contours start at 4$\sigma$ at every 10$\sigma$ for 218.440 and 243.916 GHz, and every 3$\sigma$ at 261.704 GHz. The ellipse in the bottom left corner represents the ALMA synthesised beam. Both source A and B identified by \citet{Alves2019} are indicated as white filled circles.}
    \label{fig: mom0}
\end{figure}

Figure \ref{fig: mom0} shows the moment 0 maps (in Jy~beam~km~s$^{-1}$) of the CH$_3$OH high spectral resolution lines at 218, 244 and 262 GHz using the 12m array. These three transitions are the only ones observed at high spectral resolution (0.1--0.2 km~s$^{-1}$). The contours start at 4$\sigma$ at every 10$\sigma$ for 218.440 and 243.916 GHz, and every 3$\sigma$ at 261.704 GHz. The ellipse in the bottom left corner represents the ALMA synthesised beam. The emission is compact (less than 1$^{\prime\prime}$) and centred at the same location, which means that, in this region, the gas is co-existent, from the low energy transition to the higher energy transition. The beam encompasses both 11A and 11B, separated by 0.2$^{\prime\prime}$, which therefore cannot be resolved. The other transitions where the spectral resolution is lower (0.6 -- 1.2 km~s$^{-1}$) also show a centred compact emission.

From Fig. \ref{fig: mom0} it is obvious that all methanol transitions do not necessarily trace the same extended structure. The 218 GHz and 244 GHz transitions, which have an upper energy of 45 K and 50 K respectively, are tracing a more extended region (around 1$^{\prime\prime}$) compared to the highest energy levels at 540 K for the 246.074 GHz transition tracing a seemingly more compact region (around 0.5$^{\prime\prime}$).

\subsubsection{CH$_{3}$OH line identification}

\begin{figure}
    \centering
    \includegraphics[width = 0.45\textwidth,trim=0 0 0 0 clip]{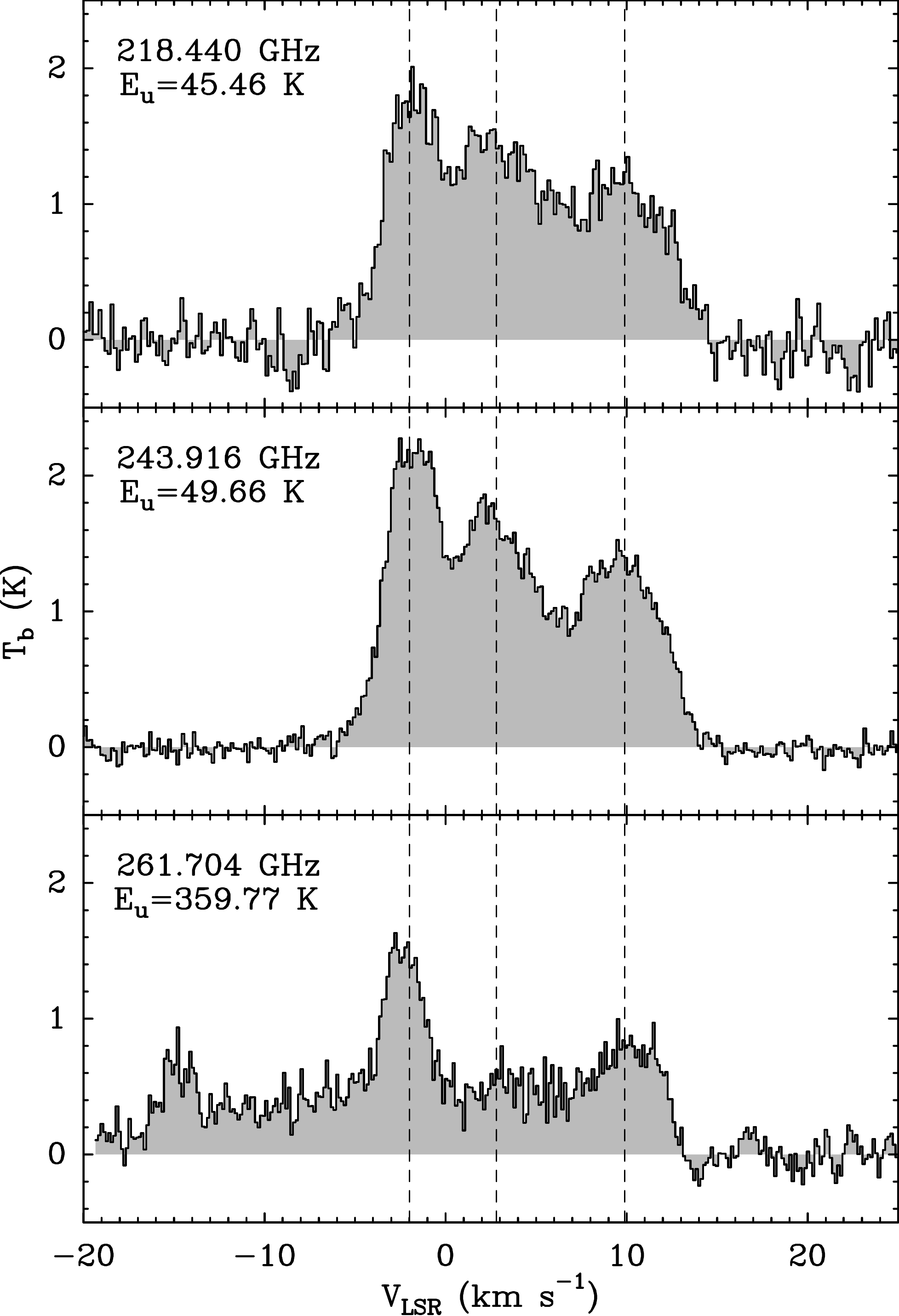}
    \caption{T$\rm _{b}$ (K) spectra for the high spectral resolution Methanol transitions at 218.440, 243.916 and 261.704 GHz. The dashed lines correspond to the results of the best-fit Gaussian at -2 km~s$^{-1}$, 2.8 km~s$^{-1}$ and 9.9 km~s$^{-1}$.  The -2 km~s$^{-1}$ component of the 261.704 GHz transition is blended with a transition of methyl formate (see text).} 
    \label{fig: spectra-HR}
\end{figure}

\begin{figure}
    \centering
    \includegraphics[width = 0.45\textwidth,trim=0 0 0 0 clip]{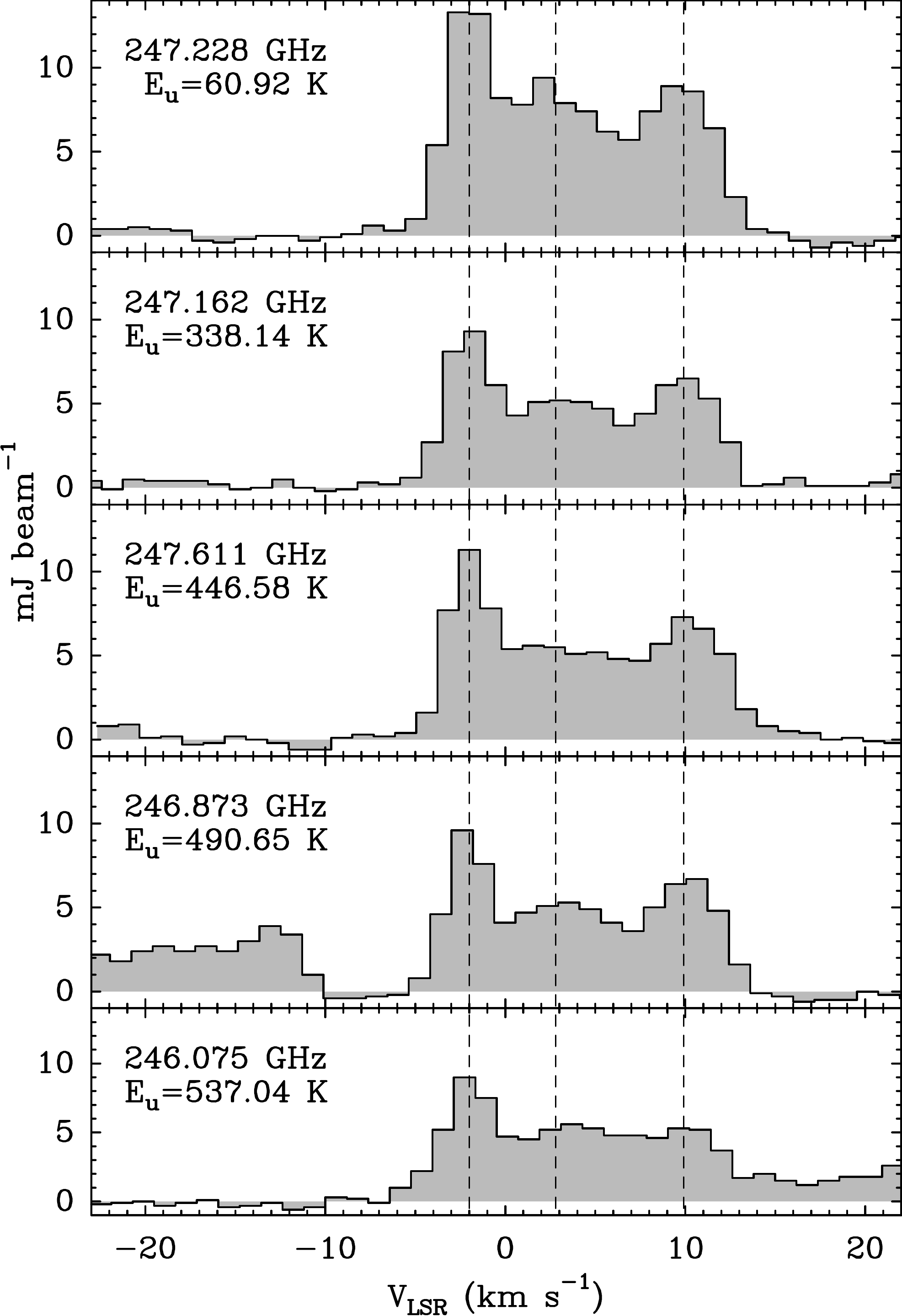}
    \caption{Spectra (in mJy~beam$^{-1}$) for the low spectral resolution Methanol transitions. The dashed lines correspond to the results of the best-fit Gaussian of the high spectral resolution transitions at -2 km~s$^{-1}$, 2.8 km~s$^{-1}$ and 9.9 km~s$^{-1}$.}
    \label{fig: spectra-LR}
\end{figure}

\begin{figure}
    \centering
    \includegraphics[width = 0.45\textwidth,trim=0 0 0 0 clip]{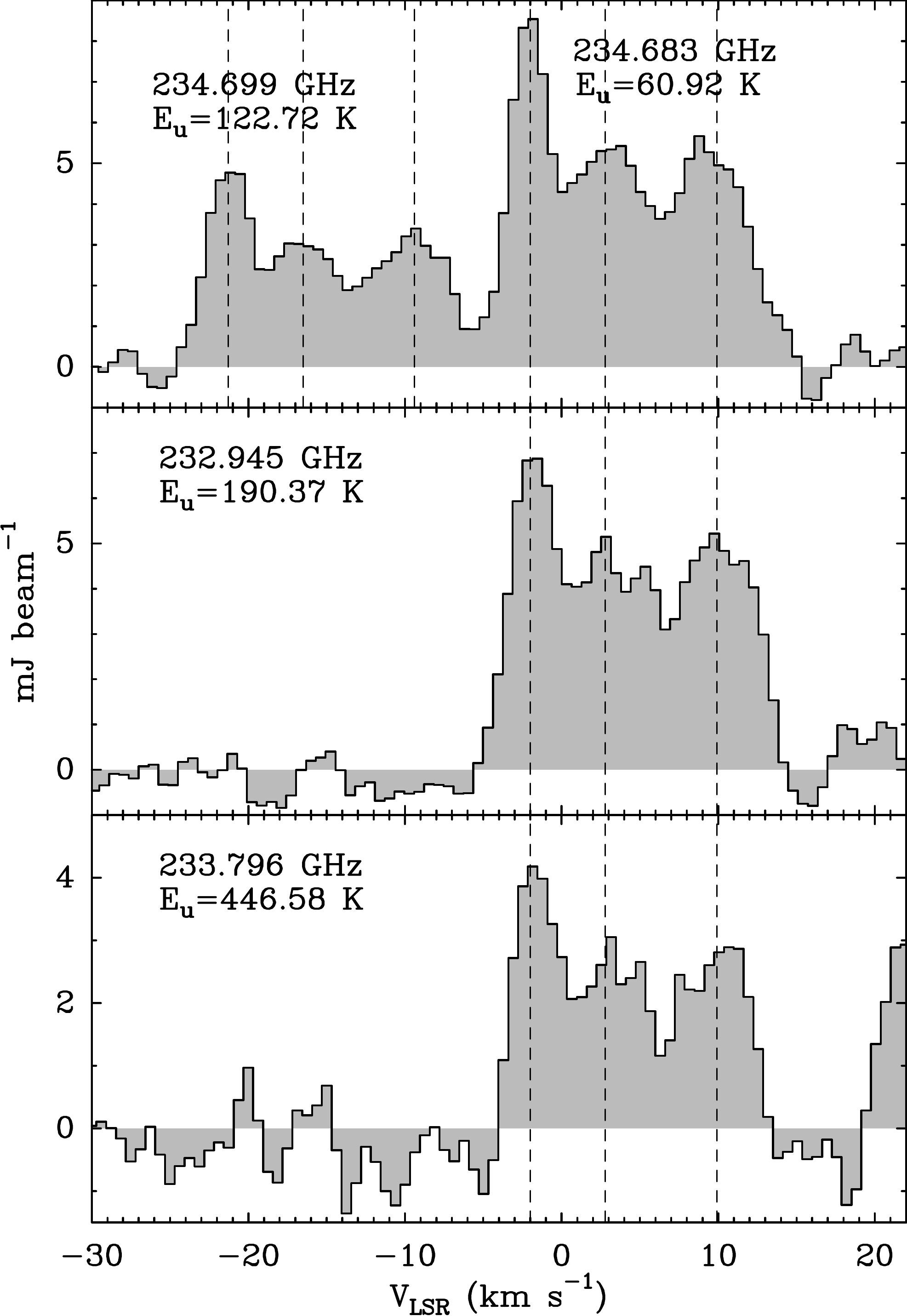}
    \caption{Spectra (in mJy~beam$^{-1}$) for the low spectral resolution methanol transitions in the 234 GHz spectral windows. The dashed lines correspond to the results of the Gaussian components obtained from each of the high spectral resolution transitions at -2 km~s$^{-1}$, 2.8 km~s$^{-1}$ and 9.9 km~s$^{-1}$.}
    \label{fig: spectra-LR234}
\end{figure}

We extracted the average spectrum over the region above 4$\sigma$ in the integrated intensity map for each molecular transition (either in Jy~beam$^{-1}$ or in Kelvin) into the CASSIS software and the high spectral resolution transitions are presented in Fig. \ref{fig: spectra-HR} while the lower spectral resolution transitions of methanol are presented in Fig. \ref{fig: spectra-LR} and \ref{fig: spectra-LR234}. From these figures we can clearly distinguish three components, which are visible at -2 km~s$^{-1}$, 2.8 km~s$^{-1}$ and 9.9 km~s$^{-1}$. Table \ref{tab: spectro} shows the results of the best-fit adjustment using the Levenberg-Marquardt fitting within CASSIS using a large number of iterations (T$\rm_{mb}$ in Kelvin, FWHM and V$\rm_{LSR}$ in km~s$^{-1}$).

\begin{figure}
    \centering
    \includegraphics[width = 0.45\textwidth,trim=0 0 0 0 clip]{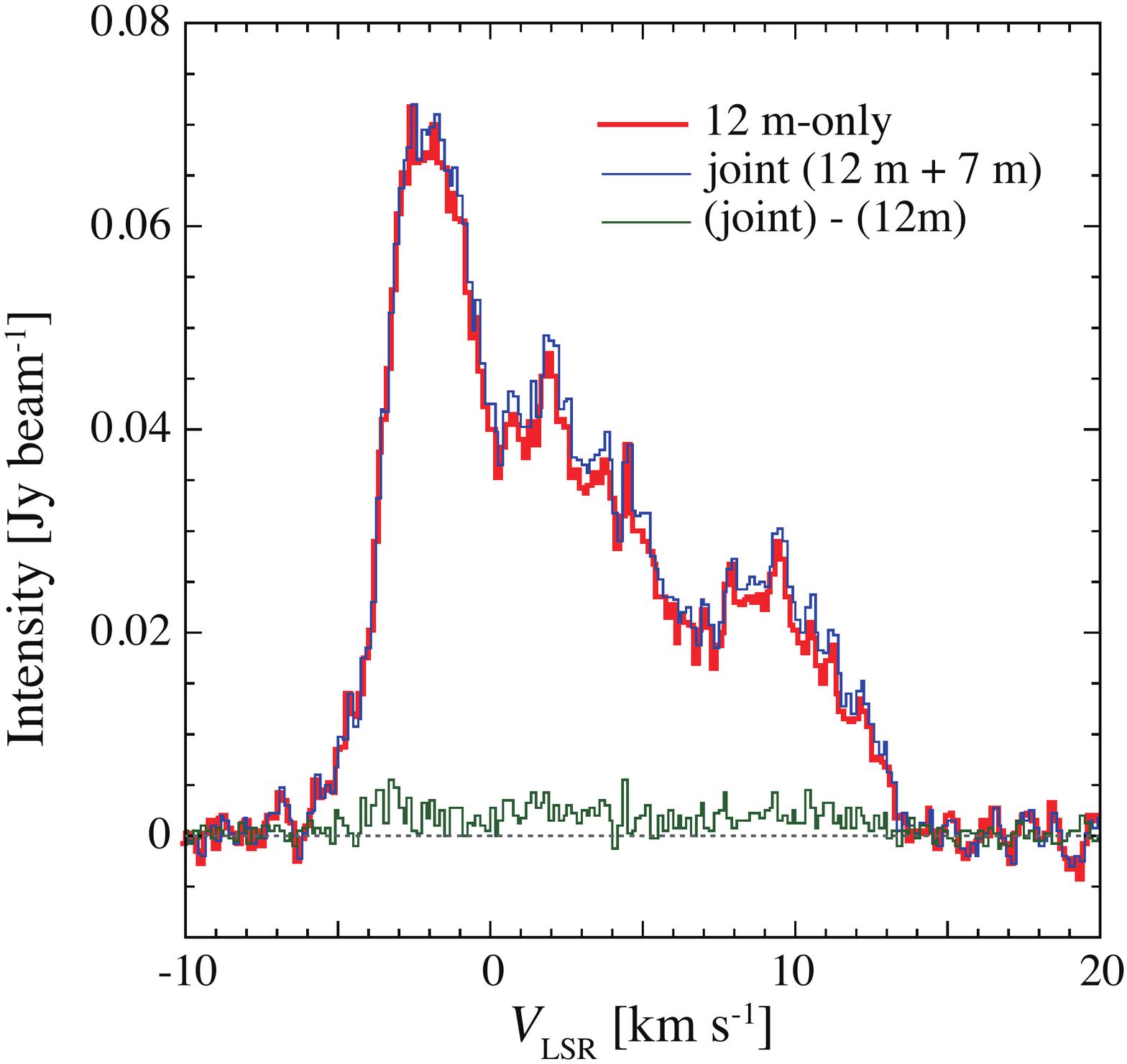}
    \caption{Difference between the 12 m-only data and the joint data for the A-CH$_{3}$OH transition at 243.916 GHz.}
    \label{fig: missing}
\end{figure}

The maximum recoverable scale of the 12-m array data is 9$^{\prime\prime}$--11$^{\prime\prime}$. Since the CH$_3$OH emission is compact and its size is less than 3$^{\prime\prime}$, the missing flux in the 12 m array data is negligible. In fact, we confirmed that the flux of the data combined with the 7-m and 12-m array data (maximum recoverable scale: 16$^{\prime\prime}$--19$^{\prime\prime}$) is comparable with that of the 12-m array data (see Fig. \ref{fig: missing} for the 243.916 GHz transition). We therefore used in the following the 12-m array data only.

\begin{figure*}
    \centering
    \includegraphics[width=\textwidth]{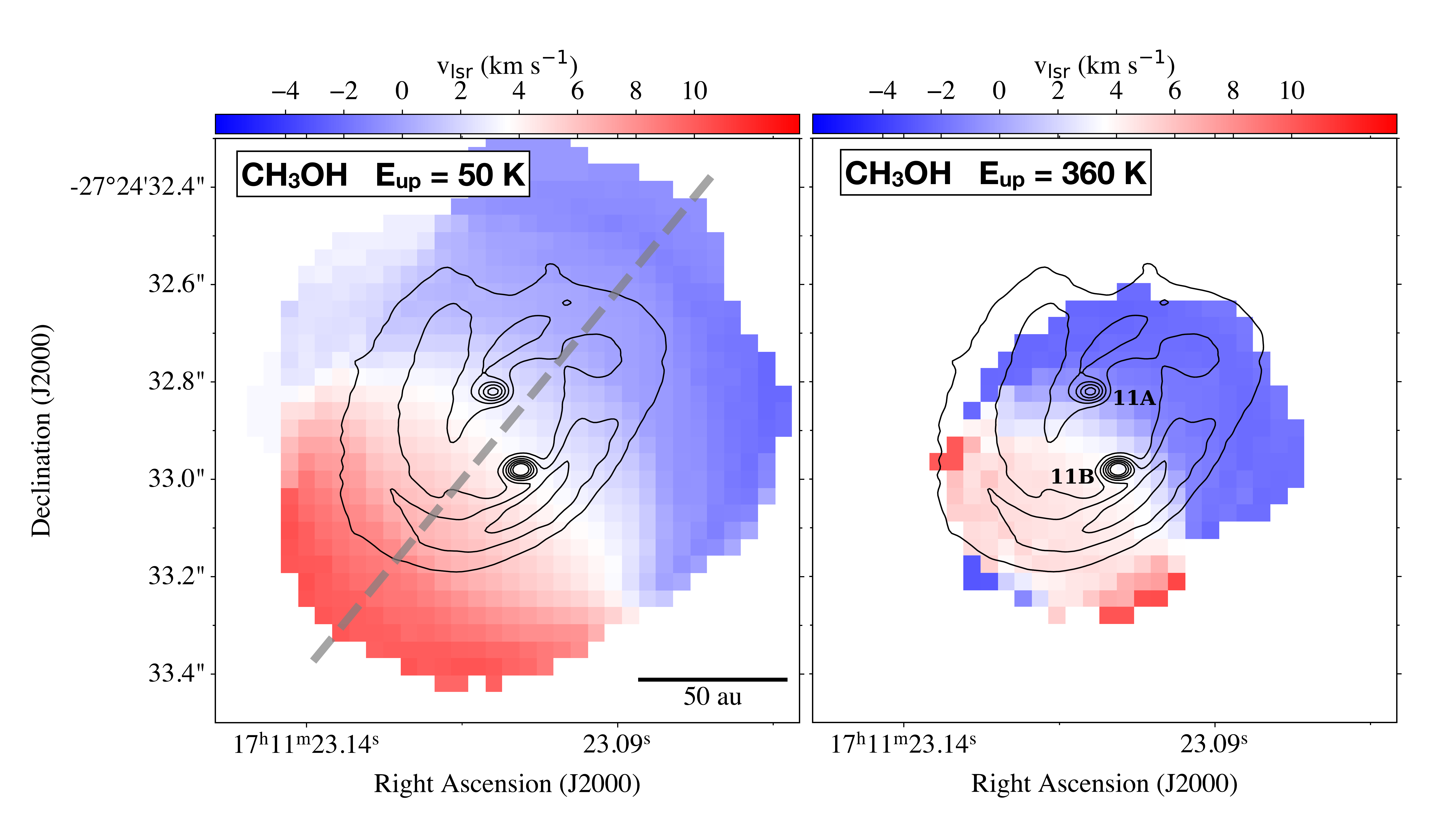}
        \caption{Velocity field determined from methanol transitions at low and high energy levels. The contours indicate the continuum intensity levels from the high spatial resolution observations of \citet{Alves2019}. The compact emission contours arise from the circumstellar disks around each protostar of the binary system: 11A (northern component) and 11B (southern component). The dashed line in the left panel shows the direction of the cut used to produce the position-velocity diagram displayed in Fig. \ref{fig:pvplot}. The cut position angle is 140$\degr$ and its width is $0.5^{\prime\prime}$, encompassing both protostars.}
    \label{fig:momcont}
\end{figure*}

Fig. \ref{fig:momcont} shows the velocity field (moment 1) of the two extreme (in upper energy values) high spectral resolution methanol transitions at $\sim 50$ K (243.916 GHz) and $\sim 350$ K (261.704 GHz) upper
energy levels (E$_{\mathrm{up}}$). The velocity structure ranges from -5 to 15 km~s$^{-1}$ in both low and high transitions, and it has a northwest-southeast gradient centred on the 11B source at the systemic velocity of the source. Although we cannot resolve the protostars in these maps, this could indicate that the methanol maps are more sensitive to the kinematics of the southern source. 

\begin{figure}
    \centering
    \includegraphics[width = 0.45\textwidth,trim=0 0 0 0 clip]{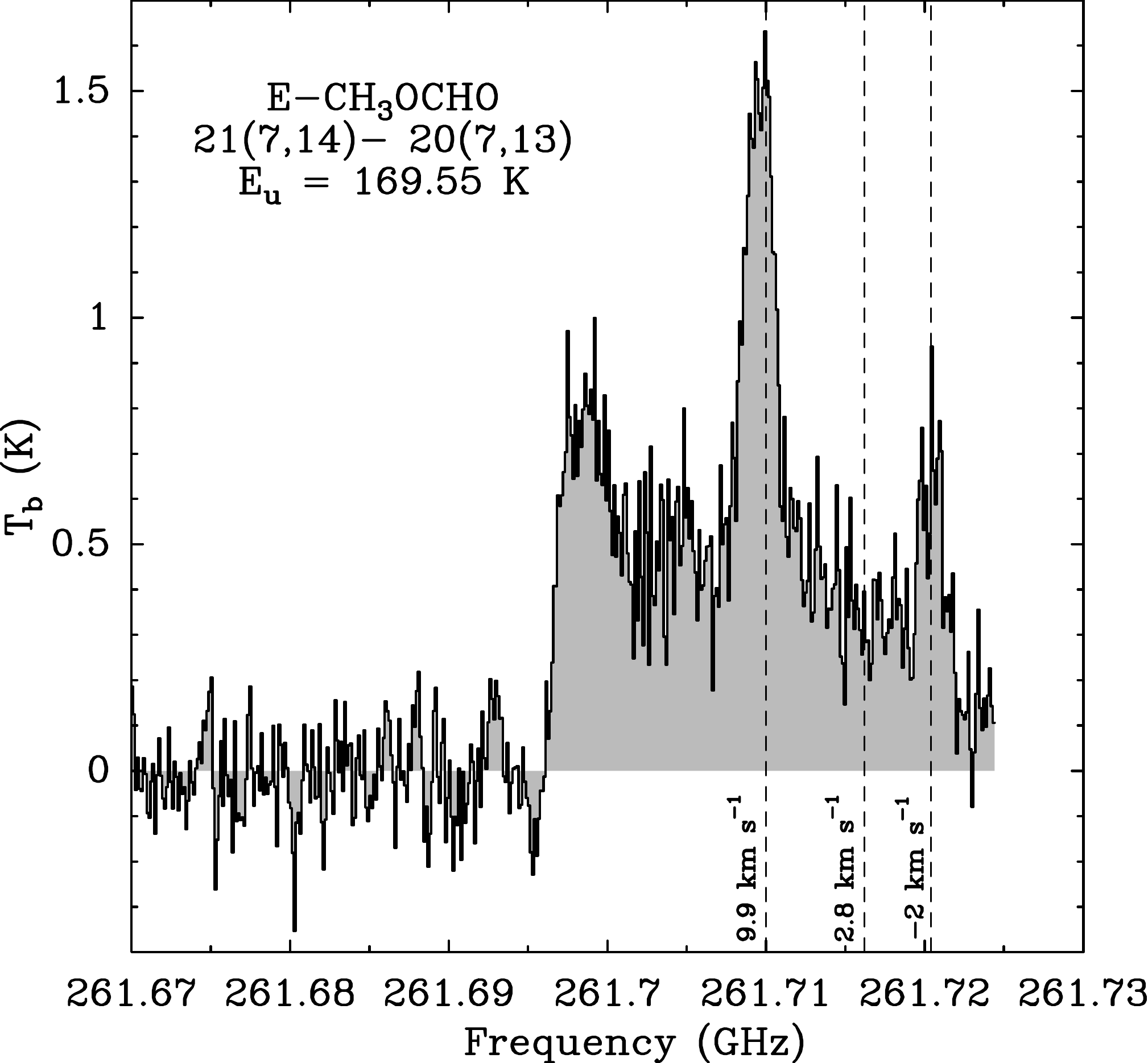}
    \caption{Spectra at the 3.6 km~s$^{-1}$ V$_{LSR}$ of the \bhb source showing the E-CH$_{3}$OCHO 21(7,14)- 20(7,13) transition at 261.715 GHz blended with the CH$_{3}$OH transition at 261.704 GHz. The dashed lines correspond to the results of the best-fit Gaussian components obtained from the high spectral resolution methanol transitions at -2 km~s$^{-1}$, 2.8 km~s$^{-1}$ and 9.9 km~s$^{-1}$ for the methyl formate transition. }
    \label{fig: CH3OCHO}
\end{figure}

Due to the molecular complexity of this source we explored the line blending from other species. The CH$_{3}$OH transition at 261.704 GHz is indeed clearly blended, as shown in Fig. \ref{fig: spectra-HR}, with the 21(7,14)- 20(7,13) E transition of methyl formate (CH$_{3}$OCHO) at 261715.5180 MHz. Figure \ref{fig: CH3OCHO} shows the methyl formate transition with the three components at -2 km~s$^{-1}$, 2.8 km~s$^{-1}$ and 9.9 km~s$^{-1}$ indicated as thin dashed lines. It is clear that the -2 km~s$^{-1}$ methanol transition is blended with the 9.9 km~s$^{-1}$ transition of methyl formate. A subsequent analysis of the iCOMs detected in our \bhb FAUST data will be published at a later time.

Within CASA ({\it imfit} command), we fitted a 2D gaussian to the integrated intensity maps of the three high spectral resolution transitions, taking into account the three velocity components. At 243.916 GHz, the image component size (deconvolved from beam) for (i) the 9.9 km~s$^{-1}$ velocity component (channel 195) is 0.27$^{\prime\prime}$ $\times$ 0.23$^{\prime\prime}$ (major axis FWHM $\times$ minor axis FWHM), (ii) the -2 km~s$^{-1}$ (channel 274)  is 0.27$^{\prime\prime}$ $\times$ 0.16$^{\prime\prime}$ (major axis FWHM $\times$ minor axis FWHM), (iii) the 2.8 km~s$^{-1}$ (channel 242) is 0.53$^{\prime\prime}$ $\times$ 0.36$^{\prime\prime}$ (major axis FWHM $\times$ minor axis FWHM). The 218.440 GHz give similar results and the 261.704 GHz non-blended components at +9.9 and 2.8 km~s$^{-1}$ tend to a point source deconvolved source size. Therefore, the two extreme velocity components seem slightly more compact than the 2.8 km~s$^{-1}$.

\subsubsection{$^{13}$CH$_{3}$OH line identification}

\begin{figure}
    \centering
    \includegraphics[width = 0.45\textwidth,trim=0 0 0 0 clip]{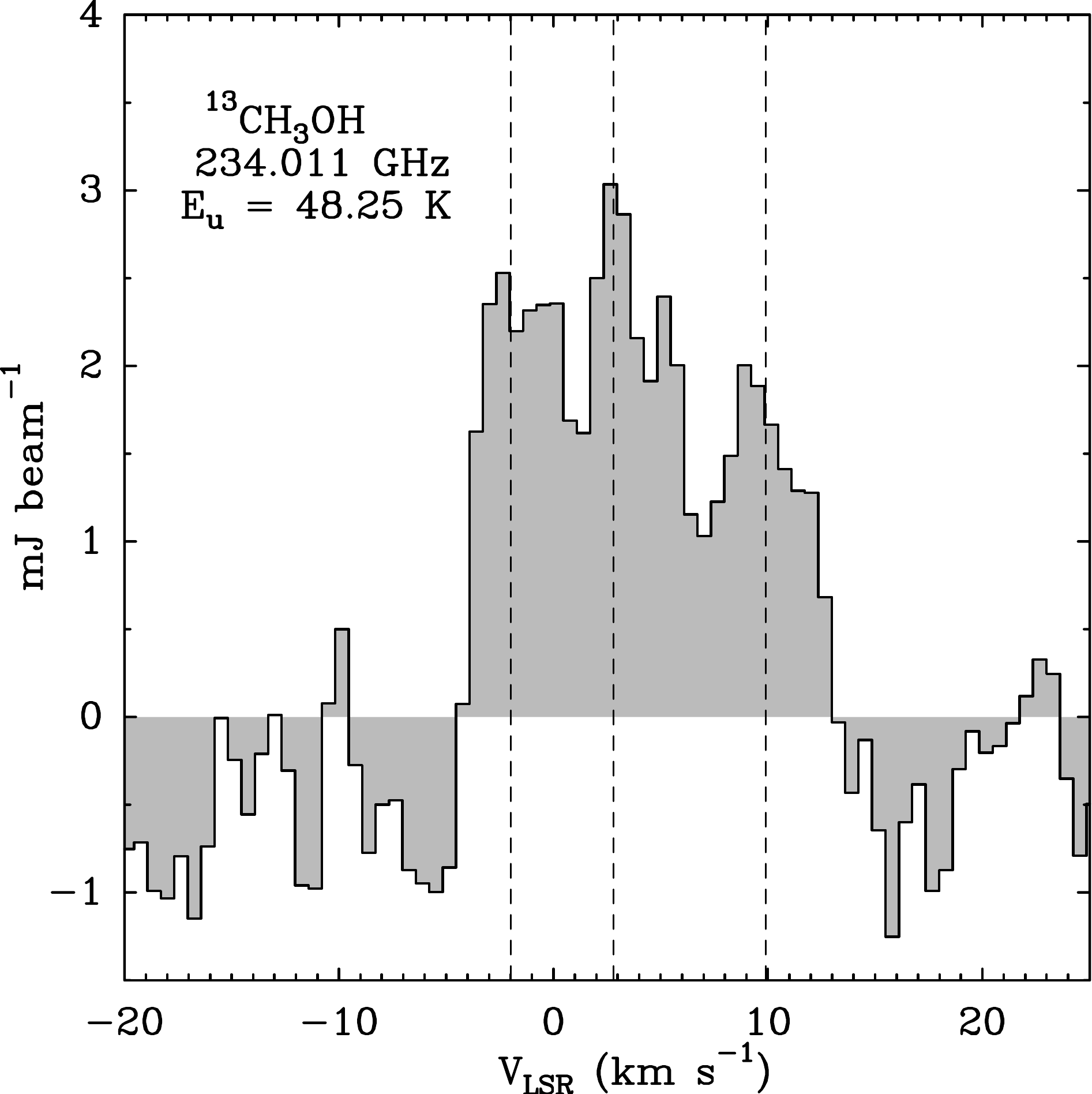}
    \caption{Spectrum (in mJ~beam$^{-1}$) of the only detected transition of $^{13}$CH$_{3}$OH at  234 GHz. The dashed lines correspond to the results of the best-fit Gaussian components obtained from the high spectral resolution transitions at -2 km~s$^{-1}$, 2.8 km~s$^{-1}$ and 9.9 km~s$^{-1}$.}
    \label{fig: 13CH3OH}
\end{figure}

\begin{figure}
    \centering
    \includegraphics[width = 0.45\textwidth,trim=0 0 0 0 clip]{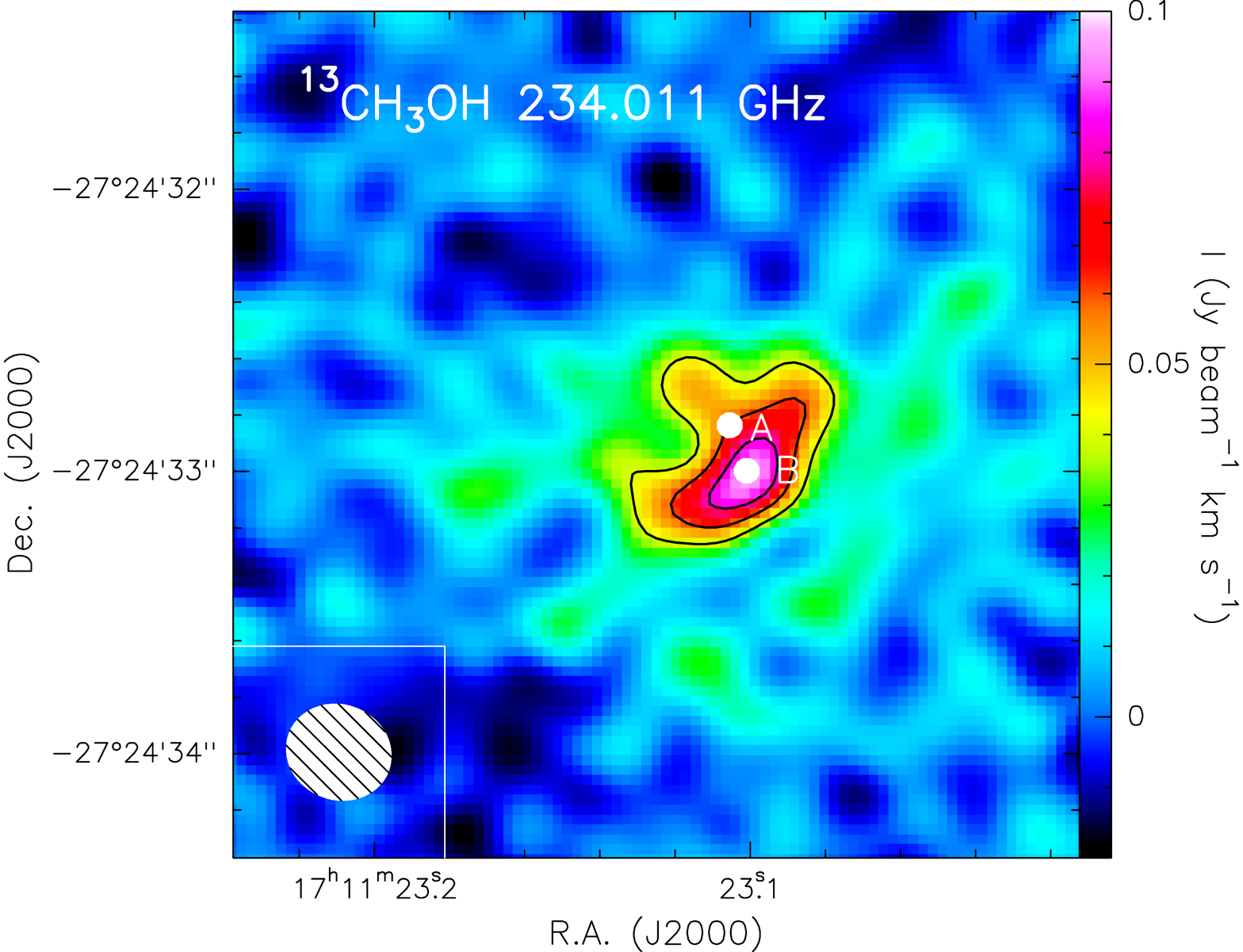}
    \caption{Moment 0 map of the $^{13}$CH$_3$OH transition at 234 GHz. Contours start at 4$\sigma$ at every 2$\sigma$. The ellipse in the bottom left corner represents the ALMA synthesised beam. Both source 11A and 11B identified by \citet{Alves2019} are indicated as white filled circles.}
    \label{fig: mom0_13}
\end{figure}

Only one transition of the $^{13}$C methanol isotopologue has been detected at 234 GHz with $\rm E_{up}$ = 48 K, and it presents the same characteristics as methanol and methyl formate with three components, as shown in Figure \ref{fig: 13CH3OH}. We present in Table \ref{tab: spectro} the results from the best-fit Gaussian adjustments for this transition. Figure \ref{fig: mom0_13} shows that the integrated line intensity is much noisier than the other methanol moment 0 maps but still compact and centred on the same position. 

\section{Modelling of the methanol and dust emission}\label{methanol-modeling}

We first performed a local thermodynamic equilibrium (LTE) analysis to estimate the excitation temperature and column density for our methanol transitions. To achieve that, we defined a polygon extracted from the average spectrum over a region above 4$\sigma$ in the integrated intensity map of the 218.440 GHz methanol transition. This polygon was then used to extract the average spectrum for each methanol transition. We then performed a best-fit of the three components and obtained a rotational diagram (RD) analysis followed by a non-LTE analysis to estimate the H$_{2}$ volume density, the column density, kinetic temperature and size of the source. The integrated line intensities used in the following are presented in Table \ref{tab: integrated}.

\begin{table}
\caption{Integrated intensities used for the LTE and non-LTE analysis, extracted over a region above 4$\sigma$ in the integrated intensity map of the 218.440 GHz methanol transition.}
\label{tab: integrated}
\centering
\begin{tabular}{c c c c c}
\hline 
                                                &  W(-2 km~s$^{-1}$)                                & W(2.8 km~s$^{-1}$)         & W(9.9 km~s$^{-1}$)   \\
     GHz                                  &  K~km~s$^{-1}$                                      &    K~km~s$^{-1}$             &  K~km~s$^{-1}$\\
\hline
\hline 
CH$_{3}$OH                           &                       &                         &                        \\   
\hline
218.440$^{*}$                           &  $5.4\pm0.2$  &   $9.5\pm0.2$  &   $6.7\pm0.2$  \\
243.916$^{*}$                            &  $5.2\pm0.1$  &   $7.9\pm0.1$  &   $5.6\pm0.1$  \\
247.229$^{*}$                            &  $3.4\pm0.1$  &   $4.3\pm0.1$  &   $3.9\pm0.1$  \\
234.683$^{*}$                            &  $3.1\pm0.1$  &   $4.0\pm0.1$  &   $3.8\pm0.1$  \\
234.698$^{*}$                            &  $1.8\pm0.1$  &   $2.5\pm0.1$  &   $2.5\pm0.1$  \\
232.946$^{*}$                            &  $2.7\pm0.1$  &   $3.6\pm0.1$  &   $4.2\pm0.1$  \\
247.162                            &  $3.2\pm0.2$  &   $3.7\pm0.2$  &   $3.8\pm0.2$                     \\
261.704$^{*}$                            &  $1.1\pm0.1$  &   $1.5\pm0.1$  &   $1.9\pm0.1$  \\
233.796                            &  $1.9\pm0.4$  &   $3.2\pm0.4$  &   $2.4\pm0.4$  \\
247.611                            &  $3.6\pm0.1$  &   $4.0\pm0.1$  &    $4.4\pm0.1$ \\
246.873                            &  $2.9\pm0.1$  &   $3.7\pm0.1$  &   $3.9\pm0.1$ \\
246.075                            &  $3.2\pm0.1$  &   $3.9\pm0.1$  &   $3.5\pm0.1$ \\
\hline
\hline
$^{13}$CH$_{3}$OH               &                       &                         &                        \\  
\hline
234.011$^{*}$                            &  $0.9\pm0.1$  &   $2.3\pm0.2$  &   $1.4\pm0.2$  \\

\hline
\end{tabular}
\tablefoot{Transitions used in the LVG are indicated by the star symbol.}
\end{table}

\subsection{LTE Rotational Diagram analysis}\label{subsec:RD}
\label{sec: RD} 

For a molecule in LTE, all excitation temperatures are the same, and the population of each level is given by:
\begin{equation}
\rm N_{u}=\frac{N_{tot}}{Q(T_{rot})}g_{u}e^{-E_u/kT_{rot}},
\end{equation}
where Q is the partition function, T$_{rot}$ is the rotational temperature, g$_{u}$ is the statistical weight in the upper level, E$_{u}$ is the energy of the upper level, k is the Boltzmann constant, N$_{u}$ is the column density in the upper level and T$_{tot}$ is the column density of the species. We can rewrite this equation to obtain:
\begin{equation}
\rm ln\frac{N_u}{g_u} = ln \frac{N_{tot}}{Q(T_{rot})} -\frac{E_u}{kT_{rot}}.
\label{rot}
\end{equation}

A rotational diagram can be useful to determine whether the emission is optically thick or thin, whether the level populations are described by LTE, and to determine what temperature describes the population distribution in the event that LTE applies \citep{Goldsmith1999}.
Equation \ref{rot}  can be written in terms of the observed integrated area W when the lines are optically thin:
\begin{equation}
\rm ln\frac{8\pi k\nu^2 W}{hc^3A_{ul}g_u} = ln \frac{N_{tot}}{Q(T_{rot})} - \frac{E_u}{kT_{rot}},
\end{equation}
where h is the Planck constant, A$_{ul}$ is the Einstein coefficient between level $up$ and level $low$, $\nu$ is the frequency for each observed transition  and c is the celerity of light.
The error bars should be taken into account for the order 1 polynomial fit, in order to obtain a reliable value for the uncertainty on the rotational temperature as well as the total column density. From the CASSIS Line Analysis module, the user can fit the lines and produce a detailed {\it .rotd} file with the integrated area and {\it rms} values for each transition. The user is required to give a value for the instrumental calibration uncertainty. The uncertainty ($\Delta$W) of the integrated area is computed though the following formula: 
\begin{equation}
\rm \Delta W = \sqrt{(\rm cal/100 \times W)^2 + (rms \sqrt{2 \times \rm FWHM \times \delta V})^2},
\label{eq: error}
\end{equation}
where cal is the calibration value ($\%$), W is the integrated area (in K~km~s$^{-1}$), rms is the noise around the selected species (in K), FWHM is the full width at half maximum (km~s$^{-1}$) and $\delta$V is the bin size (in km~s$^{-1}$). Therefore, the plotted uncertainties are simply:
\begin{equation}
\rm \Delta \left( ln\frac{N_u}{g_u} \right) = \frac{\Delta W}{W}.
\end{equation}
From the fitted straight line ($y = ax+b$) the slope $a$ is related to the rotational excitation temperature as  $T_{rot} = -1/a$. 
Then $\Delta T_{rot} = \Delta a/a^2$. The intercept $b$ is related to the total column density as N$_{tot}$ = Q(T$_{rot}$) $\times$ e$^b$. 
Therefore $\Delta N_{tot}$ = Q(T$_{rot}$) $\times \Delta b \times e^b$.\\

\begin{figure}
    \centering
    \includegraphics[width = 0.85\columnwidth,trim=2cm 0 0 0 clip]{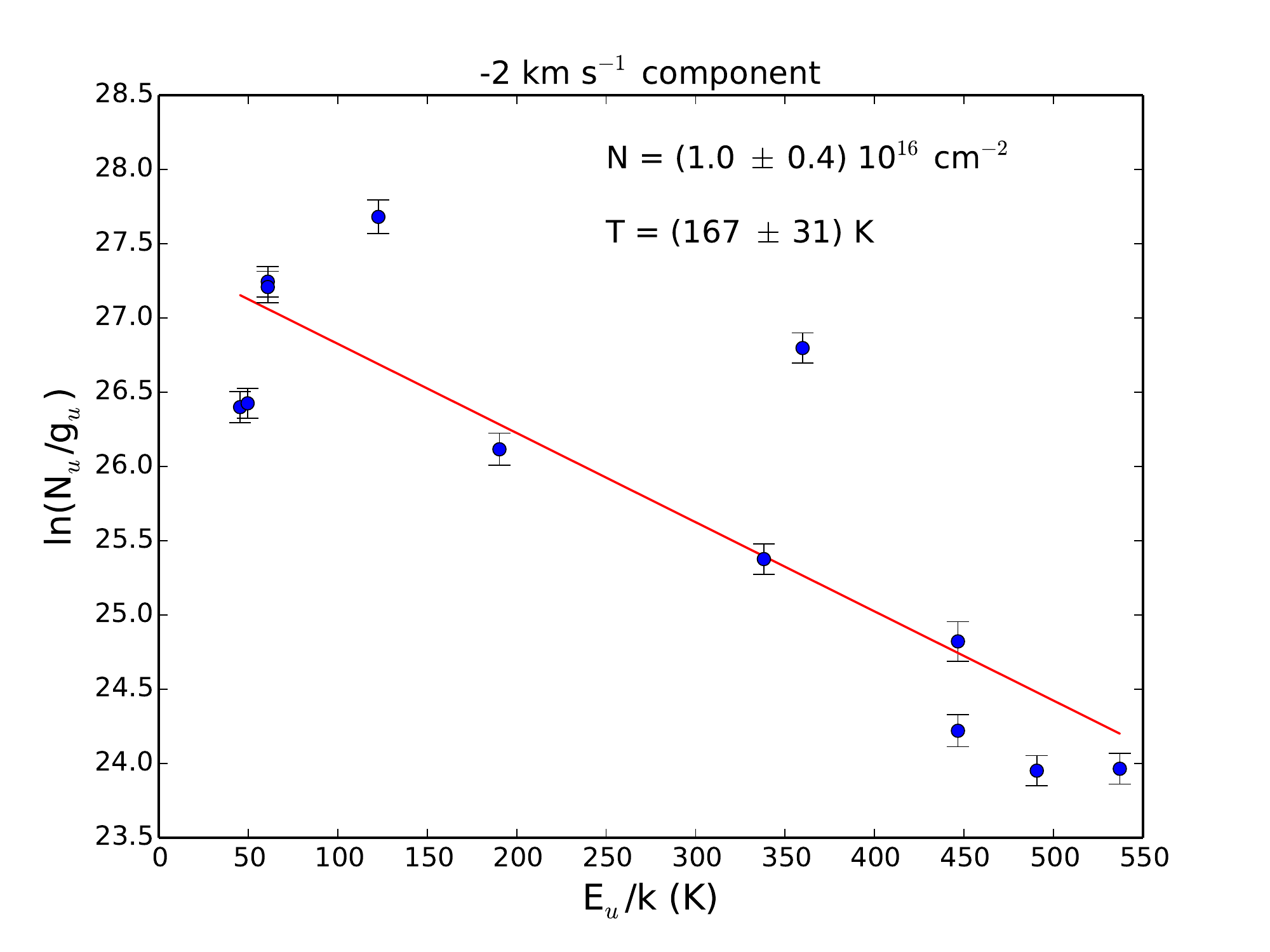}
    \includegraphics[width=0.85\columnwidth,trim=2cm 0 0 0 clip]{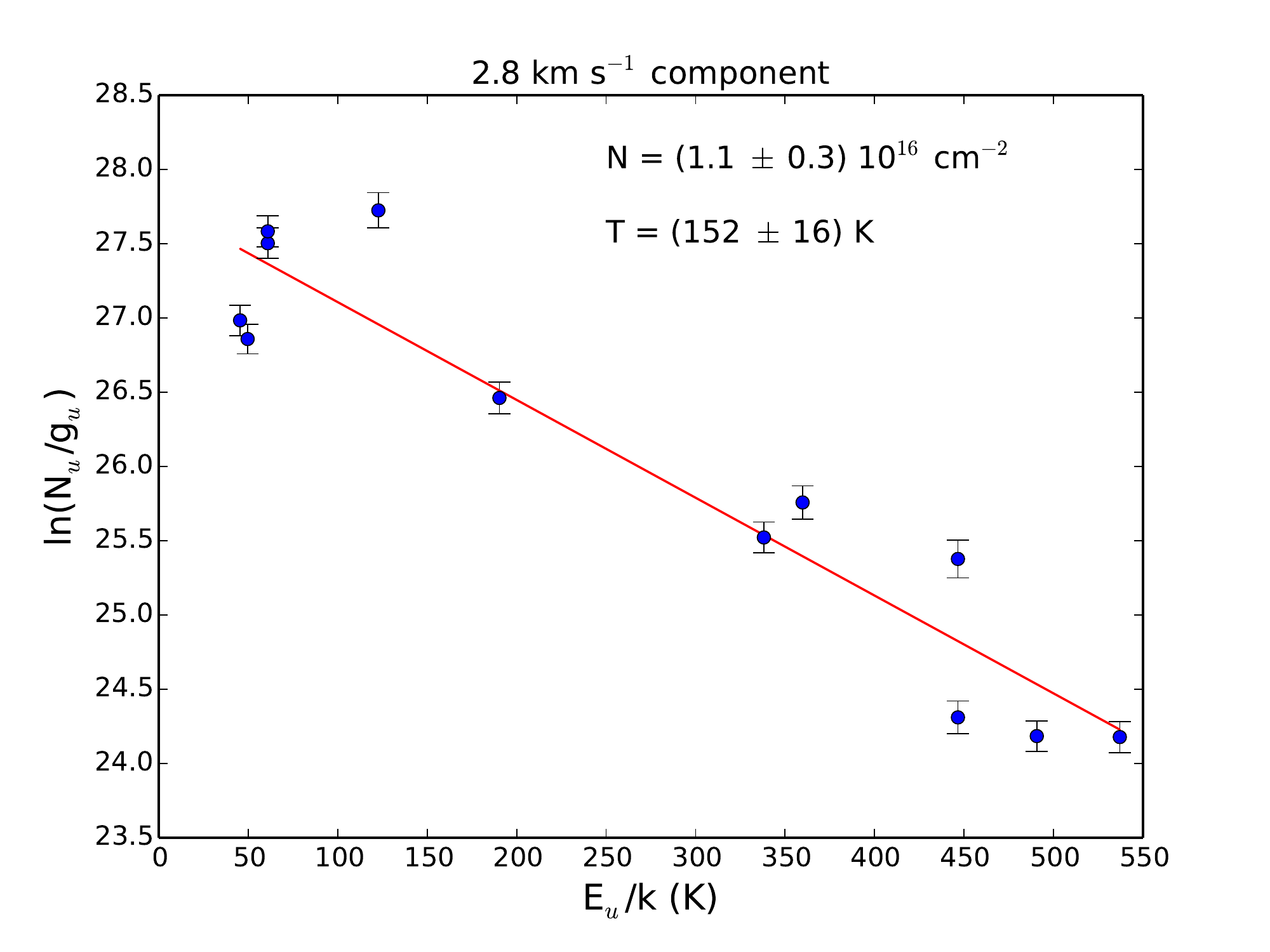} \ 
    \includegraphics[width=0.85\columnwidth,trim=2cm 0 0 0 clip]{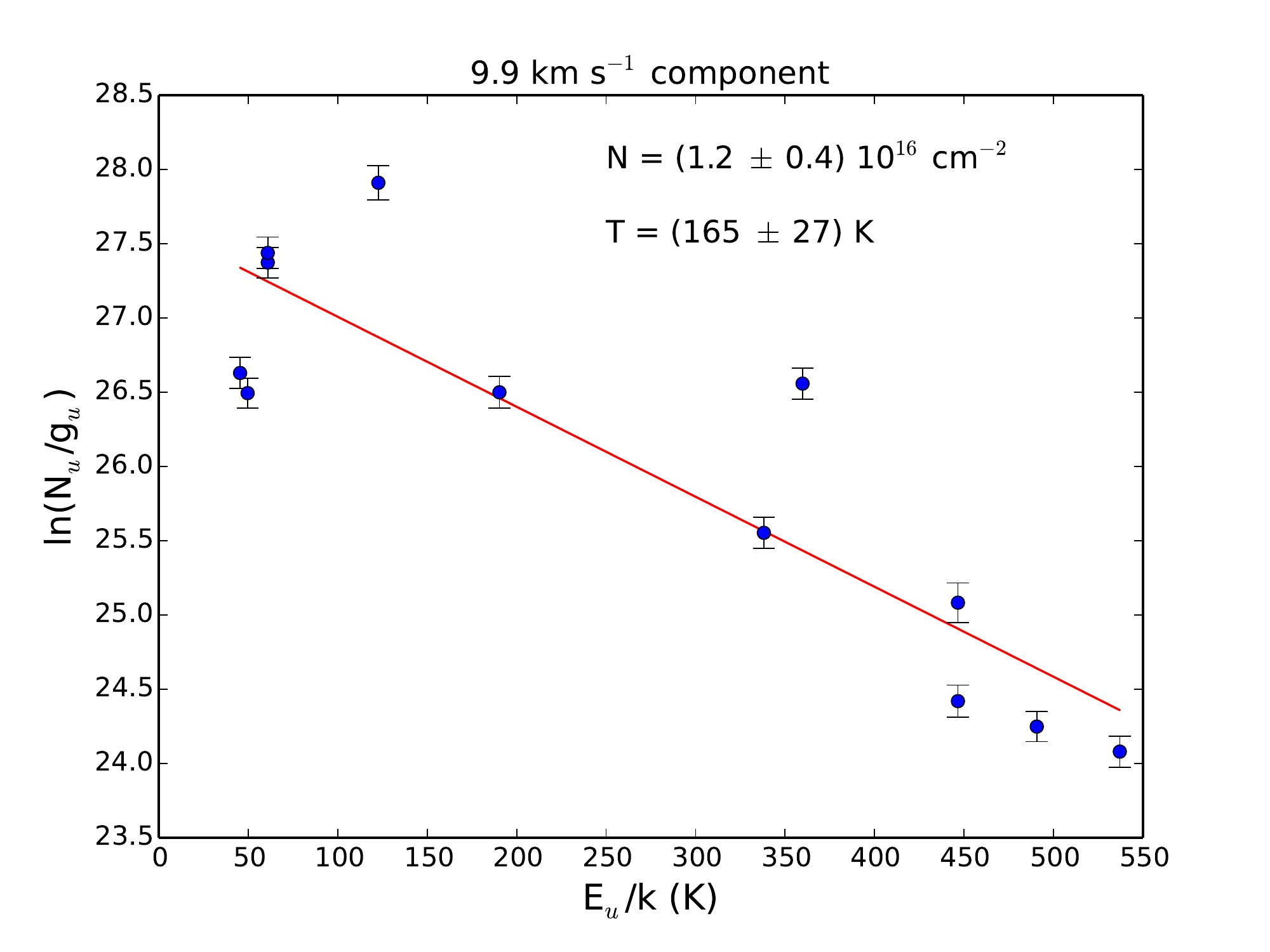} \ 
    \caption{Rotational Diagrams of the detected methanol lines for the -2 km~s$^{-1}$ (top panel), 2.8 km~s$^{-1}$ (middle panel) and 9.9 km~s$^{-1}$ (bottom panel) components, respectively.}
    \label{fig: RD}
\end{figure}

The RD analysis is presented in Figure \ref{fig: RD} along with the resulting column densities and excitation (also called rotational) temperature. All the transitions are shown, although the blended transition of the 261.7 GHz methanol transition at -2 km~s$^{-1}$ is not used in the fit. A large dispersion can be seen in the Figures. The low energy transitions are also tracing the more extended and colder envelope compared to the higher energy transitions. Removing the transitions with upper energy less than 150 K does not change the results because of the wide dispersion in these points.

The estimated parameters for the three components are similar:\\
-2 km~s$^{-1}$: N = (1.0 $\pm$ 0.4) 10$^{16}$ cm$^{-2}$, $\rm T_{rot}$ = (167 $\pm$ 31) K, \\
2.8 km~s$^{-1}$: N = (1.1 $\pm$ 0.3) 10$^{16}$ cm$^{-2}$, $\rm T_{rot}$ = (152 $\pm$ 16) K, \\
9.9 km~s$^{-1}$: N = (1.2 $\pm$ 0.4) 10$^{16}$ cm$^{-2}$, $\rm T_{rot}$ = (165 $\pm$ 27) K, \\
where the column density is the beam-averaged one and the lines are predicted to be optically thin.



\subsection{Non-LTE LVG analysis} \label{subsec:nonLTE}

In order to better take into account the possible/probable non-LTE effects, we modelled the methanol line emission via the Large Velocity Gradient (LVG) code {\texttt grelvg}, developed by \citet{Ceccarelli2003}. We used the collisional coefficients with para-H$_2$, computed by \citet{Rabli2010} between 10 and 200 K for the first 256 levels and provided by the BASECOL database \citep{Dubernet2012,Dubernet2013}. We therefore used the methanol transitions up to J=15 for the non-LTE analysis. We assumed an A-/E- CH$_3$OH ratio equal to 1 and the $^{12}$C/$^{13}$C equal to 68 \citep[e.g.][]{Milam2005}.
To compute the line escape probability as a function of the line optical depth we adopted a spherical geometry and a linewidth equal to 3.2 km~s$^{-1}$ for the -2 km~s$^{-1}$ component, 6 km~s$^{-1}$ for the 2.8 km~s$^{-1}$ component and 5 km~s$^{-1}$ for the 9.9 km~s$^{-1}$ component, following the observations.

We ran a large grid of models ($\sim 10000$) to cover the $\chi^2$ surface in the parameters space: the total (A- plus E- ) methanol column density N(CH$_3$OH) from $1\times 10^{13}$ to $2\times 10^{18}$ cm$^{-2}$, the H$_2$ density n$_{H2}$ from $1\times 10^{5}$ to $1\times 10^{9}$ cm$^{-3}$ and the temperature T from 15 to 200 K.
We then simultaneously fitted the measured $^{12}$C and $^{13}$C -A- and $^{12}$C E- line intensities extracted from the 4$\sigma$ polygon of the 218.440 GHz methanol transition, by comparing them with those predicted by the model, leaving N(CH$_3$OH), n$_{H2}$ and T.
Since the line emission is unresolved with our spatial resolution, we also left the emitting size $\theta_s$ as a free parameter.
We performed this same procedure for the three velocity components separately.

\begin{figure}
    \centering
    \includegraphics[width=0.85\columnwidth]{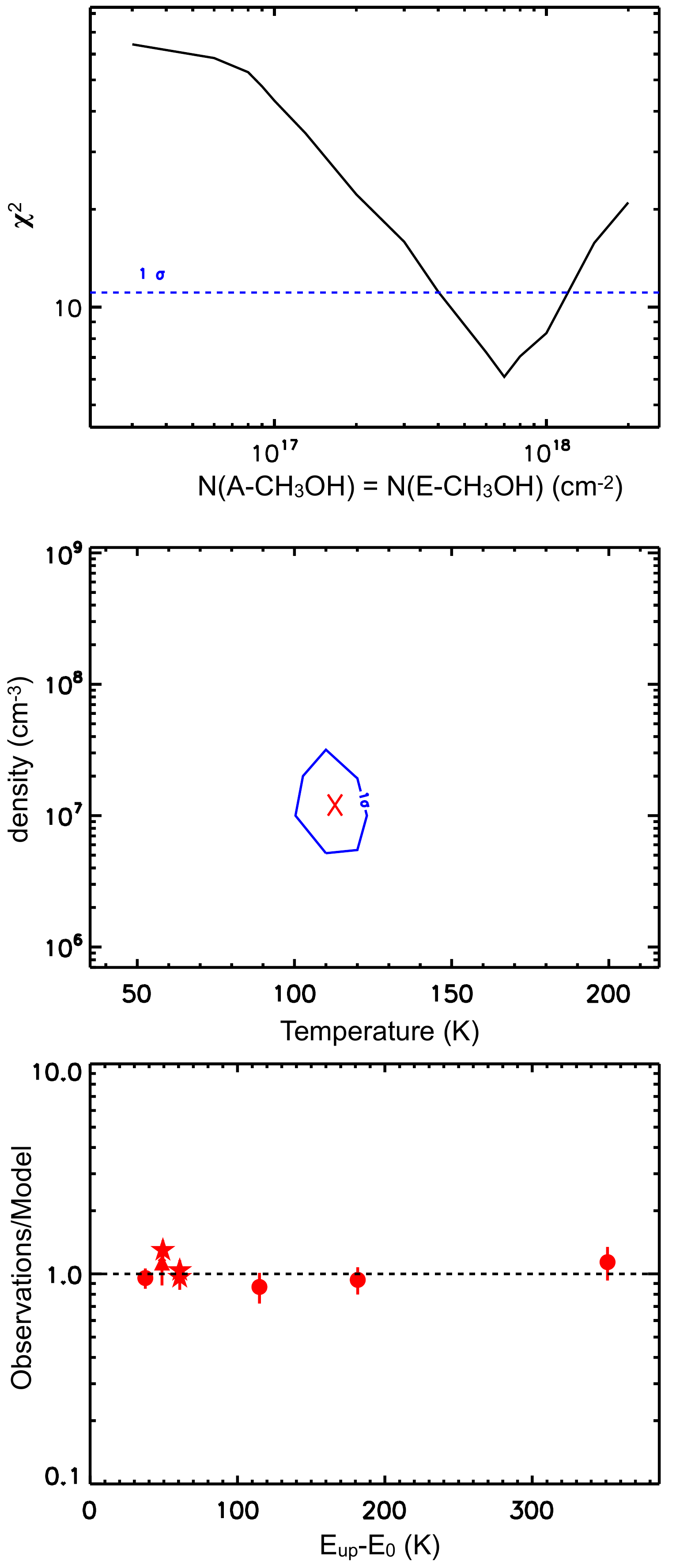}
    \caption{\textit{Upper panel}: Results from the $\chi^2$ minimisation for the -2 km~s$^{-1}$ component fitting simultaneously the A and E forms with a ratio equal to unity and the $^{13}$C E methanol transition with a ratio of 68. The minimum has to therefore be multiplied by two to obtain the A+E forms of methanol. 
    \textit{Middle panel}: density and temperature range obtained for the best fit methanol column density of 2 $\times$ (7 $\times$ 10$^{17}$) cm$^{-2}$ and source size of 0.15$^{\prime\prime}$.
    \textit{Bottom panel}: Observed integrated intensities versus modelled ones as a function of the upper energy level value with respect to the lowest value: 7.9 K for E methanol (red filled circles) and 0 K for A methanol (red filled stars). The $^{13}$C E methanol is represented as a red filled triangle.} 
    \label{fig:non-LTE-analysis}
\end{figure}

\paragraph{Component at -2 km~s$^{-1}$:} 
We obtained a good fit (reduced $\chi^2$ = 1.2 for 5 degrees of freedom) for all lines with the following values:
N(A+E CH$_3$OH) = ($1.4 \pm 0.6$) $\times 10^{18}$ cm$^{-2}$, T = ($110 \pm 10$) K and n$_{H2}$ = ($1.0^{+3}_{-0.5}$) $\times 10^{7}$ cm$^{-3}$ and $\theta_s$ = ($0.15^{\prime\prime} \pm 0.10^{\prime\prime}$).
Fig. \ref{fig:non-LTE-analysis} reports (i) the minimum $\chi^2$, with respect to the temperature T and density n$_{H2}$, as a function of the methanol column density N(CH$_3$OH-A).
The same figure also shows the temperature-density $\chi^2$ contour plot at the best fit N(CH$_3$OH) (note that the x-axis reports N(CH$_3$OH)/2).
Finally, the third panel of Fig. \ref{fig:non-LTE-analysis} shows the observed/predicted line intensity ratio as a function of the line upper level energy $E_{up}$.

The $^{12}$C lines are predicted to be from moderately (the 234.69 GHz line has the smallest value, $\tau = 0.64$) to heavily (the 243.91 GHz line has the largest value, $\tau \sim 12$) optically thick, while the $^{13}$C line is optically thin ($\tau=0.25$).

\paragraph{Component at +9.9 km~s$^{-1}$:} 
The best fit (reduced $\chi^2$ = 1.9) of the +9.9 km~s$^{-1}$ gives values very similar to those obtained for the -2 km~s$^{-1}$ component, namely:
N(CH$_3$OH) = ($3.0 \pm 0.8$) $\times 10^{18}$ cm$^{-2}$, T = ($130 \pm 10$) K and n$_{H2}$ = ($2.5 \pm 2.2$) $\times 10^{7}$ cm$^{-3}$ and $\theta_s$ = (0.12$^{\prime\prime}$ $\pm$ 0.10$^{\prime\prime}$).

\paragraph{Component at +2.8 km~s$^{-1}$:} 
The best fit (reduced $\chi^2$ = 2.7) of the +2.8 km~s$^{-1}$ gives the following values:
N(CH$_3$OH) = ($4.0 \pm 1.0$) $\times 10^{18}$ cm$^{-2}$, T = ($130 \pm 10$) K and n$_{H2}$ = ($2.0 \pm 0.5$) $\times 10^{6}$ cm$^{-3}$ and $\theta_s$ = ($0.13^{\prime\prime} \pm 0.10^{\prime\prime}$).

\paragraph{{\it Summary of the non-LTE analysis:}}
The three velocity components trace gas with similar characteristics: hot (with a temperature around 110--130 K), dense (with a H$_2$ density 2--20 $\times 10^6$ cm$^{-3}$), enriched in methanol (with a methanol column density [1--5] $\times 10^{18}$ cm$^{-2}$) and relatively compact (0.12$^{\prime\prime}$--0.15$^{\prime\prime}$). Table \ref{tab: nonLTE} lists the results of the analysis for the three components.
Assuming that the emission in each velocity component originates in a spherically symmetric region, one can estimate the approximate H$_2$ column density ($\sim 2 \times 10^{23}$ cm$^{-2}$) and, then, the methanol abundance, whose value ends up to be very high, $\sim 10^{-5}$.\\
Compared to the results of the LTE RD analysis (see \S ~ \ref{subsec:RD}), the non-LTE analysis predicts slightly lower gas temperatures and much higher methanol column densities, by about two orders of magnitude. 
These differences are due to two major effects. 
First, the column densities calculated with the RD are beam-averaged whilst the LVG analysis, where the emitting size is derived, shows that the methanol emission is relatively compact.
This explains the large difference between the two estimates.
Second, the RD in Fig.~\ref{fig: RD} show a clear scatter of the points around the best-fit straight line, which indicates optically thick lines and/or non-LTE effects.
They both lead to a misleading evaluation of the gas temperature (and, to a lesser extent, methanol column density).
Finally, even though the source size and the optical depths of the lines can be corrected a posteriori (using the rotational diagram analysis), non-LTE effects (e.g. sub-thermally populated lines), if present, cannot be corrected unless performing a non-LTE analysis. 
In summary, in the RD analysis, all of the above mentioned effects lead to an underestimation of the methanol column density and incorrect gas temperature.

\begin{table*}
\caption{Results from the non-LTE analysis. }
\label{tab: nonLTE}
\centering
\begin{tabular}{c c c c }
\hline \hline
                                                     &  -2 km~s$^{-1}$                                  & 2.8 km~s$^{-1}$ & 9.9 km~s$^{-1}$   \\
\hline \hline
N(CH$_{3}$OH) (cm$^{-2}$)        & ($1.4 \pm 0.6$) $\times 10^{18}$      &  ($4.0 \pm 1.0$) $\times 10^{18}$  & ($3.0 \pm 0.8$) $\times 10^{18}$ \\
T (K)                                             &  $110 \pm 10$                                   & $130 \pm 10$   & $130 \pm 10$ \\
n$_{H_{2}}$ (cm$^{-3}$)              & ($1.0^{+3}_{-0.5}$) $\times 10^{7}$        &($2.0 \pm 0.5$) $\times 10^{6}$ & ($2.5 \pm 2.2$) $\times 10^{7}$\\
$\theta_{S}$ ($^{\prime\prime}$)  &  $0.15^{\prime\prime} \pm 0.10^{\prime\prime}$ & $0.13^{\prime\prime} \pm 0.10^{\prime\prime}$  & 0.12$^{\prime\prime} \pm 0.10^{\prime\prime}$\\
\hline
\end{tabular}
\end{table*}

\subsection{Dust opacity towards the binary system objects, 11A and 11B} \label{subsec:tau-dust}
\label{subsec:tau-dust}

As described in Sect. \ref{sec: Sources}, \bhb is a binary system where the two objects (11A and 11B) are embedded in their respective circumstellar disks \citep{Alves2019}.
The disks are both detected in the continuum at 0.94 cm with the Karl G. Jansky Very Large Array (JVLA) and at 1.3 mm with ALMA. Following the findings by \cite{DeSimone2020} that the dust in front of the hot corino of NGC1333 IRAS4A1 completely absorbs its methanol lines in the millimeter wavelengths, here we aim to estimate the possible absorption of the methanol emission specifically related to the compact disk scales around each of the sources 11A and 11B. To this scope, we use the dust continuum observations by \citet{Alves2019} at 225 GHz, who found a flux density $F_{225}$ equal to $5.2 \pm 0.53$ mJy towards 11A and $5.62 \pm 0.45$ mJy towards 11B. 
The deconvolved sizes are $7.3(\pm 1.3) \times 5.4(\pm 1.1)$ au$^2$ and $4.9 (\pm 1.1) \times 3.7 (\pm 1.0)$ au$^2$ for 11A and 11B, respectively.
Thus, the 225 GHz continuum emission is more compact in 11B than 11A, despite having a similar flux density.

We can constrain the dust optical depth $\tau_{\nu}$ and temperature towards each of the two sources using the equation:
\begin{equation}\label{eq: flux}
    F_\nu=B_\nu (T_d)~(1-e^{-\tau_\nu})~\Omega_s~,
\end{equation}
where $B_\nu (T_d)$ is the Planck function, which depends on the dust temperature $T_d$, and $\Omega_s$ is the source solid angle, derived from the continuum observations. The source solid angle is defined as $\sim A/d^2$ where A is the source emitting area (derived from the sources size) and $d^2$ is the distance to the sources. The dust optical depth can be estimated by inverting Eq. \ref{eq: flux}, as follows:
\begin{equation}\label{eq:tau}
    \tau_\nu=-\text{ln} \left(1-\frac{F_\nu}{B_\nu(T_d)~\Omega_s}\right)~.
\end{equation}
Finally, using Eq. \ref{eq:tau}, we can estimate the possible dust absorption coefficient using the following equation:
\begin{equation}
    e^{\tau_\nu}=\frac{I_\nu^{\text{pred}}}{I_\nu^{\text{obs}}}~.
\end{equation}

\begin{figure*}
    \centering
    \includegraphics[width=16cm]{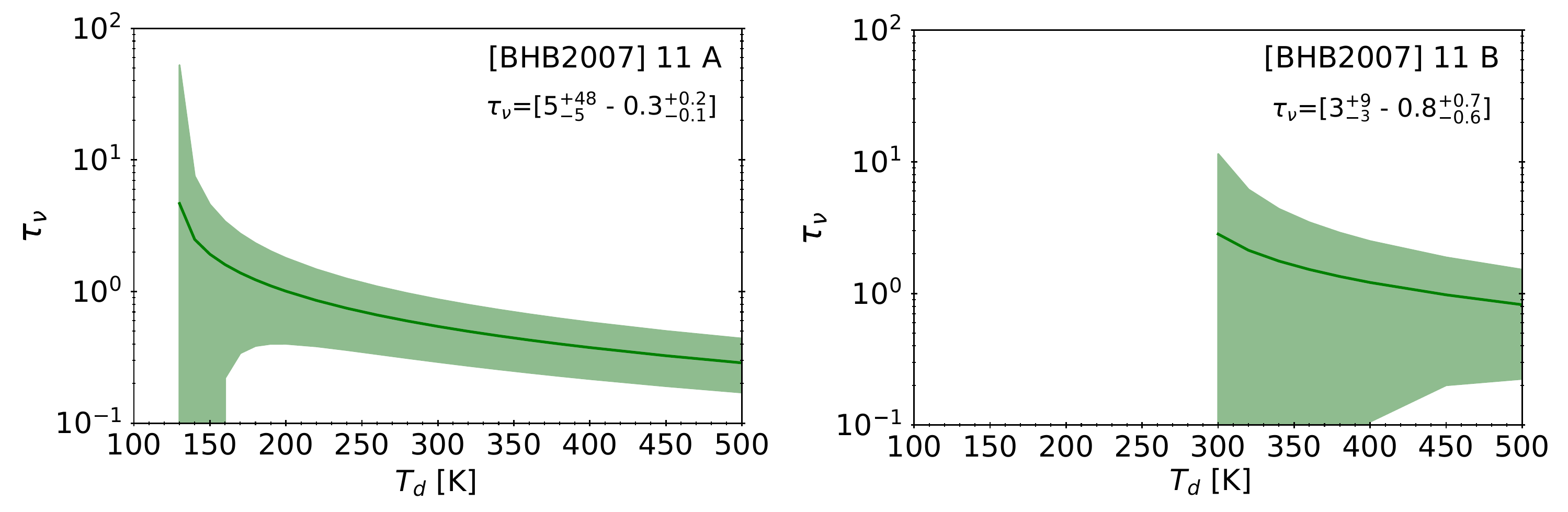}
    \caption{Dust optical depth $\tau_\nu$ at 225 GHz as a function of the dust temperature T$_{d}$, derived from the observations of \citet{Alves2019} and following the method described in \citet{Bouvier2021}. The lines show $\tau_\nu$ as derived from Eq. \ref{eq:tau}. The errors are calculated using the error propagation method described in Appendix C.2. of \citet{Bouvier2021}, and include the calibration uncertainty of 15$\%$ \citep{Alves2019} and the source size uncertainty. The left and right panels refer to 11A and 11B, respectively.}
    \label{fig:dust-tau}
\end{figure*}

As we do not know the dust temperature, $T_d$, of the sources we derived the dust optical depth for temperatures ranging between 100 and 500 K with steps of 10 K from 100 K to 200 K, and by steps of 20 K from 200 to 500 K, following the method described in \citet{Bouvier2021}.
The observed 225 GHz flux provides a lower limit to $T_d$. 
Specifically, the condition $F_\nu$/$\Omega_s \leq B_\nu (T_d)$ implies $T_d \geq 130$ K in 11A and $T_d \geq 300$ K in 11B (as it is more compact).
Then, for each dust temperature, we can derive the dust opacity at 225 GHz, reported in Fig. \ref{fig:dust-tau}.
The figure shows that $\tau_{\nu}$ is well constrained in the two sources, except around the minimum T$_d$ for each of them, where the error bars become very large.
Specifically, in 11A $\tau_{\nu}$ varies from $5^{+48}_{-5}$ at $T_d \sim 130$ K to $0.3^{+0.2}_{-0.1}$ at $T_d =500$ K; in 11B, $\tau_{\nu}$ varies from $3^{+9}_{-3}$ at $T_d \sim 300$ K to $0.8^{+0.7}_{-0.6}$ at $T_d =500$ K.
The dust opacities tend to be optically thick and, taking the largest dust optical depths gives $\tau_{\nu} \leq 53$ and $\tau_{\nu} \leq 12$ in 11A and 11B, respectively. Therefore, the possible methanol lines from the hot corinos associated with 11A and 11B, if they exist, can be completely attenuated by the foreground dust.

\section{Origin of the observed methanol components}

\begin{figure}
    \centering
    \includegraphics[width =\columnwidth]{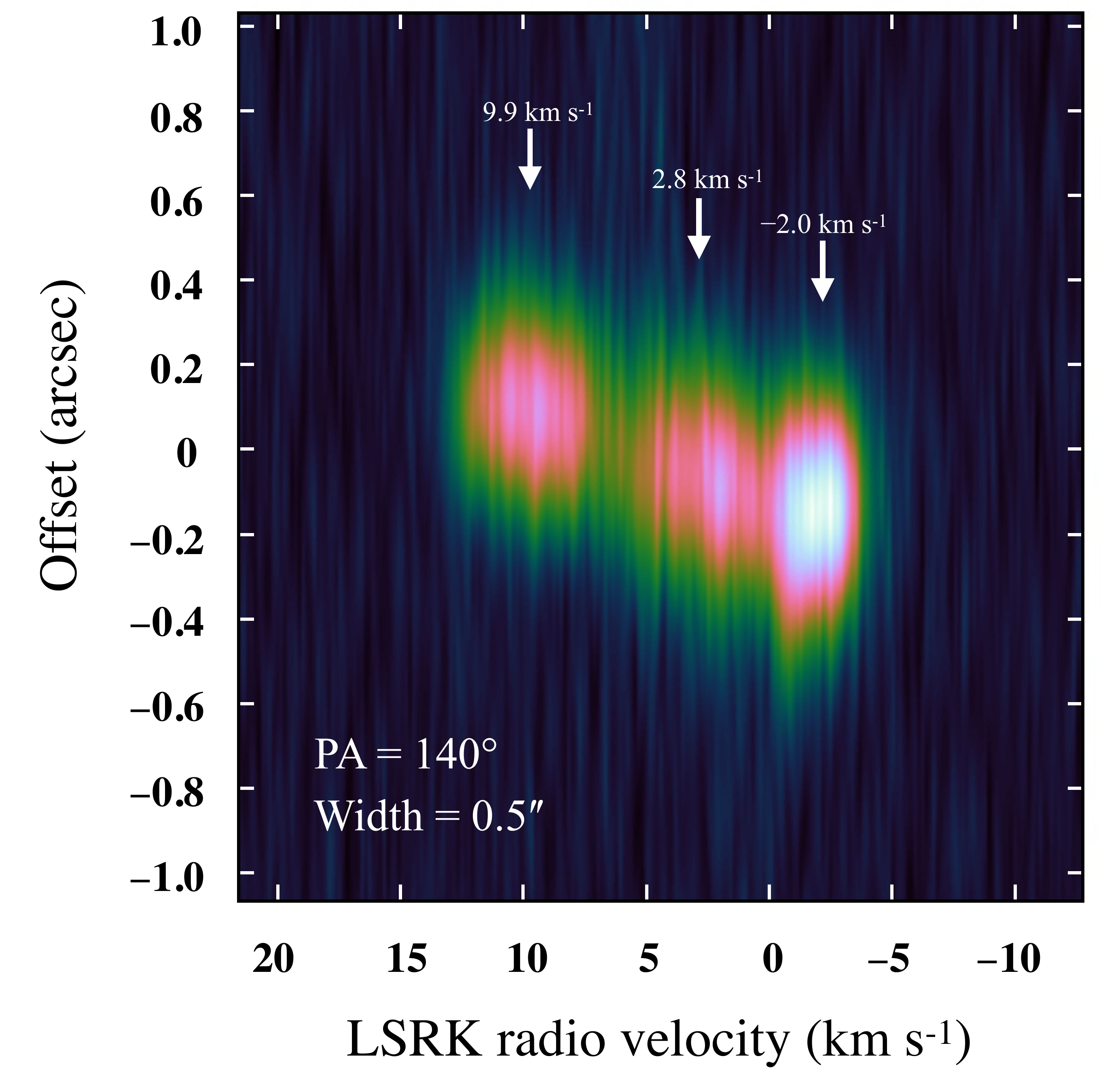}
    \caption{Position-velocity (PV) diagram obtained from the \chtoh emission at 243.916 GHz. The position angle (PA) and width of the PV cut is indicated in the figure. Note that the direction of the cut used to produce the position-velocity diagram is shown in Fig. \ref{fig:momcont} as a grey dashed line. The three velocity components observed in the methanol spectra are also indicated. }
    \label{fig:pvplot}
\end{figure}

In the following, we discuss the possible origin of the three velocity components detected in our methanol spectra.
In order to better understand this point, we created a position-velocity (PV) diagram for the methanol transition at 243.916 GHz (E$_\mathrm{up} \sim 50$ K), displayed in Fig. \ref{fig:pvplot}.
The PV plot is produced from a cut oriented parallel to the velocity gradient observed in the moment 1 map of the methanol emission (Fig. \ref{fig:momcont}), whose position angle (PA) is 140\degr\,East of North. 
The cut width is 0.5$^{\prime\prime}$, whose value is comparable to the angular resolution of the FAUST data, and encompasses both core 11A and core 11B, separated by only 0.2$^{\prime\prime}$. 
Unlike the H$_2$CO PV plot, which shows a smooth Keplerian profile over the velocity range \citep{Alves2019}, a fragmented profile  is clearly seen in the methanol PV plot. 
The three substructures correspond to the three velocity components observed in the methanol spectra (a similar distribution is observed at the higher E$_\mathrm{up}$ transition). \\

\begin{figure*}
    \centering
    \includegraphics[trim = 0 2.5cm 0 2.5cm, clip, width =\textwidth]{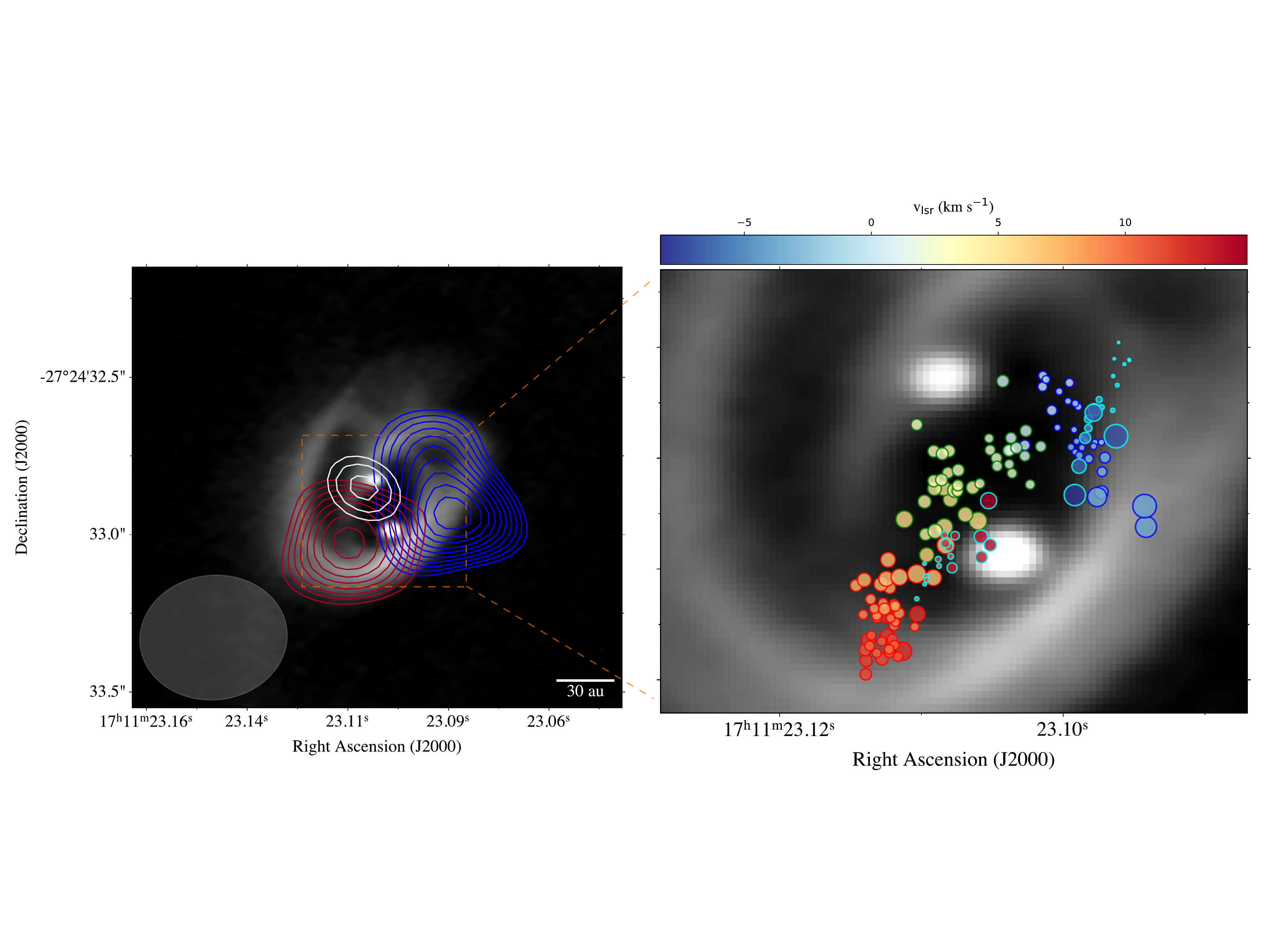}
    \caption{{\it Left panel}: Methanol intensity contours at 243.916 GHz plotted over the dust continuum map of \citet{Alves2019}. The contours show the methanol emission at extreme velocity channels detected at a 3$\sigma$ level (1$\sigma = 1.8$ mJy beam$^{-1}$). The blue and red contours correspond to emission at the $-4.89$ km s$^{-1}$ and $13.3$ km s$^{-1}$ velocity channels, respectively. White contours show emission (from a 20$\sigma$ level) from velocity channels near the systemic velocity of the source (2.91 km s$^{-1}$). The synthesised beam of the methanol (FAUST) data is shown in the lower left corner. {\it Right panel}: Position of intensity peaks (circles) from the two-dimensional Gaussian fit in each velocity channels separated by $\sim 0.15$ km s$^{-1}$ from each other . Colour code for the edges of the circle is red for the 9.9 km~s$^{-1}$ methanol component, green for the 2.8 km~s$^{-1}$ methanol component and blue for the -2 km~s$^{-1}$ methanol component, while the CO (2--1) high-velocity components have cyan. The circle size is defined by the uncertainty on the absolute position of the emission peak in each velocity component (largest circle has $\sim 0.02^{\prime\prime}$)}
    \label{fig:metvel}
\end{figure*}

Although we cannot locate the methanol emission with respect to the binary system due to our limited angular resolution, we notice that blue- and red- shifted emission in the high velocity channels where SNR is larger than 5 (-4.89 km~s$^{-1}$ in blue and 13.3 km~s$^{-1}$ in red) tends to be aligned with source 11B (Fig. \ref{fig:metvel}, left panel), alike the high-velocity components of the CO emission shown by \citet{Alves2019}. The methanol emission at velocity channels near the systemic velocity of the source ($\sim$ 3.6 km~s$^{-1}$) also tends to peak near 11A. The three methanol components appear in a compact configuration which is compatible with the compact methanol seen in the moment 0 maps (Fig. \ref{fig: mom0}).\\

\begin{figure*}
    \centering
    \includegraphics[width = 0.32\textwidth,trim=0 0 0 0 clip]{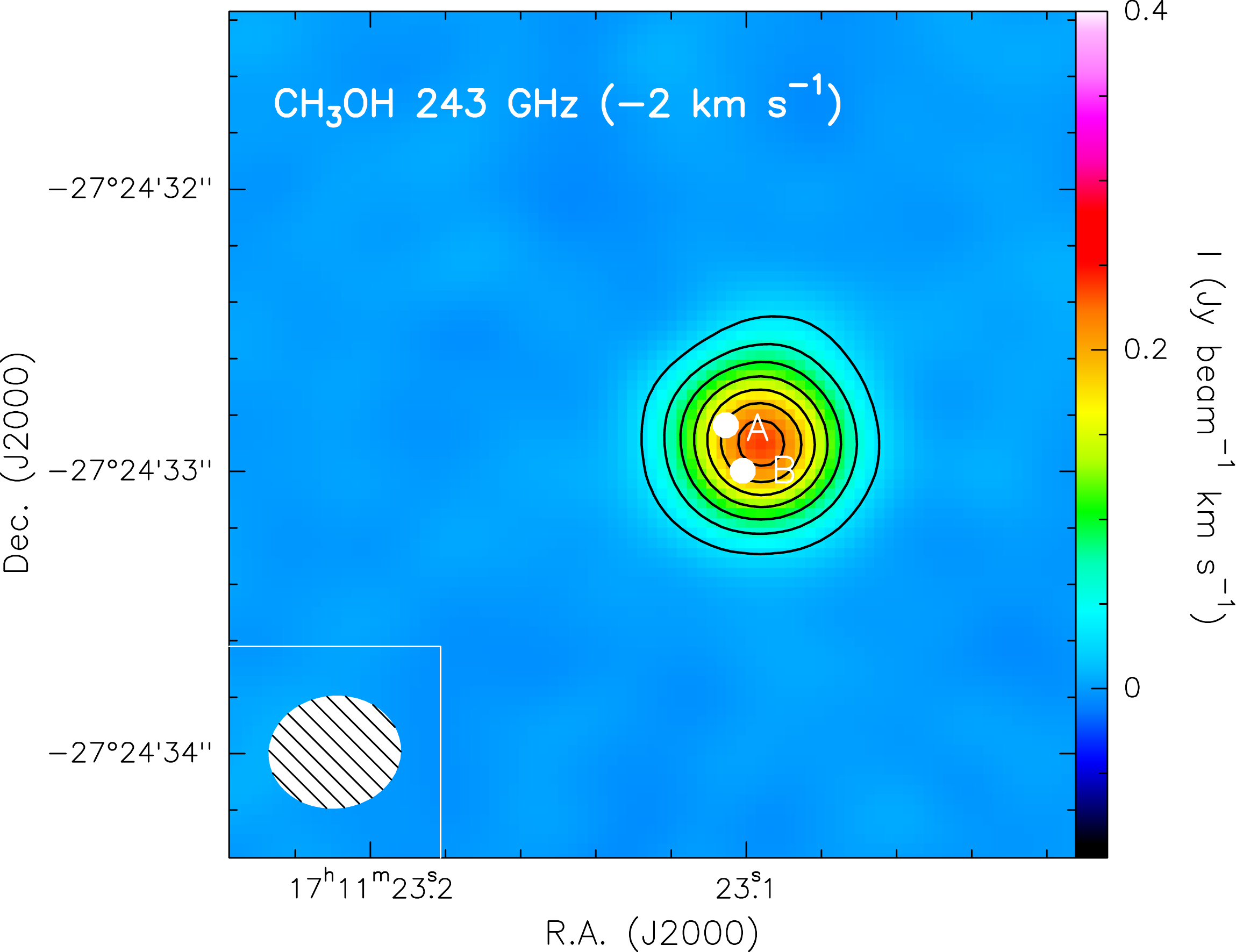}\
    \includegraphics[width=0.32\textwidth,trim=0 0 0 0 clip]{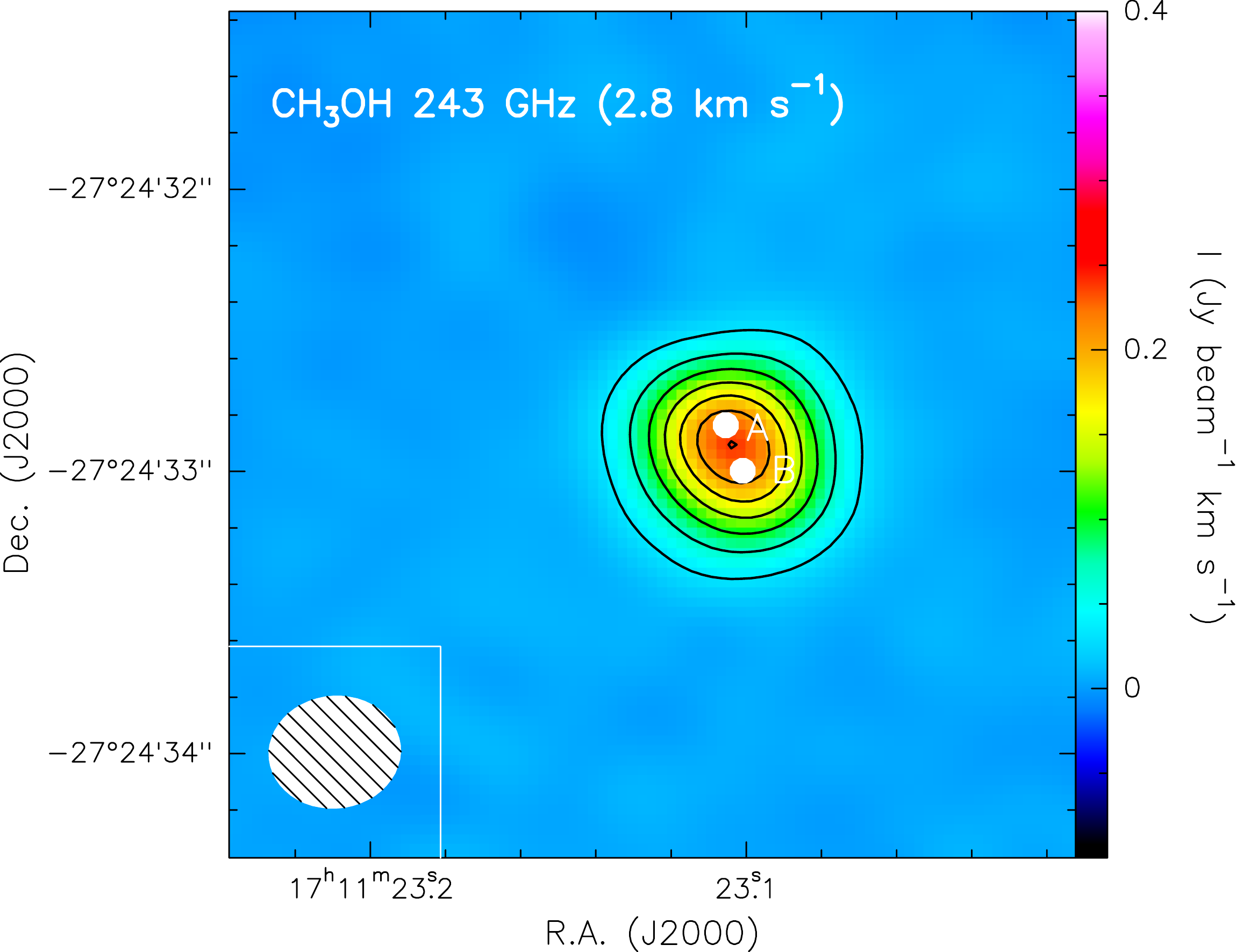} \ 
    \includegraphics[width=0.32\textwidth,trim=0 0 0 0 clip]{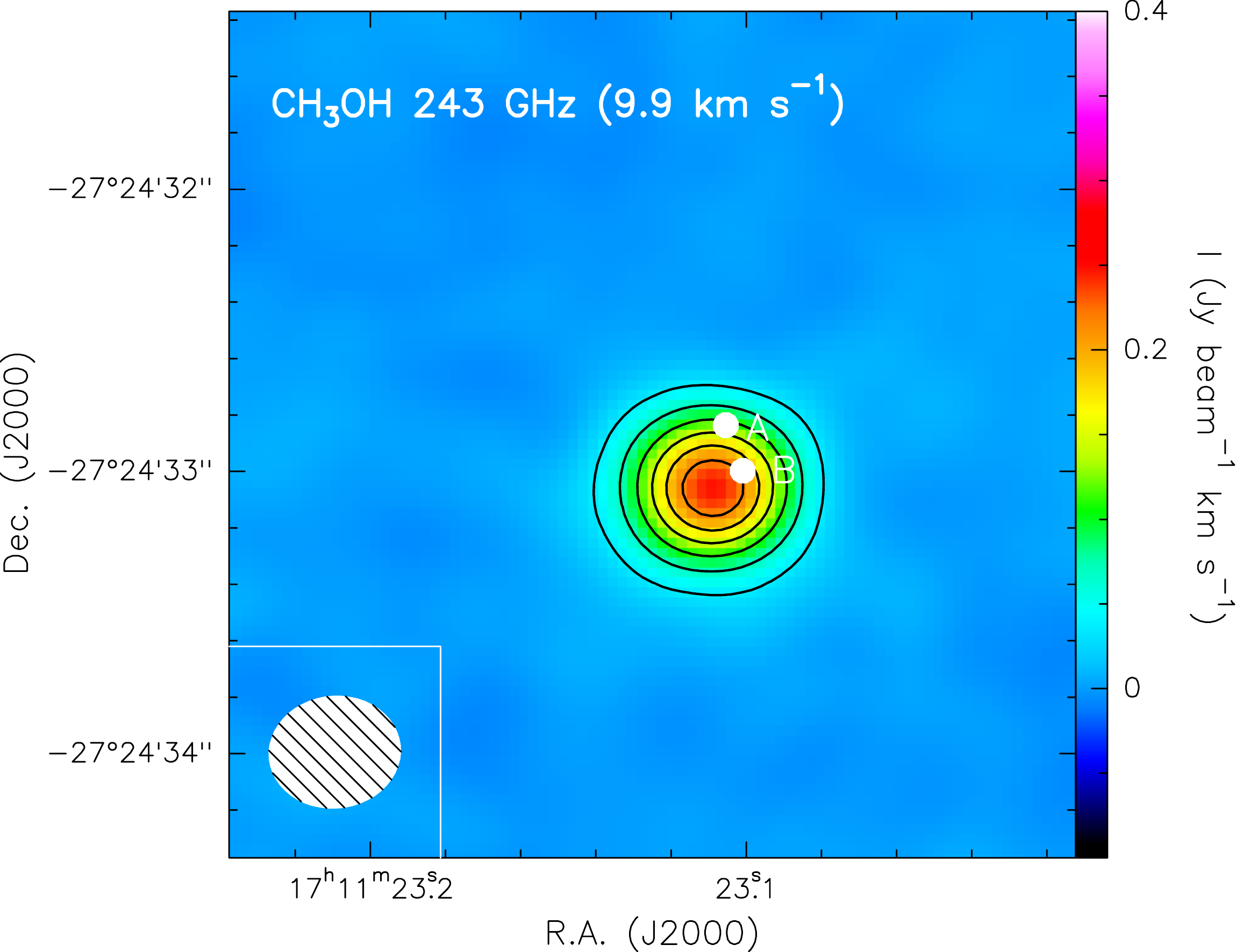} \ 
        \includegraphics[width = 0.3\textwidth]{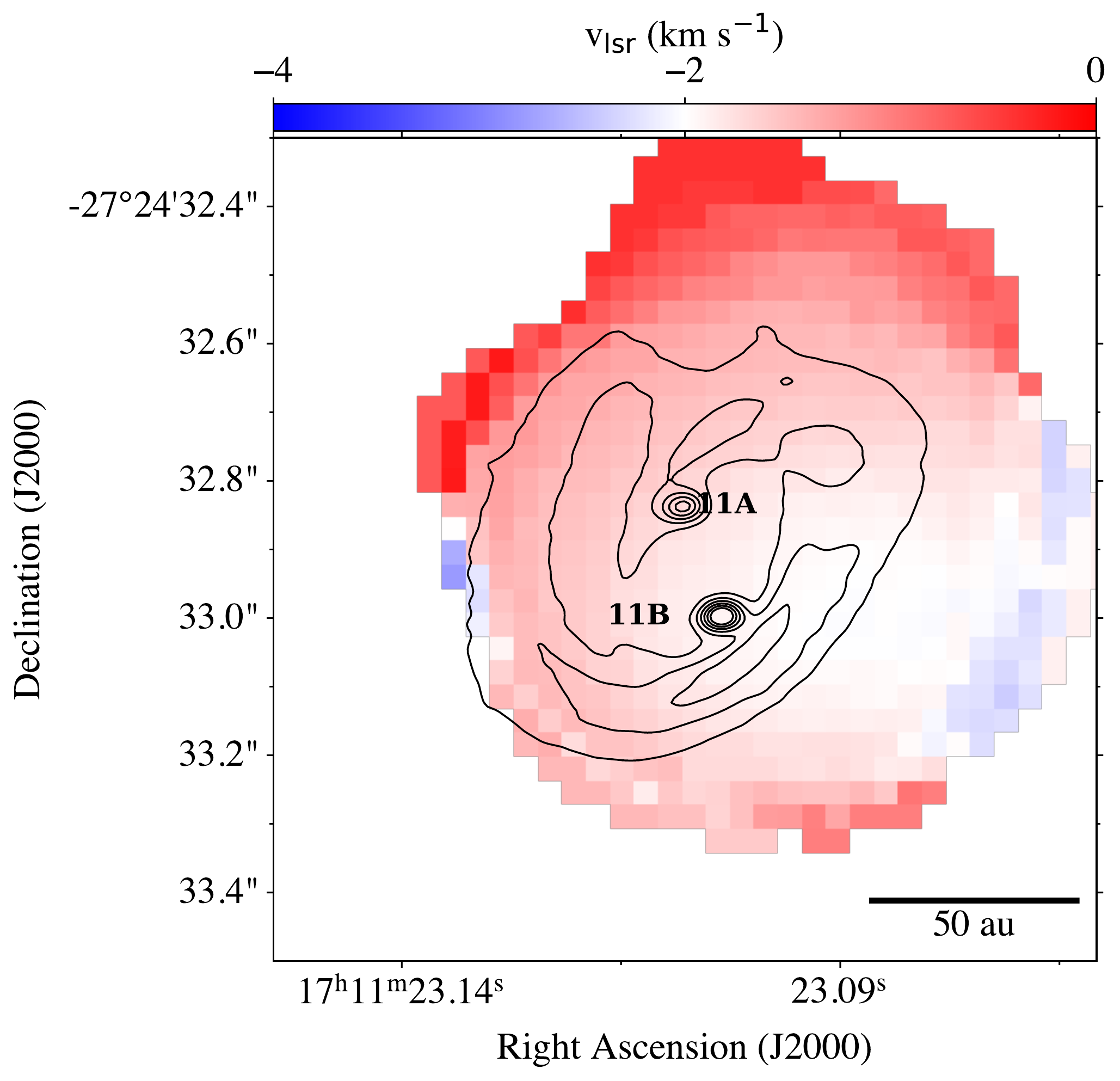}\
    \includegraphics[width=0.3\textwidth]{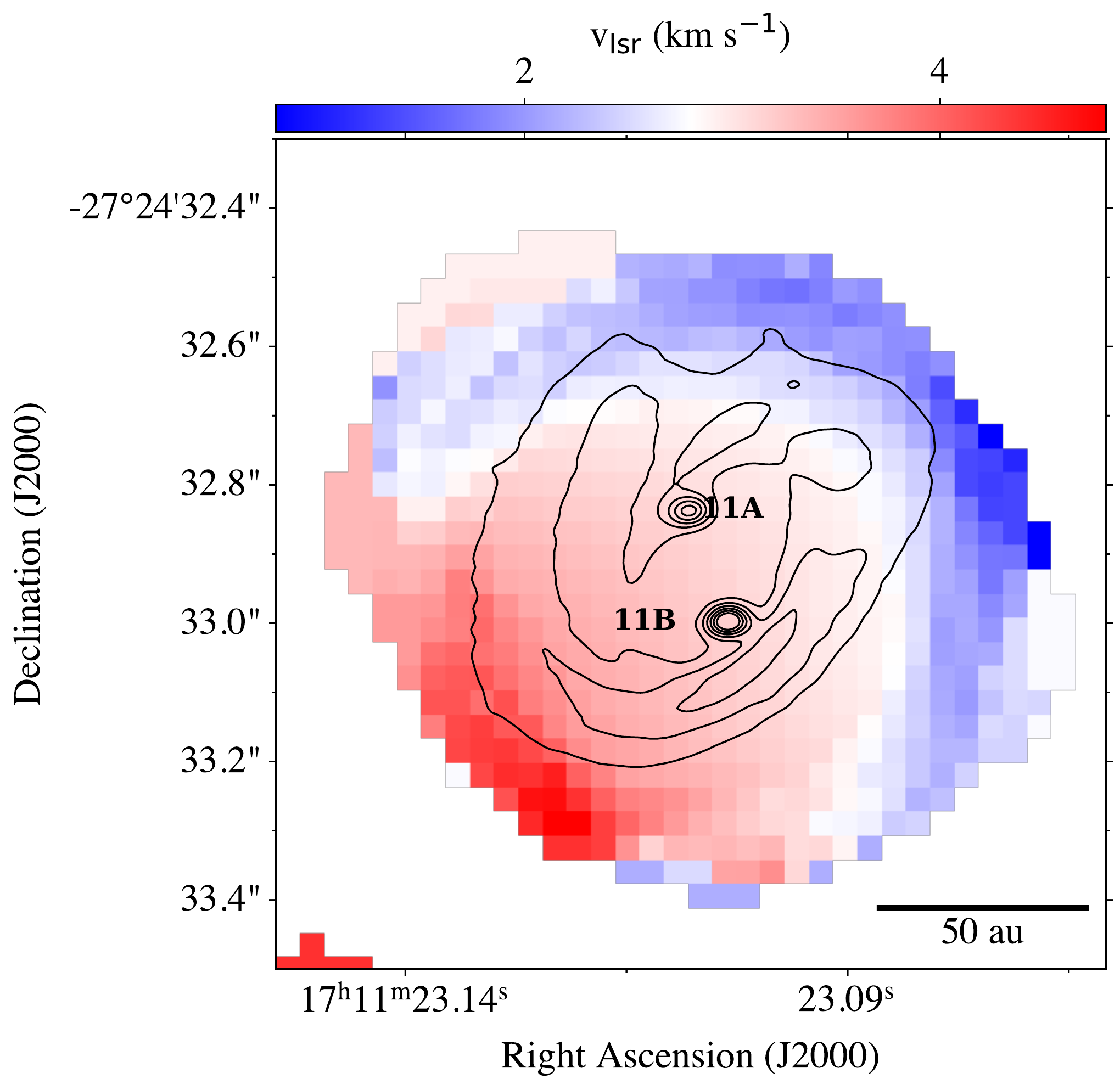} \ 
    \includegraphics[width=0.3\textwidth]{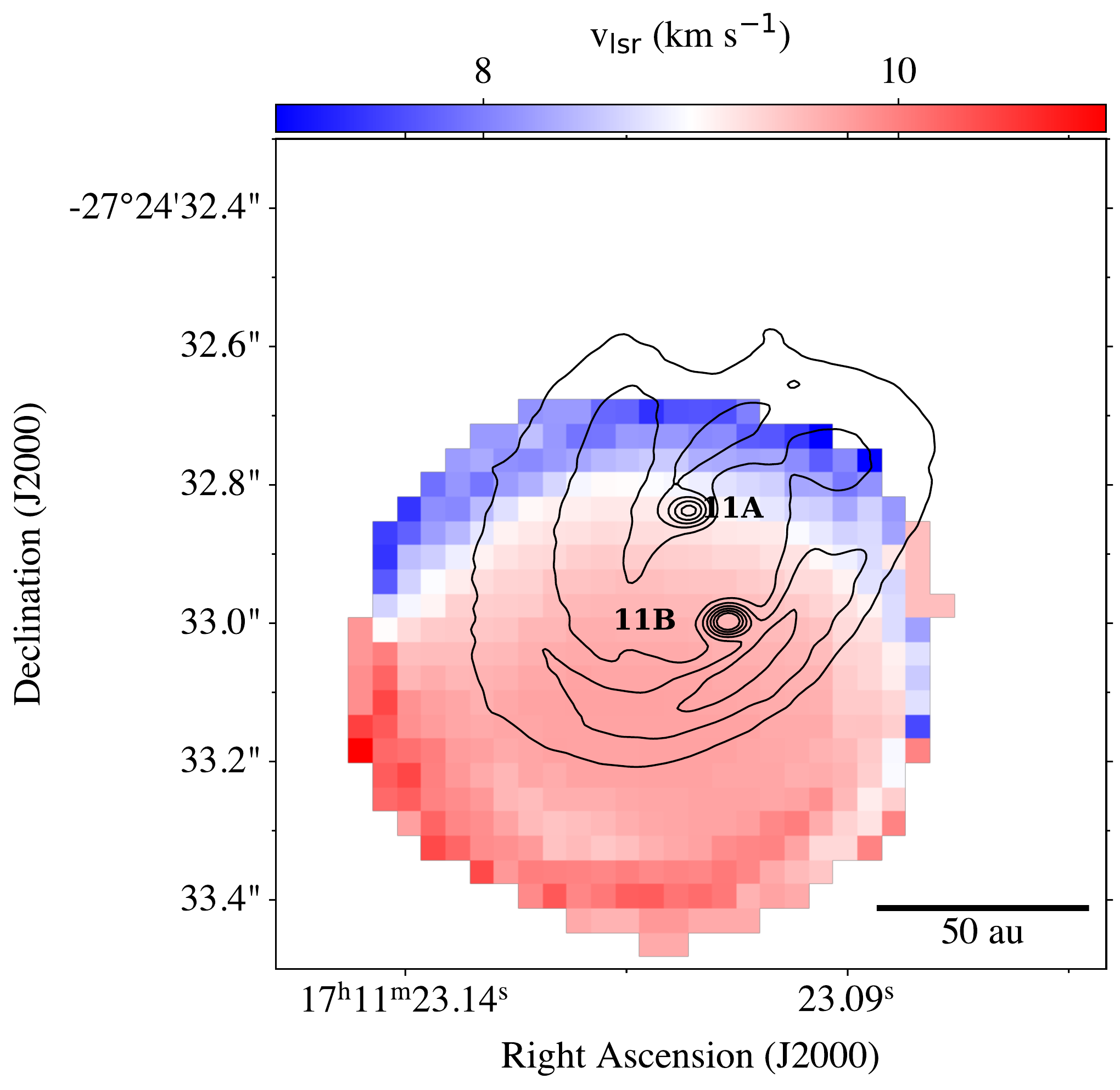} 
    \caption{Top: Moment 0 map of the high spectral resolution methanol transition at 243.916 GHz ($\rm E_{u}$=49.66 K) for the three components ([-7--0] km~s$^{-1}$ for the -2 km~s$^{-1}$ component, [0--6] km~s$^{-1}$ for the 2.8 km~s$^{-1}$ component and [6--16] km~s$^{-1}$ for the 9.9 km~s$^{-1}$ component,). Contours start at 4$\sigma$ at every 4$\sigma$. The ellipse in the bottom left corner represents the ALMA synthesised beam. Both source A and B identified by \citet{Alves2019} are indicated as white filled circles. Bottom: Moment 1 map of the high spectral resolution methanol transition at 243.916 GHz ($\rm E_{u}$=49.66 K) for the three components.}
    \label{fig: mom0+1}
\end{figure*}

It is clear from Fig. \ref{fig: spectra-HR}, \ref{fig: spectra-LR} and \ref{fig: spectra-LR234} that the three velocity components are not completely spectrally independent and that they are affected by blending from one another. We present in Fig. \ref{fig: mom0+1} the integrated intensity and velocity field maps from the 243 GHz methanol transition, for the three velocity components in the [-7--0] km~s$^{-1}$ range for the -2 km~s$^{-1}$ component, the [0--6] km~s$^{-1}$ range for the 2.8 km~s$^{-1}$ component and the [6--16] km~s$^{-1}$ range for the 9.9 km~s$^{-1}$ component. The three positions are clearly identified in the moment 0 maps:  the 2.8 km~s$^{-1}$ component in between \bhb A and \bhb B, while the -2 and 9.9 km~s$^{-1}$ are located North-West and South-East from \bhb B, respectively. There is no clear evidence of velocity structure in these two velocity range, which points to the conclusion of the results from a compact shock emission (see Sec. \ref{shock}). The 2.8  km~s$^{-1}$ seems to be more sensitive to the disk rotation \citep[like the H$_{2}$CO centroid map in][]{Alves2019} and that could explain the gradient that is seen more clearly than in the other two components. Also, Fig. \ref{fig:metvel} (right panel), must be analysed with caution as the overlap of velocity ranges from one component to another can be misleading to conclude that we have an elongated structure for the methanol emission. We added, on this Figure, coloured borders to identify the three velocity components observed in our methanol observations. However, higher spatial resolution observations of methanol are necessary to determine  the velocity structure of methanol accurately.     

Since the emission in each velocity channel is compact and centrally peaked, we performed two-dimensional (2D) Gaussian fits on the emission in each channel to obtain the position of their intensity peak. The accuracy on the relative position is given by $\Theta / \mathrm{(2\times SNR)}$, where $\Theta$ is the synthesised beam size and SNR is the signal-to-noise ratio of the peak with respect to the map noise \citep{Condon1997,Condon1998}. Thus the uncertainty in the position is inversely proportional to the SNR in each velocity channel. Centroid fitting to the compact molecular emission yields astrometry with precisions better than those of the spatial resolution in the case of high SNR. For this calculation, we considered only intensity peaks brighter than 5 times its Gaussian fit error and SNR larger than 10.0 (which corresponds to a position accuracy better than 0.02$^{\prime\prime}$). 
The right panel of Fig. \ref{fig:metvel} shows the 2D Gaussian centroid map with the location of each methanol emission peak (black circle) as a function of velocity. The velocity components satisfying the conditions above span from $-$4.1 to 12.7 km s$^{-1}$.\\
It is noticeable that the methanol emission with velocities between 0 and 8 km s$^{-1}$ tends to lie in the region in between sources 11A and 11B, while the gas with velocities $<$0 km s$^{-1}$ and $>$8 km s$^{-1}$ seems to approach source B, as also seen in the CO (2--1) observations presented in \citet{Alves2017} and \citet{Alves2020}. Their analysis shows an increasing velocity towards 11 B that seems to indicate acceleration of infalling gas from the circumbinary disk onto 11 B. This can be shown on the right side of Fig. \ref{fig:metvel} where we also present the CO emission as circles with green borders, as opposed to the methanol emission as circles with black borders. The CO emission observed by \citet{Alves2017} and shown in the left panel of our Fig. \ref{fig:dust} traces primarily large-scale outflowing gas emission, launched from a distance of $\sim$ 100 au from the disk centre and propagating perpendicular to the envelope major axis. However, velocity components faster than the outflows are detected within the circumbinary disk at only a few au. These high-velocity CO components reach up to 15 km~s$^{-1}$  and the blue- and red-shifted velocities (V$_{LSR}$ $<$ -1.5 km~s$^{-1}$ and V$_{LSR}$ $>$ 9 km~s$^{-1}$ respectively) are centred on and aligned with 11 B, the higher velocities being closest to the position of the protostar. Note that no high velocity components are seen near 11 A. The increasing velocity toward \bhb B indicates acceleration of infalling gas from the circumbinary disk onto \bhb B.\\
From our methanol analysis, an increasing velocity is not as clear as in CO, possibly due to our lower spatial resolution, but also due to the fact that the three velocity components identified in our methanol spectra are partially overlapping. The informations provided by the PV diagram and centroid maps provide constrains on the origin of the three velocity components, as follows.

\subsection{The -2.0 and +9.9 km s$^{-1}$ components: }
\label{subsec:ori-2+9comp}

Based on the arguments above, these two velocity components are likely associated with the gas streaming towards 11B. The most extreme components, $-5$ km~s$^{-1}$ and 13 km~s$^{-1}$, are consistent with the high-velocity components observed in CO lines distributed along the filaments and confined within the circumbinary disk  \citep[see also higher spatial resolution observations by ][and their Fig. 2 and S2]{Alves2019}. These components exceed the Keplerian velocities observed in the H$_2$CO maps of \citet{Alves2019}, -3 km s$^{-1} \lesssim \mathrm{v}_{Kep} \lesssim 12$ km s$^{-1}$), and are distributed around 11B.
Our non-LTE analysis of the methanol lines indicates that this gas is very warm (100--140 K), dense ($\sim 10^7$ cm$^{-3}$) and very enriched in methanol (abundance w.r.t H$_{2} \sim 10^{-5}$). We will discuss the physical origin of gaseous methanol later in this section.

\subsection{The +2.8 km s$^{-1}$ component:} 
\label{subsec:ori-2p8comp}

This component seems rather associated with 11A when considering the arguments based on the velocity field, position-velocity diagram and centroid maps above.
This gas is also warm (120--140 K), dense ($\sim 2 \times 10^6$ cm$^{-3}$) and very enriched in methanol (abundance w.r.t H$_{2} \sim 10^{-5}$), which would point to a similar origin of the methanol emission to the other two components. Moreover, the H$_2$CO (E$_{\mathrm{up}} \sim 70$ K) observations of \cite{Alves2019}, taken at comparable spectral resolution but a factor $\sim 2$ better angular resolution, show relatively warm gas associated with 11A. In fact, the emission observed in the velocity range of 2.5 -- 5.0 km s$^{-1}$ is spatially coincident with 11A, which is consistent with our findings \citep[see figure S3 in][]{Alves2019}. This component is close to the systemic velocity of 3.6 km~s$^{-1}$ and could be tracing (partially) the keplerian motion of the disk (similar to H$_{2}$CO), so a velocity gradient is expected and identified in our Fig. \ref{fig: mom0+1} (middle Figure, bottom panel).  

Nevertheless, due to the limited angular resolution of our maps, we should not completely discard that this velocity component has some contribution from 11B or from gas located in between sources.

\subsection{Origin of the gaseous methanol: shock versus thermal desorption}
\label{shock}

Methanol is known to have a very low abundance $\leq 10^{-8}$ in cold objects \citep[e.g.][]{Friberg1988,Nagy2019} while its abundance is high (10$^{-7}$--10$^{5}$) in hot cores/corinos \citep[e.g.][]{Gibb2000,Cazaux2003} and molecular outflow shocks \citep[e.g.][]{Bachiller1995,Codella2020}.
It is commonly accepted that methanol is indeed synthesised on the grain surfaces \citep[e.g.][]{Watanabe2002,Rimola2014} during the cold prestellar phase and released from the grain mantles into the gas phase by thermal and non-thermal processes.
Hot cores/corinos are therefore the results of thermal processes \citep[e.g.][]{Maret2005,Taquet2016} while molecular outflow triggers non-thermal desorption via shocks  \citep[e.g.][]{Caselli1997C,Jimenez-Serra2008}.

The temperature derived from the non-LTE analysis of the methanol lines tells us that the gas is hot and dense.
The latter would also imply that the dust and gas are thermally coupled \citep[e.g.][]{Goldsmith2001} and, hence, also the dust is warm enough for methanol to be thermally desorbed, based on the recent estimates of the methanol binding energy by \citep{Ferrero2020}. These authors found the CO binding energy ranges between $\sim$4400 and $\sim$6500 K, which would correspond to a dust sublimation temperature of about 80 to 120 K, assuming that the grain mantles have been warmed up since $\sim10^4$ yr\footnote{Note that these sublimation temperatures could be lower by less than 10 K if the heating of the ices is 1 Myr instead.}.
Of course, it is possible that methanol is trapped in the water-rich ices, so that the sublimation temperature could be dominated by the co-desorption of methanol with water.
Again based on the \cite{Ferrero2020} work, this would happen at about 110 K, namely in between the dust sublimation temperature of a pure methanol ice.

A first question to answer is whether the measured temperature, 110--130 K (\S ~\ref{subsec:nonLTE}), is compatible with the warming of the dust from the two sources or it necessitates a non-thermal mechanism.
To this end, we can estimate the approximate dust temperature of a dust grain heated by a $L_{\star}$ source at a distance $r$ from it.
When the dust is optically thin, the dust temperature profile heated by a central source can be approximated by the following equation \citep{Ceccarelli2000a}:
\begin{equation}
\rm T_{d} [K] = 75 \left[\frac{L_{\star}}{27L_{\odot}}\right]^{0.25}\left[\frac{r}{150au}\right]^{-0.5}.
\end{equation}
Considering that the two sources 11A and 11B have a total bolometric luminosity of 4.4 L$_{\odot}$ \citet{Sandell2021} and that the distance between them is 28 au (0.2$^{\prime\prime}$), the dust temperature would be $\sim$93 K. As already stated in Sec. \ref{subsec:tau-dust}, the dust is optically thick, therefore the dust temperature should be above $\sim$93 K. This means that all the dust in between 11A and 11B could be warm enough for the methanol to sublimate, within the uncertainties discussed above on the frozen methanol sublimation temperature.
A fortiori, if 11A and 11B host one or two hot corinos, the observed methanol could just originate from them.
However, the above analysis on the location of the gas (\S ~\ref{subsec:ori-2+9comp} and ~\ref{subsec:ori-2p8comp}) and the dust optical depth towards 11A and 11B (\S ~\ref{subsec:tau-dust}) is not in favour of the hypothesis that the observed hot methanol originates in the 11A or 11B cores, therefore invalidating their hot corino nature.

Alternatively, the observed methanol could be the result of the sputtering of the grain mantles in shocks \citep[e.g.][]{Caselli1997C,Jimenez-Serra2008,DeSimone2020,Codella2020} created by the streaming of the gas towards 11A and 11B and impacting the quiescent gas of the circumbinary envelope/disk or, possibly, the two circumstellar disks. The mass contained in the shocked gas traced by methanol would then be about 10$^{-6}$ M$_{\odot}$, namely a very minor fraction of the dynamical mass ($\sim$ 2.2 M$_{\odot}$) or even the mass of the larger circumbinary disk \citep[$\sim$ 0.08 M$_{\odot}$][]{Alves2019}.

Indeed, the gas kinematics in a binary system is expected to be more complex compared to a simple protostar formation, with accretion shocks and structures which are related to the transition from the circumbinary material to the circumstellar disks \citep{Matsumoto2019}. \citet {Alves2019} computed a mass accretion rate from the circumbinary disk into the circumstellar disk of $\sim 1.1 \times 10^{-5}$ M$_{\odot}$~yr$^{-1}$, which is also consistent with other Class 0/I protostars. Also, the infall motion, traced by the high velocity CO, is perfectly consistent with the dynamical mass of the system.
A similar situation as in \bhb is seen in the young binary system IRAS16293 A where some molecular tracers such as HNCO, NH$_{2}$CHO and t-HCOOH coincide with the location of the dust substructures detected in the continuum emission, away from the the two binaries \citep{Maureira2020}.  
Also, the CH$_{3}$OH emission and the HCOOCH$_{3}$ molecular emission detected by \citet{Oya2016} likely originate in the ring-like region with a radius of 50 au surrounding the protostar, based on their kinematic structures \citep{Oya2020}, with temperatures rising to more than 300 K. 
The evidence presented here (dust opacity, compactness of the three components, centroid analysis) would seem to point to the hypothesis of a shocked region due to the large scale streaming material feeding the proto-binary system (see Fig. \ref{fig:metvel}).
However, since the beam of our observations is relatively large with respect to the supposed shock structures, this hypothesis will need to be confirmed by future high-angular resolution observations.

\section{Conclusions}

We have analysed the emission of methanol measured towards the Class 0/I protostar system \bhb and obtained within the ALMA FAUST large program. The Band 3 and Band 6 continuum images show that the dust continuum emission is faint and it extends beyond the northwest-southeast envelope previously detected by \citet{Alves2019}. A series of methanol transitions are detected towards this source and reveal the presence of hot methanol gas arising from three different velocity components centred at V$_{LSR}$ = -2, 2.8 and 9.9 km s$^{-1}$. The overall emission of hot methanol peaks at the location of the circumbinary disk. The moment 1 maps of the methanol emission show a velocity gradient with the blue-shifted gas located towards the northwest and the red-shifted gas peaking towards the southeast, consistent with the gas kinematics previously detected in H$_2$CO \citep[][]{Alves2019}. Non-LTE analysis shows that the methanol emission is compact ($\sim$ 0.15$^{\prime\prime}$), the gas is hot (100--140 K) and dense (2 $\times$ 10$^6$-- 2 $\times$ 10$^7$ cm$^{-3}$) and the inferred methanol column densities are a few 10$^{18}$ cm$^{-2}$. Although the binary system, \bhb A and \bhb B,  cannot be resolved in the FAUST images, the 2D Gaussian centroid map of the hot methanol line at 243.916 GHz shows that this emission arises from the region associated with \bhb A and North-West and South-East from \bhb B. This could either be due to the fact that the dust emission is very optically thick towards the hot corinos and/or the hot methanol gas is associated with shocked gas within an accretion streamer falling onto source \bhb B. This idea is supported by the fact that the methanol emission is compact, located nearby sources \bhb A and B, within the infalling CO gas detected at extreme velocities $<-$1.5 km s$^{-1}$ and $>$9 km s$^{-1}$. However, higher angular resolution observations are needed to disentangle between these two possibilities.

\begin{acknowledgements}
This project has received funding within the European Union's Horizon 2020 research and innovation program from the European Research Council (ERC) for the project {\it The Dawn of Organic Chemistry} (DOC), grant agreement No 741002, and from the Marie Sklodowska-Curie for the project {\it Astro-Chemical Origins} (ACO), grant agreement No 811312. I.J.-S. has received partial support from the Spanish State Research Agency (project number PID2019-105552RB-C41). D.J.\ is supported by NRC Canada and by an NSERC Discovery Grant. SY thanks the support by Grant-in-Aids from Ministry of Education, Culture, Sports, Science, and Technologies of Japan (18H05222). 
This paper makes use of the following ALMA data: ADS/JAO.ALMA\#2018.1.01205.L. ALMA is a partnership of ESO (representing its member states), NSF (USA) and NINS (Japan), together with NRC (Canada), MOST and ASIAA (Taiwan), and KASI (Republic of Korea), in cooperation with the Republic of Chile. The Joint ALMA Observatory is operated by ESO, AUI/NRAO and NAOJ. 
The National Radio Astronomy Observatory is a facility of the National Science Foundation operated under cooperative agreement by Associated Universities, Inc. 

\end{acknowledgements}

\bibliographystyle{aa}
\bibliography{Methanol-BHB_CV.bbl}

\begin{thebibliography}{90}
\expandafter\ifx\csname natexlab\endcsname\relax\def\natexlab#1{#1}\fi

\bibitem[{{Agurto-Gangas} {et~al.}(2019){Agurto-Gangas}, {Pineda},
  {Sz{\H{u}}cs}, {Testi}, {Tazzari}, {Miotello}, {Caselli}, {Dunham},
  {Stephens}, \& {Bourke}}]{Agurto2019}
{Agurto-Gangas}, C., {Pineda}, J.~E., {Sz{\H{u}}cs}, L., {et~al.} 2019, \aap,
  623, A147

\bibitem[{{Aikawa} {et~al.}(2020){Aikawa}, {Furuya}, {Yamamoto}, \&
  {Sakai}}]{Aikawa2020}
{Aikawa}, Y., {Furuya}, K., {Yamamoto}, S., \& {Sakai}, N. 2020, \apj, 897, 110

\bibitem[{{Alves} {et~al.}(2019){Alves}, {Caselli}, {Girart}, {Segura-Cox},
  {Franco}, {Schmiedeke}, \& {Zhao}}]{Alves2019}
{Alves}, F.~O., {Caselli}, P., {Girart}, J.~M., {et~al.} 2019, Science, 366, 90

\bibitem[{{Alves} {et~al.}(2020){Alves}, {Cleeves}, {Girart}, {Zhu}, {Franco},
  {Zurlo}, \& {Caselli}}]{Alves2020}
{Alves}, F.~O., {Cleeves}, L.~I., {Girart}, J.~M., {et~al.} 2020, \apjl, 904,
  L6

\bibitem[{{Alves} {et~al.}(2017){Alves}, {Girart}, {Caselli}, {Franco}, {Zhao},
  {Vlemmings}, {Evans}, \& {Ricci}}]{Alves2017}
{Alves}, F.~O., {Girart}, J.~M., {Caselli}, P., {et~al.} 2017, \aap, 603, L3

\bibitem[{{Alves} {et~al.}(2018){Alves}, {Girart}, {Padovani}, {Galli},
  {Franco}, {Caselli}, {Vlemmings}, {Zhang}, \& {Wiesemeyer}}]{Alves2018}
{Alves}, F.~O., {Girart}, J.~M., {Padovani}, M., {et~al.} 2018, \aap, 616, A56

\bibitem[{{Andr\'e} {et~al.}(1993){Andr\'e}, {Ward-Thompson}, \&
  {Barsony}}]{Andre1993}
{Andr\'e}, P., {Ward-Thompson}, D., \& {Barsony}, M. 1993, \apj, 406, 122

\bibitem[{{Bachiller} {et~al.}(1995){Bachiller}, {Liechti}, {Walmsley}, \&
  {Colomer}}]{Bachiller1995}
{Bachiller}, R., {Liechti}, S., {Walmsley}, C.~M., \& {Colomer}, F. 1995, \aap,
  295, L51

\bibitem[{{Bacmann} {et~al.}(2012){Bacmann}, {Taquet}, {Faure}, {Kahane}, \&
  {Ceccarelli}}]{Bacmann2012}
{Bacmann}, A., {Taquet}, V., {Faure}, A., {Kahane}, C., \& {Ceccarelli}, C.
  2012, \aap, 541, L12

\bibitem[{{Balucani} {et~al.}(2015){Balucani}, {Ceccarelli}, \&
  {Taquet}}]{balucani2015}
{Balucani}, N., {Ceccarelli}, C., \& {Taquet}, V. 2015, \mnras, 449, L16

\bibitem[{{Bertin} {et~al.}(2016){Bertin}, {Romanzin}, {Doronin}, {Philippe},
  {Jeseck}, {Ligterink}, {Linnartz}, {Michaut}, \& {Fillion}}]{Bertin2016}
{Bertin}, M., {Romanzin}, C., {Doronin}, M., {et~al.} 2016, \apjl, 817, L12

\bibitem[{{Bizzocchi} {et~al.}(2014){Bizzocchi}, {Caselli}, {Spezzano}, \&
  {Leonardo}}]{Bizzocchi2014}
{Bizzocchi}, L., {Caselli}, P., {Spezzano}, S., \& {Leonardo}, E. 2014, \aap,
  569, A27

\bibitem[{{Bottinelli} {et~al.}(2004){Bottinelli}, {Ceccarelli}, {Lefloch},
  {Williams}, {Castets}, {Caux}, {Cazaux}, {Maret}, {Parise}, \&
  {Tielens}}]{Bottinelli2004}
{Bottinelli}, S., {Ceccarelli}, C., {Lefloch}, B., {et~al.} 2004, \apj, 615,
  354

\bibitem[{{Bouvier} {et~al.}(2021){Bouvier}, {L{\'o}pez-Sepulcre},
  {Ceccarelli}, {Sakai}, {Yamamoto}, \& {Yang}}]{Bouvier2021}
{Bouvier}, M., {L{\'o}pez-Sepulcre}, A., {Ceccarelli}, C., {et~al.} 2021, \aap,
  653, A117

\bibitem[{{Brooke} {et~al.}(2007){Brooke}, {Huard}, {Bourke}, {Boogert},
  {Allen}, {Blake}, {Evans}, {Harvey}, {Koerner}, {Mundy}, {Myers}, {Padgett},
  {Sargent}, {Stapelfeldt}, {van Dishoeck}, {Chapman}, {Cieza}, {Dunham},
  {Lai}, {Porras}, {Spiesman}, {Teuben}, {Young}, {Wahhaj}, \&
  {Lee}}]{Brooke2007}
{Brooke}, T.~Y., {Huard}, T.~L., {Bourke}, T.~L., {et~al.} 2007, \apj, 655, 364

\bibitem[{{Caselli} \& {Ceccarelli}(2012)}]{CaselliCeccarelli2012}
{Caselli}, P. \& {Ceccarelli}, C. 2012, \aapr, 20, 56

\bibitem[{{Caselli} {et~al.}(1997){Caselli}, {Hartquist}, \&
  {Havnes}}]{Caselli1997C}
{Caselli}, P., {Hartquist}, T.~W., \& {Havnes}, O. 1997, \aap, 322, 296

\bibitem[{{Cazaux} {et~al.}(2003){Cazaux}, {Tielens}, {Ceccarelli}, {Castets},
  {Wakelam}, {Caux}, {Parise}, \& {Teyssier}}]{Cazaux2003}
{Cazaux}, S., {Tielens}, A.~G.~G.~M., {Ceccarelli}, C., {et~al.} 2003, \apjl,
  593, L51

\bibitem[{{Ceccarelli} {et~al.}(2017){Ceccarelli}, {Caselli}, {Fontani},
  {Neri}, {L{\'o}pez-Sepulcre}, {Codella}, {Feng}, {Jim{\'e}nez-Serra},
  {Lefloch}, {Pineda}, {Vastel}, {Alves}, {Bachiller}, {Balucani}, {Bianchi},
  {Bizzocchi}, {Bottinelli}, {Caux}, {Chac{\'o}n-Tanarro}, {Choudhury},
  {Coutens}, {Dulieu}, {Favre}, {Hily-Blant}, {Holdship}, {Kahane}, {Jaber
  Al-Edhari}, {Laas}, {Ospina}, {Oya}, {Podio}, {Pon}, {Punanova}, {Quenard},
  {Rimola}, {Sakai}, {Sims}, {Spezzano}, {Taquet}, {Testi}, {Theul{\'e}},
  {Ugliengo}, {Vasyunin}, {Viti}, {Wiesenfeld}, \& {Yamamoto}}]{Ceccarelli2017}
{Ceccarelli}, C., {Caselli}, P., {Fontani}, F., {et~al.} 2017, \apj, 850, 176

\bibitem[{{Ceccarelli} {et~al.}(2007){Ceccarelli}, {Caselli}, {Herbst},
  {Tielens}, \& {Caux}}]{Ceccarelli2007}
{Ceccarelli}, C., {Caselli}, P., {Herbst}, E., {Tielens}, A.~G.~G.~M., \&
  {Caux}, E. 2007, in Protostars and Planets V, ed. B.~{Reipurth}, D.~{Jewitt},
  \& K.~{Keil}, 47

\bibitem[{{Ceccarelli} {et~al.}(2000{\natexlab{a}}){Ceccarelli}, {Castets},
  {Caux}, {Hollenbach}, {Loinard}, {Molinari}, \& {Tielens}}]{Ceccarelli2000a}
{Ceccarelli}, C., {Castets}, A., {Caux}, E., {et~al.} 2000{\natexlab{a}}, \aap,
  355, 1129

\bibitem[{{Ceccarelli} {et~al.}(2000{\natexlab{b}}){Ceccarelli}, {Loinard},
  {Castets}, {Faure}, \& {Lefloch}}]{Ceccarelli2000b}
{Ceccarelli}, C., {Loinard}, L., {Castets}, A., {Faure}, A., \& {Lefloch}, B.
  2000{\natexlab{b}}, \aap, 362, 1122

\bibitem[{{Ceccarelli} {et~al.}(2003){Ceccarelli}, {Maret}, {Tielens},
  {Castets}, \& {Caux}}]{Ceccarelli2003}
{Ceccarelli}, C., {Maret}, S., {Tielens}, A.~G.~G.~M., {Castets}, A., \&
  {Caux}, E. 2003, \aap, 410, 587

\bibitem[{{Chahine} {et~al.}(2022){Chahine}, {L{\'o}pez-Sepulcre}, {Neri},
  {Ceccarelli}, {Mercimek}, {Codella}, {Bouvier}, {Bianchi}, {Favre}, {Podio},
  {Alves}, {Sakai}, \& {Yamamoto}}]{Chahine2022}
{Chahine}, L., {L{\'o}pez-Sepulcre}, A., {Neri}, R., {et~al.} 2022, \aap, 657,
  A78

\bibitem[{{Charnley} {et~al.}(1992){Charnley}, {Tielens}, \&
  {Millar}}]{Charnley1992}
{Charnley}, S.~B., {Tielens}, A.~G.~G.~M., \& {Millar}, T.~J. 1992, \apjl, 399,
  L71

\bibitem[{{Codella} {et~al.}(2020){Codella}, {Ceccarelli}, {Bianchi},
  {Balucani}, {Podio}, {Caselli}, {Feng}, {Lefloch}, {L{\'o}pez-Sepulcre},
  {Neri}, {Spezzano}, \& {De Simone}}]{Codella2020}
{Codella}, C., {Ceccarelli}, C., {Bianchi}, E., {et~al.} 2020, \aap, 635, A17

\bibitem[{{Codella} {et~al.}(2021){Codella}, {Ceccarelli}, {Chandler}, {Sakai},
  {Yamamoto}, \& {FAUST Team}}]{Codella2021}
{Codella}, C., {Ceccarelli}, C., {Chandler}, C., {et~al.} 2021, Frontiers in
  Astronomy and Space Sciences, 8, 227

\bibitem[{{Condon}(1997)}]{Condon1997}
{Condon}, J.~J. 1997, \pasp, 109, 166

\bibitem[{{Condon} {et~al.}(1998){Condon}, {Cotton}, {Greisen}, {Yin},
  {Perley}, {Taylor}, \& {Broderick}}]{Condon1998}
{Condon}, J.~J., {Cotton}, W.~D., {Greisen}, E.~W., {et~al.} 1998, \aj, 115,
  1693

\bibitem[{{Covey} {et~al.}(2010){Covey}, {Lada}, {Rom{\'a}n-Z{\'u}{\~n}iga},
  {Muench}, {Forbrich}, \& {Ascenso}}]{Covey2010}
{Covey}, K.~R., {Lada}, C.~J., {Rom{\'a}n-Z{\'u}{\~n}iga}, C., {et~al.} 2010,
  \apj, 722, 971

\bibitem[{{De Simone} {et~al.}(2020){De Simone}, {Ceccarelli}, {Codella},
  {Svoboda}, {Chandler}, {Bouvier}, {Yamamoto}, {Sakai}, {Caselli}, {Favre},
  {Loinard}, {Lefloch}, {Liu}, {L{\'o}pez-Sepulcre}, {Pineda}, {Taquet}, \&
  {Testi}}]{DeSimone2020}
{De Simone}, M., {Ceccarelli}, C., {Codella}, C., {et~al.} 2020, \apjl, 896, L3

\bibitem[{{Duarte-Cabral} {et~al.}(2012){Duarte-Cabral}, {Chrysostomou},
  {Peretto}, {Fuller}, {Matthews}, {Schieven}, \& {Davis}}]{Duarte2012}
{Duarte-Cabral}, A., {Chrysostomou}, A., {Peretto}, N., {et~al.} 2012, \aap,
  543, A140

\bibitem[{{Dubernet} {et~al.}(2012){Dubernet}, {Nenadovic}, \&
  {Doronin}}]{Dubernet2012}
{Dubernet}, M., {Nenadovic}, L., \& {Doronin}, N. 2012, in Astronomical Society
  of the Pacific Conference Series, Vol. 461, Astronomical Data Analysis
  Software and Systems XXI, ed. P.~{Ballester}, D.~{Egret}, \& N.~P.~F.
  {Lorente}, 335

\bibitem[{{Dubernet} {et~al.}(2013){Dubernet}, {Alexander}, {Ba},
  {Balakrishnan}, {Balan{\c{c}}a}, {Ceccarelli}, {Cernicharo}, {Daniel},
  {Dayou}, {Doronin}, {Dumouchel}, {Faure}, {Feautrier}, {Flower}, {Grosjean},
  {Halvick}, {K{\l}os}, {Lique}, {McBane}, {Marinakis}, {Moreau}, {Moszynski},
  {Neufeld}, {Roueff}, {Schilke}, {Spielfiedel}, {Stancil}, {Stoecklin},
  {Tennyson}, {Yang}, {Vasserot}, \& {Wiesenfeld}}]{Dubernet2013}
{Dubernet}, M.~L., {Alexander}, M.~H., {Ba}, Y.~A., {et~al.} 2013, \aap, 553,
  A50

\bibitem[{{Dzib} {et~al.}(2018){Dzib}, {Loinard}, {Ortiz-Le{\'o}n},
  {Rodr{\'\i}guez}, \& {Galli}}]{Dzib2018}
{Dzib}, S.~A., {Loinard}, L., {Ortiz-Le{\'o}n}, G.~N., {Rodr{\'\i}guez}, L.~F.,
  \& {Galli}, P. A.~B. 2018, \apj, 867, 151

\bibitem[{{Favre} {et~al.}(2018){Favre}, {Fedele}, {Semenov}, {Parfenov},
  {Codella}, {Ceccarelli}, {Bergin}, {Chapillon}, {Testi}, {Hersant},
  {Lefloch}, {Fontani}, {Blake}, {Cleeves}, {Qi}, {Schwarz}, \&
  {Taquet}}]{Favre2018}
{Favre}, C., {Fedele}, D., {Semenov}, D., {et~al.} 2018, \apjl, 862, L2

\bibitem[{{Ferrero} {et~al.}(2020){Ferrero}, {Zamirri}, {Ceccarelli}, {Witzel},
  {Rimola}, \& {Ugliengo}}]{Ferrero2020}
{Ferrero}, S., {Zamirri}, L., {Ceccarelli}, C., {et~al.} 2020, \apj, 904, 11

\bibitem[{{Flower} \& {Pineau des Forets}(1995)}]{Flower1995}
{Flower}, D.~R. \& {Pineau des Forets}, G. 1995, \mnras, 275, 1049

\bibitem[{{Forbrich} {et~al.}(2009){Forbrich}, {Lada}, {Muench}, {Alves}, \&
  {Lombardi}}]{Forbrich2009}
{Forbrich}, J., {Lada}, C.~J., {Muench}, A.~A., {Alves}, J., \& {Lombardi}, M.
  2009, \apj, 704, 292

\bibitem[{{Forbrich} {et~al.}(2010){Forbrich}, {Posselt}, {Covey}, \&
  {Lada}}]{Forbrich2010}
{Forbrich}, J., {Posselt}, B., {Covey}, K.~R., \& {Lada}, C.~J. 2010, \apj,
  719, 691

\bibitem[{{Friberg} {et~al.}(1988){Friberg}, {Madden}, {Hjalmarson}, \&
  {Irvine}}]{Friberg1988}
{Friberg}, P., {Madden}, S.~C., {Hjalmarson}, A., \& {Irvine}, W.~M. 1988,
  \aap, 195, 281

\bibitem[{{Garrod} {et~al.}(2008){Garrod}, {Widicus Weaver}, \&
  {Herbst}}]{Garrod2008}
{Garrod}, R.~T., {Widicus Weaver}, S.~L., \& {Herbst}, E. 2008, \apj, 682, 283

\bibitem[{{Gibb} {et~al.}(2000){Gibb}, {Nummelin}, {Irvine}, {Whittet}, \&
  {Bergman}}]{Gibb2000}
{Gibb}, E., {Nummelin}, A., {Irvine}, W.~M., {Whittet}, D.~C.~B., \& {Bergman},
  P. 2000, \apj, 545, 309

\bibitem[{{Goldsmith}(2001)}]{Goldsmith2001}
{Goldsmith}, P.~F. 2001, \apj, 557, 736

\bibitem[{{Goldsmith} \& {Langer}(1999)}]{Goldsmith1999}
{Goldsmith}, P.~F. \& {Langer}, W.~D. 1999, \apj, 517, 209

\bibitem[{{Hara} {et~al.}(2013){Hara}, {Shimajiri}, {Tsukagoshi}, {Kurono},
  {Saigo}, {Nakamura}, {Saito}, {Wilner}, \& {Kawabe}}]{Hara13}
{Hara}, C., {Shimajiri}, Y., {Tsukagoshi}, T., {et~al.} 2013, \apj, 771, 128

\bibitem[{{Jaber} {et~al.}(2014){Jaber}, {Ceccarelli}, {Kahane}, \&
  {Caux}}]{Jaber2014}
{Jaber}, A.~A., {Ceccarelli}, C., {Kahane}, C., \& {Caux}, E. 2014, \apj, 791,
  29

\bibitem[{{Jim{\'e}nez-Serra} {et~al.}(2008){Jim{\'e}nez-Serra}, {Caselli},
  {Mart{\'\i}n-Pintado}, \& {Hartquist}}]{Jimenez-Serra2008}
{Jim{\'e}nez-Serra}, I., {Caselli}, P., {Mart{\'\i}n-Pintado}, J., \&
  {Hartquist}, T.~W. 2008, \aap, 482, 549

\bibitem[{{Jim{\'e}nez-Serra} {et~al.}(2016){Jim{\'e}nez-Serra}, {Vasyunin},
  {Caselli}, {Marcelino}, {Billot}, {Viti}, {Testi}, {Vastel}, {Lefloch}, \&
  {Bachiller}}]{Jimenez2016}
{Jim{\'e}nez-Serra}, I., {Vasyunin}, A.~I., {Caselli}, P., {et~al.} 2016,
  \apjl, 830, L6

\bibitem[{{Jim{\'e}nez-Serra} {et~al.}(2021){Jim{\'e}nez-Serra}, {Vasyunin},
  {Spezzano}, {Caselli}, {Cosentino}, \& {Viti}}]{Jimenez-Serra2021}
{Jim{\'e}nez-Serra}, I., {Vasyunin}, A.~I., {Spezzano}, S., {et~al.} 2021,
  \apj, 917, 44

\bibitem[{{Jin} \& {Garrod}(2020)}]{Jin2020}
{Jin}, M. \& {Garrod}, R.~T. 2020, \apjs, 249, 26

\bibitem[{{J{\o}rgensen} {et~al.}(2016){J{\o}rgensen}, {van der Wiel},
  {Coutens}, {Lykke}, {M{\"u}ller}, {van Dishoeck}, {Calcutt}, {Bjerkeli},
  {Bourke}, {Drozdovskaya}, {Favre}, {Fayolle}, {Garrod}, {Jacobsen},
  {{\"O}berg}, {Persson}, \& {Wampfler}}]{Jorgensen2016}
{J{\o}rgensen}, J.~K., {van der Wiel}, M.~H.~D., {Coutens}, A., {et~al.} 2016,
  \aap, 595, A117

\bibitem[{{Lee} {et~al.}(2019){Lee}, {Codella}, {Li}, \& {Liu}}]{Lee2019}
{Lee}, C.-F., {Codella}, C., {Li}, Z.-Y., \& {Liu}, S.-Y. 2019, \apj, 876, 63

\bibitem[{{Maret} {et~al.}(2005){Maret}, {Ceccarelli}, {Tielens}, {Caux},
  {Lefloch}, {Faure}, {Castets}, \& {Flower}}]{Maret2005}
{Maret}, S., {Ceccarelli}, C., {Tielens}, A.~G.~G.~M., {et~al.} 2005, \aap,
  442, 527

\bibitem[{{Matsumoto} {et~al.}(2019){Matsumoto}, {Saigo}, \&
  {Takakuwa}}]{Matsumoto2019}
{Matsumoto}, T., {Saigo}, K., \& {Takakuwa}, S. 2019, \apj, 871, 36

\bibitem[{{Maureira} {et~al.}(2020){Maureira}, {Pineda}, {Segura-Cox},
  {Caselli}, {Testi}, {Lodato}, {Loinard}, \&
  {Hern{\'a}ndez-G{\'o}mez}}]{Maureira2020}
{Maureira}, M.~J., {Pineda}, J.~E., {Segura-Cox}, D.~M., {et~al.} 2020, \apj,
  897, 59

\bibitem[{{Milam} {et~al.}(2005){Milam}, {Savage}, {Brewster}, {Ziurys}, \&
  {Wyckoff}}]{Milam2005}
{Milam}, S.~N., {Savage}, C., {Brewster}, M.~A., {Ziurys}, L.~M., \& {Wyckoff},
  S. 2005, \apj, 634, 1126

\bibitem[{{Minissale} {et~al.}(2016){Minissale}, {Dulieu}, {Cazaux}, \&
  {Hocuk}}]{Minissale2016}
{Minissale}, M., {Dulieu}, F., {Cazaux}, S., \& {Hocuk}, S. 2016, \aap, 585,
  A24

\bibitem[{{M{\"u}ller} {et~al.}(2005){M{\"u}ller}, {Schl{\"o}der}, {Stutzki},
  \& {Winnewisser}}]{Muller2005}
{M{\"u}ller}, H. S.~P., {Schl{\"o}der}, F., {Stutzki}, J., \& {Winnewisser}, G.
  2005, Journal of Molecular Structure, 742, 215

\bibitem[{{Nagy} {et~al.}(2019){Nagy}, {Spezzano}, {Caselli}, {Vasyunin},
  {Tafalla}, {Bizzocchi}, {Prudenzano}, \& {Redaelli}}]{Nagy2019}
{Nagy}, Z., {Spezzano}, S., {Caselli}, P., {et~al.} 2019, \aap, 630, A136

\bibitem[{{{\"O}berg} {et~al.}(2015){{\"O}berg}, {Guzm{\'a}n}, {Furuya}, {Qi},
  {Aikawa}, {Andrews}, {Loomis}, \& {Wilner}}]{Oberg2015}
{{\"O}berg}, K.~I., {Guzm{\'a}n}, V.~V., {Furuya}, K., {et~al.} 2015, \nat,
  520, 198

\bibitem[{{{\"O}berg} {et~al.}(2021){{\"O}berg}, {Guzm{\'a}n}, {Walsh},
  {Aikawa}, {Bergin}, {Law}, {Loomis}, {Alarc{\'o}n}, {Andrews}, {Bae},
  {Bergner}, {Boehler}, {Booth}, {Bosman}, {Calahan}, {Cataldi}, {Cleeves},
  {Czekala}, {Furuya}, {Huang}, {Ilee}, {Kurtovic}, {Le Gal}, {Liu}, {Long},
  {M{\'e}nard}, {Nomura}, {P{\'e}rez}, {Qi}, {Schwarz}, {Sierra}, {Teague},
  {Tsukagoshi}, {Yamato}, {van't Hoff}, {Waggoner}, {Wilner}, \&
  {Zhang}}]{Oberg2021}
{{\"O}berg}, K.~I., {Guzm{\'a}n}, V.~V., {Walsh}, C., {et~al.} 2021, \apjs,
  257, 1

\bibitem[{{Onishi} {et~al.}(1999){Onishi}, {Kawamura}, {Abe}, {Yamaguchi},
  {Saito}, {Moriguchi}, {Mizuno}, {Ogawa}, \& {Fukui}}]{Onishi1999}
{Onishi}, T., {Kawamura}, A., {Abe}, R., {et~al.} 1999, \pasj, 51, 871

\bibitem[{{Oya} {et~al.}(2016){Oya}, {Sakai}, {L{\'o}pez-Sepulcre}, {Watanabe},
  {Ceccarelli}, {Lefloch}, {Favre}, \& {Yamamoto}}]{Oya2016}
{Oya}, Y., {Sakai}, N., {L{\'o}pez-Sepulcre}, A., {et~al.} 2016, \apj, 824, 88

\bibitem[{{Oya} {et~al.}(2017){Oya}, {Sakai}, {Watanabe}, {Higuchi}, {Hirota},
  {L{\'o}pez-Sepulcre}, {Sakai}, {Aikawa}, {Ceccarelli}, {Lefloch}, {Caux},
  {Vastel}, {Kahane}, \& {Yamamoto}}]{Oya2017}
{Oya}, Y., {Sakai}, N., {Watanabe}, Y., {et~al.} 2017, \apj, 837, 174

\bibitem[{{Oya} \& {Yamamoto}(2020)}]{Oya2020}
{Oya}, Y. \& {Yamamoto}, S. 2020, \apj, 904, 185

\bibitem[{{Pickett} {et~al.}(1998){Pickett}, {Poynter}, {Cohen}, {Delitsky},
  {Pearson}, \& {M{\"u}ller}}]{Pickett1998}
{Pickett}, H.~M., {Poynter}, R.~L., {Cohen}, E.~A., {et~al.} 1998, \jqsrt, 60,
  883

\bibitem[{Pineda {et~al.}(2022)Pineda, Arzoumanian, André, Friesen, Zavagno,
  Clarke, Inoue, Chen, Lee, Soler, \& Kuffmeier}]{Pineda2022}
Pineda, J., Arzoumanian, D., André, P., {et~al.} 2022, To appear in Protostars
  and Planets VII [\eprint[arXiv]{2205.03935}]

\bibitem[{{Pineda} {et~al.}(2020){Pineda}, {Segura-Cox}, {Caselli},
  {Cunningham}, {Zhao}, {Schmiedeke}, {Maureira}, \& {Neri}}]{Pineda2020}
{Pineda}, J.~E., {Segura-Cox}, D., {Caselli}, P., {et~al.} 2020, Nature
  Astronomy, 4, 1158

\bibitem[{{Punanova} {et~al.}(2022){Punanova}, {Vasyunin}, {Caselli}, {Howard},
  {Spezzano}, {Shirley}, {Scibelli}, \& {Harju}}]{Punanova2021}
{Punanova}, A., {Vasyunin}, A., {Caselli}, P., {et~al.} 2022, \apj, 927, 213

\bibitem[{{Rabli} \& {Flower}(2010)}]{Rabli2010}
{Rabli}, D. \& {Flower}, D.~R. 2010, \mnras, 406, 95

\bibitem[{{Rathborne} {et~al.}(2008){Rathborne}, {Lada}, {Muench}, {Alves}, \&
  {Lombardi}}]{Rathborne2008}
{Rathborne}, J.~M., {Lada}, C.~J., {Muench}, A.~A., {Alves}, J.~F., \&
  {Lombardi}, M. 2008, \apjs, 174, 396

\bibitem[{{Redaelli} {et~al.}(2017){Redaelli}, {Alves}, {Caselli}, {Pineda},
  {Friesen}, {Chac{\'o}n-Tanarro}, {Matzner}, {Ginsburg}, {Rosolowsky},
  {Keown}, {Offner}, {Di Francesco}, {Kirk}, {Myers}, {Hacar}, {Cimatti},
  {Chen}, {Chen}, {Lee}, \& {Seo}}]{Redaelli2017}
{Redaelli}, E., {Alves}, F.~O., {Caselli}, P., {et~al.} 2017, \apj, 850, 202

\bibitem[{{Riaz} {et~al.}(2009){Riaz}, {Mart{\'\i}n}, {Bouy}, \&
  {Tata}}]{Riaz2009}
{Riaz}, B., {Mart{\'\i}n}, E.~L., {Bouy}, H., \& {Tata}, R. 2009, \apj, 700,
  1541

\bibitem[{{Rimola} {et~al.}(2014){Rimola}, {Taquet}, {Ugliengo}, {Balucani}, \&
  {Ceccarelli}}]{Rimola2014}
{Rimola}, A., {Taquet}, V., {Ugliengo}, P., {Balucani}, N., \& {Ceccarelli}, C.
  2014, \aap, 572, A70

\bibitem[{{Rom{\'a}n-Z{\'u}{\~n}iga} {et~al.}(2010){Rom{\'a}n-Z{\'u}{\~n}iga},
  {Alves}, {Lada}, \& {Lombardi}}]{Roman2010}
{Rom{\'a}n-Z{\'u}{\~n}iga}, C.~G., {Alves}, J.~F., {Lada}, C.~J., \&
  {Lombardi}, M. 2010, \apj, 725, 2232

\bibitem[{{Sakai} \& {Yamamoto}(2013)}]{Sakai2013}
{Sakai}, N. \& {Yamamoto}, S. 2013, Chemical Reviews, 113, 8981

\bibitem[{{Sandell} {et~al.}(2021){Sandell}, {Reipurth}, {Vacca}, \&
  {Bajaj}}]{Sandell2021}
{Sandell}, G., {Reipurth}, B., {Vacca}, W.~D., \& {Bajaj}, N.~S. 2021, \apj,
  920, 7

\bibitem[{{Scibelli} \& {Shirley}(2020)}]{Scibelli2020}
{Scibelli}, S. \& {Shirley}, Y. 2020, \apj, 891, 73

\bibitem[{{Shingledecker} {et~al.}(2018){Shingledecker}, {Tennis}, {Le Gal}, \&
  {Herbst}}]{Shingledecker2018}
{Shingledecker}, C.~N., {Tennis}, J., {Le Gal}, R., \& {Herbst}, E. 2018, \apj,
  861, 20

\bibitem[{{Spezzano} {et~al.}(2020){Spezzano}, {Caselli}, {Pineda},
  {Bizzocchi}, {Prudenzano}, \& {Nagy}}]{Spezzano2020}
{Spezzano}, S., {Caselli}, P., {Pineda}, J.~E., {et~al.} 2020, \aap, 643, A60

\bibitem[{{Spezzano} {et~al.}(2022){Spezzano}, {Fuente}, {Caselli}, {Vasyunin},
  {Navarro-Almaida}, {Rodr{\'\i}guez-Baras}, {Punanova}, {Vastel}, \&
  {Wakelam}}]{Spezzano2022}
{Spezzano}, S., {Fuente}, A., {Caselli}, P., {et~al.} 2022, \aap, 657, A10

\bibitem[{{Taquet} {et~al.}(2016){Taquet}, {Wirstr{\"o}m}, \&
  {Charnley}}]{Taquet2016}
{Taquet}, V., {Wirstr{\"o}m}, E.~S., \& {Charnley}, S.~B. 2016, \apj, 821, 46

\bibitem[{{Testi} {et~al.}(2014){Testi}, {Birnstiel}, {Ricci}, {Andrews},
  {Blum}, {Carpenter}, {Dominik}, {Isella}, {Natta}, {Williams}, \&
  {Wilner}}]{Testi2014}
{Testi}, L., {Birnstiel}, T., {Ricci}, L., {et~al.} 2014, in Protostars and
  Planets VI, ed. H.~{Beuther}, R.~S. {Klessen}, C.~P. {Dullemond}, \&
  T.~{Henning}, 339

\bibitem[{{Vastel} {et~al.}(2014){Vastel}, {Ceccarelli}, {Lefloch}, \&
  {Bachiller}}]{Vastel2014}
{Vastel}, C., {Ceccarelli}, C., {Lefloch}, B., \& {Bachiller}, R. 2014, \apjl,
  795, L2

\bibitem[{{Vasyunin} {et~al.}(2017){Vasyunin}, {Caselli}, {Dulieu}, \&
  {Jim{\'e}nez-Serra}}]{Vasyunin2017}
{Vasyunin}, A.~I., {Caselli}, P., {Dulieu}, F., \& {Jim{\'e}nez-Serra}, I.
  2017, \apj, 842, 33

\bibitem[{{Wakelam} {et~al.}(2021){Wakelam}, {Dartois}, {Chabot}, {Spezzano},
  {Navarro-Almaida}, {Loison}, \& {Fuente}}]{Wakelam2021}
{Wakelam}, V., {Dartois}, E., {Chabot}, M., {et~al.} 2021, \aap, 652, A63

\bibitem[{{Walsh} {et~al.}(2016){Walsh}, {Loomis}, {{\"O}berg}, {Kama}, {van 't
  Hoff}, {Millar}, {Aikawa}, {Herbst}, {Widicus Weaver}, \&
  {Nomura}}]{Walsh2016}
{Walsh}, C., {Loomis}, R.~A., {{\"O}berg}, K.~I., {et~al.} 2016, \apjl, 823,
  L10

\bibitem[{{Watanabe} \& {Kouchi}(2002)}]{Watanabe2002}
{Watanabe}, N. \& {Kouchi}, A. 2002, \apjl, 571, L173

\bibitem[{{Yang} {et~al.}(2021){Yang}, {Sakai}, {Zhang}, {Murillo}, {Zhang},
  {Higuchi}, {Zeng}, {L{\'o}pez-Sepulcre}, {Yamamoto}, {Lefloch}, {Bouvier},
  {Ceccarelli}, {Hirota}, {Imai}, {Oya}, {Sakai}, \& {Watanabe}}]{Yang2021}
{Yang}, Y.-L., {Sakai}, N., {Zhang}, Y., {et~al.} 2021, \apj, 910, 20

\end{thebibliography}

\end{document}